\documentclass[aps,pra,nofootinbib,twocolumn,amsfonts,showpacs,longbibliography,superscriptaddress]{revtex4-1}
\pdfoutput=1 
\usepackage{hyperref}
\usepackage[T1]{fontenc}
\usepackage[normalem]{ulem}
\usepackage{standalone}
\usepackage{graphicx}			
\usepackage{amsmath}
\usepackage{amssymb}
\usepackage{amsthm}
\usepackage[capitalize]{cleveref}
\usepackage{color}
\usepackage{mathtools}
\usepackage{tikz-cd}
\usepackage{xfrac}
\usepackage{bm}
\usetikzlibrary{arrows}
\usetikzlibrary{decorations.markings}
\usetikzlibrary{shapes.geometric}
\usetikzlibrary{decorations.pathmorphing}
\tikzset{snake it/.style={decorate, decoration=snake}}

\newcommand{\ket}[1]{\left| #1 \right>} 
\newcommand{\bra}[1]{\left< #1 \right|}

\newcommand{\outerproduct}[2]{|#1\rangle\langle #2|}

\newcommand{\rmi}{\mathrm{i}}
\newcommand{\rmd}{\mathrm{d}}

\newcommand{\rme}{\mathrm{e}}
\newcommand{\tr}{\mathrm{tr}}

\newcommand{\cpt}{\mathbb{Z}_2^{\mathrm{CPT}}}
\newcommand{\ah}{{\hat{a}}}

\usepackage{framed}
\definecolor{shadecolor}{gray}{0.95}

\hypersetup{
    colorlinks=true,
    linkcolor=blue,
    citecolor=blue,
    filecolor=blue,
    urlcolor=blue,
}

\newtheorem{result}{Result}

\newtheorem{consequence}{Consequence}

\newtheorem{definition}{Definition}

\begin{document}

\title{Charge pumps, pivot Hamiltonians and symmetry-protected topological phases}

 	\author{Nick G. Jones}
     \altaffiliation{\href{mailto:nick.jones@maths.ox.ac.uk}{nick.jones@maths.ox.ac.uk} \\ The published version of this article is Phys. Rev. B 112, 165123 (2025); \href{ https://doi.org/10.1103/rtq1-pplf}{https://doi.org/10.1103/rtq1-pplf}.}
     \affiliation{St John’s College and Mathematical Institute, University of Oxford, UK}	
	\author{Ryan Thorngren}
    \affiliation{Mani L. Bhaumik Institute for Theoretical Physics, Department of Physics and Astronomy, University of California, Los Angeles, CA 90095, USA}
    \author{Ruben Verresen}
    \affiliation{Pritzker School of Molecular Engineering, University of Chicago, Chicago, IL 60637, USA}
    \author{Abhishodh Prakash}
    \affiliation{Harish-Chandra Research Institute, Prayagraj (Allahabad) - 211019, India}
\begin{abstract}
Generalised charge pumps are topological obstructions to trivialising loops in the space of symmetric gapped Hamiltonians.
We show that given mild conditions on such pumps, the associated loop has high-symmetry points which must be in distinct symmetry-protected topological (SPT) phases.
To further elucidate the connection between pumps and SPTs, we focus on closed paths, `pivot loops', defined by two Hamiltonians, where the first is unitarily evolved by the second `pivot' Hamiltonian.
While such pivot loops have been studied as entanglers for SPTs, here we explore their connection to pumps.
We construct families of pivot loops which pump charge for various symmetry groups, often leading to SPT phases---including dipole SPTs.
Intriguingly, we find examples where non-trivial pumps do not lead to genuine SPTs but still entangle representation-SPTs (RSPTs).
We use the anomaly associated to the non-trivial pump to explain the a priori `unnecessary' criticality between these RSPTs.
We also find that particularly nice pivot families form circles in Hamiltonian space, which we show is equivalent to the Hamiltonians satisfying the Dolan-Grady relation---known from the study of integrable models.
This additional structure allows us to derive more powerful constraints on the phase diagram.
Natural examples of such circular loops arise from pivoting with the Onsager-integrable chiral clock models, containing the aforementioned RSPT example.
In fact, we show that these Onsager pivots underlie general group cohomology-based pumps in one spatial dimension.
Finally, we recast the above in the language of equivariant families of Hamiltonians and relate the invariants of the pump to the candidate SPTs. We also highlight how certain SPTs arise in cases where the equivariant family is labelled by spaces that are not manifolds.

\vspace{10pt}

\end{abstract}\maketitle

\tableofcontents

\section{Introduction}
The classification of phases of a given family of gapped Hamiltonians corresponds to dividing this family into connected regions. These regions, each corresponding to a single phase of matter, may themselves have a rich topological structure. The simplest probe of this structure comes from loops in the space, corresponding to closed paths of Hamiltonians---non-contractible loops are generalised Thouless charge pumps \cite{Thouless83,Hsin20,Cordova20,Hermele_talk,Shiozaki22,Wen23}. Obstructions to contractibility are gapless loci, that are referred to as diabolical \cite{Hsin20}.  In one spatial dimension, these loops act as pumps since, despite the
bulk state being periodic, there is a net flow of charge that may be observed when the system has boundaries. In higher dimensions it is conjectured that loops are classified by the appropriate generalisations of such charge pumps \cite{Kitaev_talk,Gaiotto19,debray2024longexactsequencesymmetry}.

Non-contractible loops within a single phase arise also in the context of \emph{pivot Hamiltonians} \cite{Tantivasadakarn23,Tantivasadakarn23b,Jones25} and symmetry-protected topological (SPT) phases \cite{Gu09,Pollmann10,Fidkowski10,Turner11,Schuch11,Chen11,Pollmann12,Senthil13,Verresen17}. These loops are unitary paths generated by a pivot Hamiltonian, $\tilde{H}$, taking the form $H_\theta = \rme^{-\rmi \theta \tilde{H}}H_0  \rme^{\rmi \theta \tilde{H}}$. Here $H_0$ is a trivial Hamiltonian and $H_{2\pi}=H_0$. If $H_0$ and $\tilde{H}$ share a symmetry group $\tilde{G}$, then this loop remains in the space of $\tilde{G}$-symmetric gapped Hamiltonians. Of particular interest are cases where $H_0$ and $H_{\pi}$ share a larger symmetry group, $G$, and where $H_\pi$ has non-trivial SPT order for that enhanced symmetry. In such cases, $U=\rme^{-\rmi \pi \tilde{H}}$ is an \emph{SPT entangler} \cite{Chen14,Verresen17,Verresen21,Tantivasadakarn23,Zhang22}. In this work we explore the interplay of pivot loops, pumps and SPT entanglers, and are led to some connections that apply more generally.

\subsection{Motivating example and first result: pumps and SPTs with the Ising pivot}\label{sec:firstresult}
The simplest example of such a pivot loop $H_\theta = \rme^{-\rmi \theta \tilde{H}}H_0  \rme^{\rmi \theta \tilde{H}}$ is constructed by taking the following two spin-$1/2$ Hamiltonians: \begin{align}
H_0=-\frac{1}{4}\sum_j X_j \qquad
\tilde{H} = -\frac{1}{4}\sum_j Z_j Z_{j+1} \ ;
\end{align}
the usual spin-1/2 Ising paramagnet and ferromagnet respectively. We show below that the pivot loop, $H_\theta$, traces out a circle in the space of Hamiltonians given by
\begin{align}
H_\theta &= \frac{H_0 +H_\pi}{2} + \frac{H_0 -H_\pi}{2} \cos(\theta) + \rmi[H_0,\tilde{H}] \sin(\theta)\ ,\label{eq:circleintro}\end{align} 
where \begin{align}
H_{\pi} &= \frac{1}{4} \sum_j Z_{j-1}X_jZ_{j+1}
 \ .
\end{align}
For all $\theta$, the family $H_\theta$ has a $\tilde{G}=\mathbb{Z}_2^P$ spin-flip symmetry generated by $P=\prod_j X_j$. We also have that $H_{2\pi}=H_0$ and that the cluster model \cite{Briegel01} $H_{\pi}$ is an SPT for the enhanced $G=\tilde{G}\times\mathbb{Z}_2^{P_\mathrm{even}}$ symmetry that is shared by $H_0$ and $H_{\pi}$ (the second generator is $P_{\mathrm{even}}=\prod_j X_{2j}$ and is explicitly broken by $\tilde{H}$) \cite{Son11}.

The unitary operator that generates a $2\pi$ pivot is given by
\begin{align}
U_{2\pi}= \rme^{\rmi \frac{\pi}{2}  \sum_j Z_j Z_{j+1} }    = \prod_j  \rme^{\rmi \frac{\pi}{2} Z_j Z_{j+1} }=\prod_j \left(\rmi Z_j Z_{j+1}\right) .
\end{align}
We see a non-trivial dependence on the boundary: if we take $\tilde{H}$ to be periodic then this product is, up to an unimportant phase, the identity operator. If we instead remove the `periodic boundary' term $Z_LZ_1$ in $\tilde{H}$, we end up with  $U_{2\pi}\propto Z_1 Z_L$. Since $Z_j$ is $P$-odd, we see that the pivot Hamiltonian, applied to a state on an open chain, will pump a $\mathbb{Z}_2$-charged operator to each boundary. This is one way to characterise the non-trivial charge pump around the non-contractible pivot loop $H_\theta$ \cite{Hsin20,Hermele_talk,Shiozaki22}.  

In Ref.~\onlinecite{Tantivasadakarn23}, the fact that $H_\pi$ is a $G$-SPT is argued to imply the non-triviality of the pump around the pivot loop. 
A natural question is whether this is an equivalence---does the pump imply a non-trivial SPT at $H_\pi$? In fact, in this particular case we can show the other direction as follows.

\begin{figure}
\includegraphics[]{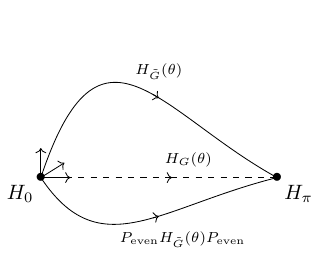}
\caption{Paths used in the argument for showing that a pump implies an SPT in the Ising pivot.
We consider paths
in the space of $\tilde{G}=\mathbb{Z}_2$-symmetric Hamiltonians between the $G=\mathbb{Z}_2\times\mathbb{Z}_2$-symmetric points $H_0$ and $H_\pi$. Assuming the existence of a $G$-symmetric path, then consistency implies that the loop constructed by following first $H_{\tilde{G}}(\theta)$ then the reverse of $P_{\mathrm{even}}H_{\tilde{G}}(\theta)P_{\mathrm{even}}$ must be a trivial $\tilde{G}$-charge pump. The pivot loop $H_\theta$ in \Cref{eq:circleintro} is of this form, which means we can conclude from the non-trivial $\mathbb{Z}_2$ pump that $H_0$ and $H_\pi$ are in distinct $G$-SPT phases.}
\label{fig:paths0}
\end{figure}

For the $G=\mathbb{Z}_2\times\mathbb{Z}_2$ symmetry group under consideration, we can show that if $H_\theta$ pumps a charge whilst going around a full cycle, then $H_0$ and $H_\pi$ must be in distinct SPT phases. Equivalently, if $H_0$ and $H_\pi$ are in the same SPT phase, the pump around $H_\theta$ must be trivial. Indeed, if there were to exist a gapped $G$-symmetric path connecting them, we could rewrite the loop $H_\theta$ as the sum of two loops (see \cref{fig:paths0}). The symmetry properties in turn imply that these loops must pump equal and opposite charges, meaning that the original loop had to be a trivial pump. We first flesh out this argument for product groups, and then in greater generality, in \Cref{sec:pumpSPT,sec:equivariant}. This relatively simple analysis is a hint of the deep connections between SPTs and pumps---laying these out is a goal of the present manuscript.

One unifying feature is the appearance of anomalous symmetries in the study of SPT phases and charge pumps. The Hamiltonian $H_\star = \frac{1}{2}(H_0 +H_\pi)$ is at the centre of the loop $H_\theta$ and is gapless, described by the compactified free boson CFT \cite{Levin12}.
The pump around the loop can be related to an anomalous $\tilde{G}\times U(1)$ symmetry at $H_\star$, while the SPT nature of $H_\pi$ implies an anomalous  $G\times\mathbb{Z}_2$ symmetry \cite{Bultinck19,Tantivasadakarn23}. We will study relations between these anomalies in the context of equivariant families below.

\subsection{Onsager-integrable clock models and Representation-SPTs}
The argument stemming from \cref{fig:paths0} works essentially without modification for cases where the larger symmetry group $G$ contains an element that commutes with $\tilde{G}$. However, the general relationship between pivot loops that pump charge and SPTs is more involved. The family of Onsager-integrable clock models studied in Ref.~\onlinecite{Jones25} provides examples of $\mathbb{Z}_N$-symmetric pivot loops that (as we will show) pump $\mathbb{Z}_N$-charge, yet $H_0$ and $H_\pi$ are not distinct SPTs for odd $N$ (for any symmetry group). 

These models are $N$-state generalisations of the transverse field Ising model considered above, with a $\mathbb{Z}_N$ clock symmetry and an anti-unitary $\cpt$ symmetry \cite{Jones25}. Using the usual $\mathbb{Z}_N$ clock operators (see Section \ref{sec:onsagerpump} for definitions), the Onsager paramagnet and ferromagnet are given by
\begin{align}
H_0 &= -\frac{1}{N}\sum_j \sum_{m=1}^{N-1}\alpha_m X_j^m, \qquad \alpha_m = \frac{1}{1-\rme^{2\pi \rmi m/N}}\  \nonumber\\ \tilde{H} &= -\frac{1}{N}\sum_j \sum_{m=1}^{N-1}\alpha_m Z_{j-1}^{-m}Z_j^m \ .  \label{eq:Onsager}
\end{align} Taking commutators, these Hamiltonians generate a representation of the Onsager algebra \cite{Onsager44}.
Pivoting the paramagnet $H_0$ with the ferromagnet $\tilde{H}$ leads to a model $H_\pi$ that is an analogue of the cluster model in this context\footnote{Indeed it reduces to it for $N=2$. Note that this is not the $\mathbb{Z}_N\times\mathbb{Z}_N$ cluster model that we discuss in \cref{sec:ZNZN} below \cite{Geraedts14,Santos15}.}. Although $H_\pi$ is indeed a $G=\mathbb{Z}_N\rtimes \cpt$ SPT for even values of $N$, it is a $G$-representation-SPT (RSPT)~\cite{OBrien2020,Verresen25} for odd values of $N$. 

The RSPT is characterised by degenerate dominant Schmidt values in the ground state---this degeneracy is due to symmetry-fractionalisation (a greater than one-dimensional \emph{linear} irreducible representation of $G$ at the boundary) and is parametrically stable despite being in the trivial SPT phase (since non-trivial SPT phases require projective representations). Moreover, as we would expect for distinct SPT phases, but not for two models in the trivial phase, tuning between $H_\pi$ and $H_0$ contains a gapless point (or region) for $N=3$ (this was observed numerically in Ref.~\onlinecite{Jones25}). In this RSPT case, the gaplessness is `unnecessary' from considerations of the SPT classification. Another curious feature is that the SPT is protected by an anti-unitary CPT symmetry, which, due to the inversion symmetry, will not have stable gapless boundary modes. Nevertheless for all $N$, taking a $2\pi$-periodic and symmetric boundary condition, we find that $H_\pi$ has edge modes. 

The non-trivial charge pump explains the above features of the phase diagram that are not explained by SPT considerations alone. We also will see that an analogous argument to \Cref{fig:paths0}, taking into account the nature of the $\cpt$ symmetry, allows for a non-trivial pump as well as a symmetric path between $H_0$ and $H_\pi$ for $N$ odd.

\subsection{Key notions and outline of paper}
Motivated by the various connections between pivot loops, charge pumps, anomalous symmetries, gapless diabolical points, and SPT phases that we have seen so far, in this work we seek to understand this interplay in greater detail. Based on the examples above, we know that there cannot be a general rule that a non-contractible loop must encounter an SPT at a high symmetry point (if such a point exists). Nevertheless, we aim to clarify what constraints the non-trivial (i.e., non-contractible) loops place on the SPT phase diagram. We emphasise that we are interested in relating SPTs and pumps in \emph{the same spatial dimension}; and, in particular, contrast this with the notion that a non-trivial $d$-dimensional loop pumps a $(d-1)$-dimensional SPT phase \cite{Kitaev11,Wen23}. 

Although most of our concrete examples are one-dimensional chains, our results are often applicable in any spatial dimension. For example, we explore the consequences of $d$-dimensional pivot Hamiltonians that generate \emph{strict circular loops} in the space of gapped Hamiltonians. These have the same form as \cref{eq:circleintro}, and
mirror the particular simplicity of the Ising pivot in any dimension (see \Cref{fig:circularloop}). We moreover expand on the argument of \Cref{sec:firstresult} to analyse constraints placed by non-contractible loops on SPTs in any dimension, where the loop pumps a group cohomology SPT. We go further and put this in a general context via analysis of $G$-equivariant families (where $G$ preserves the family rather than individual Hamiltonians within the family) \cite{debray2024longexactsequencesymmetry}.  

In this work, we make use of the notion of short-range entangled (SRE) states\footnote{Although we will not require a concrete definition, in the broadest sense of the term these are also called invertible states, for each such state there is a corresponding state that tensors to give the trivial phase \cite{Kapustin21}.  Their appearance is natural as we expect a pump to be reversible.} \cite{Kitaev11,Kitaev_talk}.
For our purposes, a generalised Thouless pump is any non-contractible closed loop\footnote{Such a loop is a family of Hamiltonians parameterised by a circle. One can further generalise to families parameterised by other spaces \cite{Wen23}, see also \Cref{sec:equivariant}.} in the space of gapped Hamiltonians with SRE ground states that respect some fixed symmetry \cite{Thouless83,Hsin20,Hermele_talk,Wen23,Inamura24}. Gapless diabolical points, or more generally diabolical loci, are obstructions to trivialising the loop \cite{Hsin20}. 
There is a conjectural classification \cite{Kitaev11,Wen23} that tells us that a non-trivial $d$-dimensional pump corresponds to a $2\pi$ periodic family of symmetric SRE states, where, on placing this family on a system with boundary, going through a periodic cycle the final boundary state differs from the initial one by a non-trivial SRE boundary state. More loosely speaking, we will describe a non-trivial pump as a family of $d$-dimensional Hamiltonians over the circle that pumps a $(d-1)$-dimensional SPT to the boundary. Lattice models for pumps of group cohomology SPTs can be constructed explicitly, this is reviewed in Appendix \ref{app:cohomology}.

Note that in the setting of one-dimensional chains, loops of symmetric\footnote{Strictly speaking, this result is for compact symmetry groups $\tilde{G}$---we will focus on finite symmetry groups in this work.} SRE states have been classified \cite{Bachmann24}. These loops are indeed classified by zero-dimensional SPTs, which are one-dimensional irreducible representations (irreps) of the symmetry group respected by the family of Hamiltonians. Note also that the non-triviality of the pump can manifest in different ways, depending on the choice of boundary termination; see Ref.~\onlinecite{Shiozaki22} for a detailed analysis of the Ising pivot discussed above.
One important point is that if we take a $2\pi$-periodic boundary termination around the loop, then a non-trivial pump excludes the possibility of having a unique symmetric boundary state for every Hamiltonian on the loop.

The outline of the paper is as follows. In the next section, we summarise our main results, and outline some of the key connections between the concepts discussed above. We then have a series of sections focused on non-trivial pumps, followed by the proofs of our results on going from non-trivial pumps to SPTs. These are mostly independent, so the reader interested in the latter can skip ahead to \Cref{sec:pumpSPT,sec:equivariant}. Following our main results, in \Cref{sec:pumps,sec:circular} we discuss strict circular loops and their pump properties in general dimensions. We then focus on exactly solvable models for one-dimensional charge pumps. This constitutes an analysis of the Onsager-integrable clock models in \Cref{sec:onsagerpump} and group cohomology pumps in \Cref{sec:cohomology}. We discuss a family of exactly solvable pumps in one-dimensional $\mathbb{Z}_N\times\mathbb{Z}_N$ SPTs in \Cref{sec:ZNZN} that go beyond strict circular loops, and then in \Cref{sec:DG} a family of strict circular loops that, unlike the generators in \cref{eq:Onsager}, do not obey the Onsager algebra. After this we return to our analysis of when pumps can imply non-trivial SPTs. In \Cref{sec:pumpSPT} we give an argument based on \Cref{fig:paths0} for group cohomology SPTs, giving results for when pumps imply SPTs. In \Cref{sec:equivariant} we take a more abstract approach and show how these connections arise in the setting of equivariant families of Hamiltonians.

\section{Summary of key results}
\label{sec:Summary}
\subsection{Strict circular loops and anomalies}
\begin{figure}
    \centering
    \includegraphics[width=.8\linewidth]{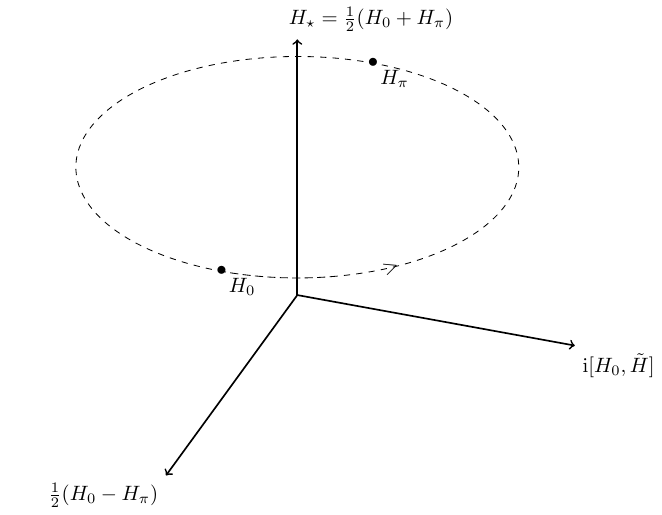}
        \caption{Visualising a strict circular loop \eqref{eq:circle} in the space of Hamiltonians. Here the path is generated by pivoting an initial Hamiltonian $H_0$ with a `pivot' Hamiltonian $\tilde H$, giving rise to $H_\theta = \rme^{-\rmi \theta \tilde H} H_0 \rme^{\rmi \theta \tilde H}$. At the half-way point, we have $H_\pi = \rme^{-\rmi \pi \tilde H} H_0 \rme^{\rmi \pi \tilde H}$, which is sometimes in a distinct SPT phase from $H_0$. If this loop is a non-trivial pump, then the centre $H_\star = \frac{H_0 + H_\pi}{2}$ axis cannot be SRE and there will be a diabolical locus inside the loop. Hamiltonians equidistant from the $H_\star$ axis are isospectral (related by a unitary pivot). In this work we relate the structure of such pivot loops, pumps, and SPTs at high-symmetry points.}
    \label{fig:circularloop}
\end{figure}
The form of the Ising pivot \cref{eq:circleintro} is particularly simple, and we use it to define the notion of
a strict circular loop generated by a Hamiltonian $\tilde{H}$. This is defined by the second equality in
\begin{align}
H_\theta &= \rme^{-\rmi \theta \tilde{H}} H_0 \rme^{\rmi \theta \tilde{H}}\nonumber\\&= \frac{H_0 +H_\pi}{2} + \frac{H_0 -H_\pi}{2} \cos(\theta) + H' \sin(\theta) \label{eq:circle}\ ,
\end{align}
which, as we will see, is a strong constraint on $H_0$ and $\tilde H$. Such a loop is illustrated in \Cref{fig:circularloop}.

Typically we take $H_0$ to be a trivial paramagnet, and we denote by $\tilde G$ the group of unitary operators commuting with $H_0$ and $\tilde H$ and thus with each of the $H_\theta$.
A number of examples of such strict circular loops generated by pivot Hamiltonians are known \cite{Tantivasadakarn23,Jones25}, and they often allow us to make a direct connection between anomalies at the centre of the circle and topologically non-trivial families around the circle. 
\begin{shaded}
\begin{result}[Strict circular loops]\label{result:circ}
We have a strict circular loop of the form \eqref{eq:circle} if and only if the Hamiltonians satisfy the Dolan-Grady relation \begin{align}
    [H_0,\tilde{H}]=\Big[\big[[H_0,\tilde{H}],\tilde{H} \big],\tilde{H}\Big] \ .
\end{align} This relation implies that we have a foliation of the plane into circular loops of radius $\lambda\geq0$:
\begin{align}
 \rme^{-\rmi \theta \tilde{H}} \Big(H_\star &+ \lambda \frac{H_0-H_\pi}{2}\Big)  \rme^{\rmi \theta\tilde{H}}\nonumber\\ &= H_\star+ \lambda \cos(\theta) \frac{H_0-H_\pi}{2}+\lambda \sin(\theta)  H'\ ,\end{align}
 where $H' = \rmi [H_0,\tilde{H}]$, $H_\star =\frac{1}{2}{(H_0+H_\pi)} $ and $[H_\star,\tilde{H}]=0$.
\end{result}
\end{shaded}
The proof is given in Section \ref{sec:circular}, and follows simply from expanding the action of the exponentials as nested commutators \cite{Magnus54}. This formula also appears in the context of certain generalised symmetries\footnote{This is a different notion of generalised symmetries to that found in Ref.~\onlinecite{Gaiotto15}.} of lattice Hamiltonians in Refs.~\onlinecite{Naudts09,Naudts12}. 
In these works, it is shown that the Dolan-Grady relation implies that the spectrum of $\tilde{H}$ consists of sectors with integer spacings. We say that 
$\tilde{H}$ generates a $U(1)$ symmetry if and only if $\rme^{- 2\pi \rmi \tilde{H}}\propto \mathbb{I}$. Hence, if the sectors of $\tilde{H}$ are commensurate\footnote{This is not necessarily the case (we could have an $\mathbb{R}$-symmetry), but all examples we are aware of give rise to a $U(1)$ symmetry on rescaling.}, then there is a constant $R$ such that $\tilde{H}/R$ generates a $U(1)$ symmetry of $H_\star$. Fixing the normalisation of the $U(1)$ generator to $R=1$ is important if we want to understand $U(1)$ anomalies, see further discussion in \Cref{app:periodicity}. Note that \Cref{result:circ} applies in any spatial dimension, and Refs.~\onlinecite{Naudts09,Naudts12} include zero-dimensional examples along with the Ising and Hubbard chains. Our key example comes from the Onsager-integrable chiral clock chain \cite{Davies90,Jones25}, defined explicitly below. 

For strict circular loops that act as pumps we can immediately draw the following conclusion.
\begin{shaded}
\begin{consequence} \rm{{(Anomalies and the Hamiltonian at the centre of a strict circular pump)}}\label{remark:circ}\it{
If we have a strict circular loop of $d$-dimensional Hamiltonians that pumps a non-trivial $(d-1)$-dimensional $\tilde{G}$-SPT, then the ground state of $H_\star$ and the ground state of $\tilde{H}$ are not SRE.

In the case that our pivot Hamiltonian generates a $U(1)$ symmetry, a non-trivial pump is equivalent to a $\tilde{G}\times U(1)^{\mathrm{pivot}}$ anomaly (in the lattice sense of Refs.~\onlinecite{Else_Nayak_SPT_PhysRevB.90.235137,Kapustin24}). This gives us an anomalous symmetry of $H_\star$, and so we again see that $H_\star$ cannot have an SRE ground state.}
\end{consequence}
\end{shaded}
To see the first statement: if $H_\star$ were SRE we would have a non-contractible loop that is also a point, which is a contradiction. An immediate corollary is that there is a phase transition along the line $(1-\lambda)H_0 +\lambda H_\pi $. We can similarly show that $\tilde{H}$ cannot have an SRE ground state. See further discussion in \cref{sec:anomaliesonloops}. 

The second statement is essentially the definition of the lattice anomaly \cite{Else_Nayak_SPT_PhysRevB.90.235137}; the fact that we have a non-trivial pump means that the symmetry algebra on a system with boundary is not the same as in the bulk. We make the connection to an anomalous family around the circle in \Cref{sec:equivariant}. It is important that $\tilde{H}$ (with no rescaling) generates a $U(1)$ to make this conclusion (see \Cref{app:periodicity}). Note also that the strict circular loop is important: taking a non-trivial unitary pump of the form $\rme^{-\rmi \theta \tilde{H}} H_0 \rme^{\rmi \theta \tilde{H}}$ will not necessarily allow us to make the same conclusions. In particular, there is no guarantee that $\tilde{H}$ commutes with $H_\star$ in such a case.

\subsection{Group cohomology pumps and the Onsager-integrable chiral clock models}
In Section \ref{sec:onsagerpump} we show that taking $H_0$, the $N$-state Onsager paramagnet, and $\tilde{H}$, the $N$-state Onsager ferromagnet (defined in \cref{eq:Onsager}), we have a strict circular loop that pumps a $\mathbb{Z}_N$ charge. A priori unrelated,
for group cohomology SPTs there is a standard construction to find representative ground states in $d$-dimensions \cite{Wen_GroupCohomology2013_PhysRevB.87.155114,BUERSCHAPER_SptTensorNetwork2014}, and this naturally generalises to pumps
\cite{RoyHarperPhysRevB.95.195128,Shiozaki22}. Remarkably, applying the group cohomology construction to one-dimensional pumps leads to the following result that connects these two concepts.
\begin{shaded}
\begin{result} \rm{{(1D group cohomology charge pumps are (locally) reducible to Onsager pivots)}}\label{result:cohomology}\it{
Using group cohomology, for any finite group $\tilde{G}$, we can construct models that pump any $\tilde{G}$-charge to the boundary of a one-dimensional chain. We then have the following connection to the Onsager-integrable chiral clock model. }
\begin{enumerate}\it{
\item
If the group $\tilde{G}$ is abelian, there exists a basis where the non-contractible loop decomposes as a stack of strict circular loops. Each chain in the stack has site-dimension $N_j$ (where the $N_j$ depend on $\tilde{G}$). For each chain, the system begins in the ground-state of the appropriate Onsager paramagnet, and the pump corresponds to pivoting with the corresponding Onsager ferromagnet applied to each chain. 
\item For non-abelian groups the group cohomology pump naturally decomposes into a spin ladder, where one chain corresponds to an abelian subgroup (the abelianization of the group) and is itself decomposable as a stack. The corresponding pump is generated by a pivot Hamiltonian formed of products of two-site operators acting on each part of the spin ladder. This operator is either trivial or acts as a \emph{local} Onsager-ferromagnetic pivot on the chain corresponding to the abelian subgroup. Globally this does not correspond to the Onsager-ferromagnet generated loop.
\item For a non-abelian group, we take the same spin ladder with initial Hamiltonian given by a sum of Onsager paramagnets on the chain corresponding to the abelian subgroup, and the Onsager ferromagnet on the second chain. Then applying the group cohomology pivot gives us a strict circular loop in a spontaneous symmetry-breaking phase.
}
\end{enumerate}

\end{result}
\end{shaded}
This is derived in \Cref{sec:cohomology}.
The first result means that for abelian groups, we can apply \Cref{result:circ,remark:circ}. It also follows that, $H_\pi$, half-way around a 1D group cohomology pump for an abelian group, has an enhanced anti-unitary $\cpt$ symmetry. In fact, we show that this symmetry holds for a general non-abelian $\tilde{G}$.

\subsection{SPT phases implied by pumps}
We now turn to the question of when a pump implies an SPT at high-symmetry points around the loop. First we state the general form of the result outlined in \Cref{sec:firstresult}.

\begin{figure}
\includegraphics[]{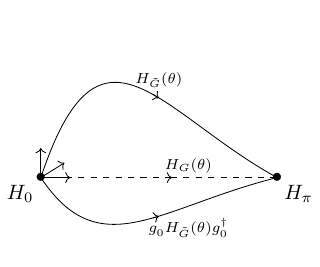}
\caption{Generalisation of \Cref{fig:paths0}: paths in the space of $\tilde{G}$-symmetric Hamiltonians between $H_0$ and $H_\pi$. If $H_0$ and $H_\pi$ are $G$-symmetric, where $\tilde{G}$ is a proper normal subgroup of $G$, we can take a $\tilde{G}$-symmetric path $H_{\tilde{G}}(\theta)$ to $g_0^{}H_{\tilde{G}}(\theta)g_0^{\dagger}$. Assuming that $H_0$ and $H_\pi$ are in the same $G$-SPT phase implies the existence of a $G$-symmetric path between them. Using this path we can find constraints on the charge pumped by the loop constructed from $H_{\tilde{G}}(\theta)$ followed by the reverse of $g_0^{}H_{\tilde{G}}(\theta)g_0^{\dagger}$. If this charge does not satisfy the constraint, we can conclude that $H_0$ and $H_\pi$ are in different $G$-SPT phases.  }
\label{fig:paths}
\end{figure}

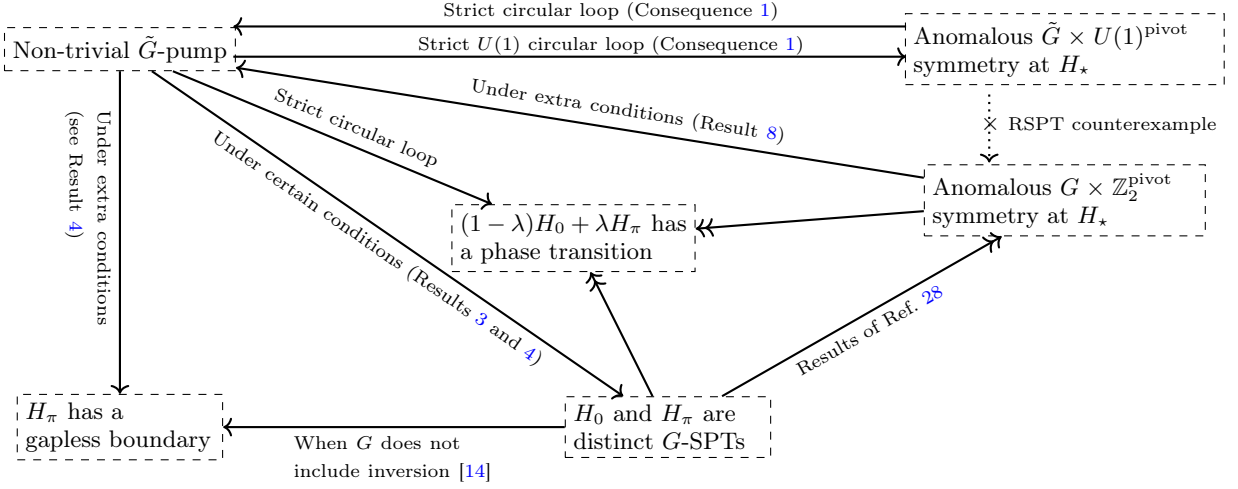
\begin{figure*}[ht]
\begin{tikzpicture}
\node (node1) at (-.5,0) [dashed,rectangle,draw] {Non-trivial $\tilde{G}$-pump};
\node (node25) at (12,0) [dashed, text width=4cm,rectangle,draw] {Anomalous $\tilde{G}\times U(1)^{\textrm{pivot}}$ symmetry at $H_\star$ };
\node (node2) at (12,-2) [dashed, text width=3.5cm,rectangle,draw] {Anomalous ${G}\times\mathbb{Z}_2^{\textrm{pivot}}$ symmetry at $H_\star$ };
\draw [->, thick] (node2)--(node1) node [midway,text width = 7cm, above, sloped] {\scriptsize{\hspace{2cm}Under extra conditions (\Cref{proposition})}};
\node (node3) at (6.75,-5) [dashed, text  width=2.5cm,rectangle,draw] {$H_0$ and $H_\pi$ are distinct $G$-SPTs };
\node (node6) at (-.5,-5) [dashed, text width=2.5cm, rectangle,draw] {$H_\pi$ has a \\gapless boundary };
\draw [->, thick] (node1)--(node6) node [midway, text width = 3.25cm, below, sloped] {\scriptsize{Under extra conditions \\(see \Cref{result:pumpantiunitary})}};
\node (node0) at (5.5,-2.5) [dashed, text  width=3cm,rectangle,draw] {$(1-\lambda)H_0+\lambda H_\pi$ has a phase transition };
\draw [->, thick] (node3)--(node6) node [midway,text width = 2.65cm, below, sloped] {\scriptsize{When $G$ does not include inversion \cite{Pollmann10}}};
\draw [->, thick] (node1)--(node3) node [midway, below, sloped] {\scriptsize{Under certain conditions (\Cref{result:pumpspt,result:pumpantiunitary})}};
\draw [->>, thick] (node3)--(node2) node [midway, below, sloped]  {\scriptsize{Results of Ref.~\onlinecite{Bultinck19}}};
\draw [->, dotted, thick] (11,-.6)--(11,-1.5);
\node (node10) at (12.45,-1) {$\times$ \scriptsize{RSPT counterexample}};
\draw [->>,  thick] (node3)--(node0);
\draw [->>,  thick] (node2)--(node0);
\draw [->, thick] (node1)--(node0) node [midway, above, sloped]  {\scriptsize{~\hspace{0.4cm}Strict circular loop}};
\draw [<-, thick] (01.,0.3)--(9.9,.3);
\node (node5) at (6,.5) {\scriptsize{Strict circular loop (\Cref{remark:circ})}};
\draw [->, thick] (01.,-0.1)--(9.9,-.1);
\node (node7) at (6,.05) {\scriptsize{Strict $U(1)$ circular loop (\Cref{remark:circ})}};
\end{tikzpicture}
\caption{Relation between pumps and SPTs for loops generated by a pivot Hamiltonian $H_\theta = \rme^{-\rmi \theta \tilde{H}} H_0 \rme^{\rmi \theta \tilde{H}}$. $H_0$ and $H_\pi$ are $G$-symmetric, while $\tilde{H}$ (and hence the family $H_\theta$) is $\tilde{G}\subsetneq G$-symmetric. The Hamiltonian $H_\star = (H_0+H_\pi)/2$ is the midpoint of the interpolation between $H_0$ and $H_\pi$. Double arrows are immediate consequences of the appropriate definition. }
\label{fig:results}
\end{figure*}

\begin{shaded}
\begin{result}[Pumps can imply distinct SPTs]\label{result:pumpspt}
Consider $d$-dimensional Hamiltonians $H_0$ and $H_\pi$ which are $G$-symmetric, where $G$ is finite and represented by on-site unitaries. Suppose we have a proper normal subgroup $\tilde{G}\subsetneq G$ and a $\tilde{G}$-symmetric path $H_{\tilde{G}}(\theta)$ from $H_0$ to $H_\pi$. Suppose there exists a $g_0$ outside $\tilde{G}$ and inside the centralizer $C_{G}(\tilde{G})$, and consider the loop
\begin{align}
H(\theta) = \begin{cases}
    H_{\tilde{G}}(\theta) \qquad & 0\leq \theta \leq \pi  \\
    g_0^{}H_{\tilde{G}}(2\pi-\theta)g_0^\dagger & \pi \leq \theta \leq 2\pi
\end{cases}  
\end{align}
(see \Cref{fig:paths}).
Then, if $H(\theta)$ pumps a non-trivial $(d-1)$-dimensional $\tilde{G}$-SPT,  
$H_0$ and $H_\pi$ are distinct $d$-dimensional $G$-SPTs. In fact, they are distinct $d$-dimensional $\tilde{G}\times \mathbb{Z}_n$ SPTs, where $n$ is the order of $g_0$. In this setting, the boundary transition implied by the pump corresponds to at least one of $H_0$ and $H_\pi$ being a non-trivial SPT with boundary degeneracy.
\end{result}
\end{shaded}
The proof for group cohomology SPTs is given in Section \ref{sec:pumpSPT}, and is summarised in \Cref{fig:paths}. 
This is a direct analogue of the argument given for the pump implying an SPT for the Ising pivot. One can also view this result as a different perspective on the decorated domain wall construction for $  \tilde{G}\times \mathbb{Z}_n$ SPTs \cite{Chen14}---we discuss this in \Cref{sec:decorateddomainwall}. Note also that while we analyse pumps corresponding to group cohomology SPTs, we expect that this can be generalised to any suitably defined pump invariant\footnote{The key general notions are that the invariant is additive if we encircle two diabolical points, and that applying $g_0$ that commutes with $\tilde{G}$ does not change the value of the invariant.}. Indeed, for the case where $g_0$ is a $\mathbb{Z}_2$ generator, we prove a general analogue using the Mayer-Vietoris sequence, given by \Cref{result:Mayer-Vietoris} below.
Our methodology can, moreover, be generalised to cases beyond $g_0 \in C_{G}(\tilde{G})\setminus\tilde{G}$; one such generalisation is the next result.
\begin{shaded}
\begin{result} \rm{{(Pumps can imply distinct SPTs---antiunitary or charge conjugation case)}}\label{result:pumpantiunitary}\it{
Consider $d$-dimensional Hamiltonians $H_0$ and $H_\pi$ which are $G$-symmetric, where $G=\tilde{G}\times\mathbb{Z}_2^T$. $\tilde{G}$ has a (pseudo)-real on-site unitary representation, and $\mathbb{Z}_2^T$ acts as complex conjugation, $\mathcal{K}$, in the basis where $\tilde{G}$ is real. Suppose we have a $\tilde{G}$-symmetric path $H_{\tilde{G}}(\theta)$ from $H_0$ to $H_\pi$ and consider the loop
\begin{align}
H(\theta) = \begin{cases}
    H_{\tilde{G}}(\theta) \qquad & 0\leq \theta \leq \pi  \\
    \mathcal{K} H_{\tilde{G}}(2\pi-\theta)\mathcal{K} & \pi \leq \theta \leq 2\pi \ .
\end{cases}  
\end{align}
Then, if $H(\theta)$ pumps a $(d-1)$-dimensional $\tilde{G}$-SPT that cannot be decomposed into a stack of two identical such $\tilde{G}$-SPTs,
$H_0$ and $H_\pi$ are distinct $G$-SPTs. }

If $H(\theta)$ pumps a $(d-1)$-dimensional $\tilde{G}$-SPT then, irrespective of whether it is decomposable, a boundary transition will occur \cite{Hsin20}.  If $H_0$ is trivial and there is a unique boundary transition around the loop, then it must occur at $H_\pi$. 

The same conclusions hold when we have a a one-dimensional chain with unitary charge conjugation, or an antiunitary $\cpt$ symmetry.
\end{result}
\end{shaded}
For group cohomology SPTs, this result can be proved in the same way as Result \ref{result:pumpspt}; this also falls into the more general framework of \Cref{sec:equivariant} (in particular, see \Cref{result:Mayer-Vietoris}). The case of a $\cpt$-symmetric chain is relevant to the Onsager-integrable chiral clock chains. Moreover, we can apply this result to the group cohomology pumps of \Cref{result:cohomology}. Depending on the representation pumped (and flexibility in the choice of Hamiltonian $H_0$), we can use this reasoning to identify a possible $\tilde{G}\rtimes\cpt$ or $\tilde{G}\times \mathbb{Z}_2^T$ SPT at $H_\pi$ for such a pump. 

\subsection{Connections between concepts}
These results allow us to understand many connections between pumps and SPTs that were hinted at in the introduction. In \Cref{fig:results} we illustrate some of the relationships that this work solidifies, while Table \ref{table:results} contains a list of the lattice models studied in the paper. We also note the results of \Cref{sec:ZNZN}, where we introduce pivot Hamiltonians and study charge pumps for $\mathbb{Z}_N\times\mathbb{Z}_N$ (dipolar) cluster chains. These are not strict circular loops, and in fact, the pump is over a space of intersecting circles, which is not a manifold. Nevertheless, there is still a natural Hamiltonian at the centre of the pivot loop, and this Hamiltonian has a large continuous symmetry group.

Complementing our results we fill in several arrows that appear elsewhere.  
Since $H_0$ is trivial, if $H_\pi$ is a non-trivial SPT then it `squares to the trivial phase' (since applying the entangler twice takes us back to $H_{2\pi}=H_0$). In Ref.~\onlinecite{Bultinck19}, Bultinck argues that the $\mathbb{Z}_2$ pivot entangler has a mixed anomaly with $G$ (this becomes an anomalous symmetry at $H_\star$). For group cohomology SPTs the relevant cohomology group is made explicit.

The Hamiltonian $H_\pi$ often has a gapless boundary, and, in the case it is a non-trivial SPT, this follows from the usual bulk-boundary correspondence \cite{Zeng19, Cirac21}. An exception is the case where the SPT is protected by inversion symmetry \cite{Pollmann10}, which is broken by the boundary. A non-trivial pump tell us that we must have a non-trivial boundary somewhere around the pivot circle (assuming a periodic boundary termination \cite{Shiozaki22}), and, under certain conditions (for example, demanding that the boundary transition occurs at a unique value of $\theta$) it must appear at $H_\pi$. This is the case even when $H_\pi$ is not an SPT, or is an SPT protected by inversion symmetry.

Note that if $H_0$ and $H_\pi$ are distinct $G$-SPTs then the line $(1-\lambda)H_0 +\lambda H_\pi$ is a $G$-symmetric path that connects them, and hence must contain a phase transition. This also follows from an anomalous $G\times \mathbb{Z}_2^{\mathrm{pivot}}$ symmetry at $H_\star$, since $H_\star$ cannot be SRE in that case. Indeed, $H_\star$ is either a phase transition point, or tuning from $H_0$ to $H_\star$ we encounter a phase transition into a non-SRE phase. However, in the case of strictly circular loops, these conclusions hold even if $H_0$ and $H_\pi$ are in the same SPT phase (per \Cref{remark:circ} above).

Finally, given an anomalous $G\times \mathbb{Z}_2^{\mathrm{pivot}}$ symmetry at $H_\star$, it was argued in Ref.~\onlinecite{Tantivasadakarn23} that this anomaly often implies a non-trivial charge pump around the pivot circle. See also Ref.~\cite{Kuno20} for further examples. We give a precise characterisation of this connection in terms of equivariant families and the symmetry breaking long exact sequence \cite{debray2024longexactsequencesymmetry} in \Cref{sec:equivariant}.

\begin{table*}
\begin{center}\begin{tabular}{c|c|c|c|c|c|c} & $\bm{H_0}$ & \textbf{Pivot}  $\bm{\tilde{H}}$ & \textbf{Pivot loop } & \textbf{SPT at} $\bm H_\pi$ & \textbf{Strict} & \textbf{Section} \\ &  && \textbf{pump} &  & \textbf{circular} &  \\\hline \textbf{Ising chain}  & Ising PM   & Ising FM & Unit $\mathbb{Z}_2$ charge & Cluster model & Yes &  \ref{sec:firstresult}\\
&  &  &  &  &  & \\
\hline \textbf{Onsager chiral} & Onsager PM  & Onsager FM  & Unit $\mathbb{Z}_N$ charge & SPT $N$ even  & Yes  & \ref{sec:onsagerpump} \\\textbf{clock model} &  & `$H\sim\alpha Z^\dagger Z$'  &  &RSPT $N$ odd  &  & 
 \\\hline \textbf{Abelian group} & Trivial  & Stack of & Any 1D irrep & Sometimes & Yes & \ref{sec:abelian}\\ \textbf{cohomology}  & PM & Onsager FM &  &  & &    \\\hline \textbf{Non-abelian group}  & See results  & See results & Any 1D irrep & Sometimes & See &  \ref{sec:nonabelian}\\ \textbf{cohomology}  & & &  &  & results & \\
 \hline \textbf{Dipolar cluster}  & Potts PM  & Dipolar pivot   & Unit $\mathbb{Z}_2 $ charge  & SPT $N$ even  & No  & \ref{sec:dipolar} \\ \textbf{chain}&  & `$H\sim|\alpha|^2 Z^\dagger Z$' & $N$ even &  &  & 
  \\\hline  
    $\bm{\mathbb Z_N \times \mathbb Z_N}$ \textbf{cluster} & Potts PM  & Cluster pivot   & Unit $\mathbb{Z}_2$ charge  & SPT $N$ even  & No  & \ref{sec:cluster} \\ \textbf{chain} &  & `$H\sim(-1)^j |\alpha|^2 Z^\dagger Z$'  & $N$ even &  &  & 
  \\\hline  \textbf{Potts/Onsager} & Potts PM & Onsager FM & Unit $\mathbb{Z}_2$ charge  &  SPT $N$ even & Yes &   \ref{sec:potts}
   \\ \textbf{chain} &  &  & $N$ even &  &  & \\ \end{tabular}\caption{Overview of the lattice models studied in this work. Paramagnets and ferromagnets are denoted PM and FM respectively. For pivot Hamiltonians we give an indicative local term, where for $N$-state models the complex coupling $\alpha = (1-\rme^{2\pi\rmi/N})^{-1}$. For the (dipolar) cluster chains, there is a more intricate SPT and pump structure when pivoting through angles $2\pi/N$---see \Cref{sec:ZNZN}.}\label{table:results} 
\end{center}
\end{table*}
 \section{Review: characterising pumps} \label{sec:pumps}
Following our introductory discussion on characterising and classifying non-trivial loops in Hamiltonian space, we give some further notions here that will be useful in our discussion below. In particular, we review two ways of characterising a non-trivial pump.

\subsection{Unitary loops}\label{sec:pumpMPU}
In this work we often consider $\ket{\psi(\theta)} =U(\theta) \ket{\psi(0)}$, or the corresponding parent Hamiltonians $H(\theta)=U(\theta)H(0)U(\theta)^\dagger$, where $U(\theta)$ is generated by a $\tilde{G}$-symmetric pivot Hamiltonian\footnote{Note that for any path of gapped local Hamiltonians, one can use the quasi-adiabatic continuation to find a local unitary evolution acting on the ground state; and for a $\tilde{G}$-symmetric path, the corresponding generating Hamiltonian can always be chosen $\tilde{G}$-symmetric \cite{Zeng19,Bachmann24}.}. To close the loop we need that $U(2\pi)$ is a symmetry of $H_0$ (in the simplest case $U(2\pi)=\mathbb{I}$). While a non-trivial pump corresponds to a loop in the space of Hamiltonians, restricting the unitary operator $U(2\pi)$ to an open or semi-infinite chain can indicate when to expect a pump. 

Indeed, when $U(2\pi) = \mathbb{I}$ (for periodic boundary conditions) this links back to our discussion of lattice anomalies \cite{Else_Nayak_SPT_PhysRevB.90.235137,Kapustin24} (we will absorb any overall phase by shifting the pivot Hamiltonian by a constant). Focusing on one-dimensional chains, $U(2\pi)$ acting on an SRE state on the half-infinite chain will give a localised symmetry charge at the open end of the chain \cite{Bachmann24}. Let us consider the case where $U(2\pi)$ is a matrix-product unitary\footnote{This means that we can write this operator as a tensor network with finite bond-dimension. All of our pivot Hamiltonian examples are of this form, since they have commuting local terms.} \cite{Cirac21}. The condition $U(2\pi)=\mathbb{I}$ means that we can use the fixed-point equations given in \cite{Sahinoglu18} to see that restricting to a finite region we have $U(2\pi)^{{[L,R]}}=\mathcal{O}_L \overline{\mathcal{O}}_R$, where $\mathcal{O}_L$ and $ \overline{\mathcal{O}}_R$ are localised on a finite number of sites at the left and right edge of the region. Since our loop is $\tilde{G}$-symmetric, these operators have opposite $\tilde{G}$-charge. By formally considering the half-infinite limit, and applying $U(2\pi)^{[L,\infty]}$ to an SRE state, the charge of $\mathcal{O}_L$ can be identified with the rigorously-defined charge acting on the left edge identified in Ref.~\onlinecite{Bachmann24}.

If we do not act on an SRE state then we will not necessarily find a pump by applying $U(2\pi)$; an example would be a $\tilde{G}$-symmetry-breaking state that is an eigenstate of $\mathcal{O}_L$. In the symmetry-breaking case one can instead find non-trivial domain-wall pumps \cite{Hermele_talk}---these pump a symmetry operator that would act trivially on a symmetric state. Examples of this arise in the Onsager-integrable chiral clock models below, see \Cref{sec:domainwall}. 

 In higher dimensions, we can similarly identify potential $\tilde{G}$-SPT pumps by unitaries $U(2\pi)$ that act trivially in the bulk and as an SPT entangler on boundary degrees of freedom. Acting on an SRE state will then give a non-trivial pump, although we are not aware of a similar mathematically rigorous classification in this case.

\subsection{MPS approach to one-dimensional pumps}
\label{sec:MPS}
 Loops of symmetric SRE states in one-dimension have been classified by irreps pumped to the boundary \cite{Bachmann24}.  Here we show how these 
irreps arise in the matrix-product state (MPS) description of ground states of these spin chains \cite{Cirac21}.

\subsubsection{Appearance of the invariant} Recall that
MPS take the form
\begin{align}
    \ket{\psi(\theta)} &= \sum_{\{j_k\}}\tr(\mathcal{A}_{j_1} (\theta)\cdots \mathcal{A}_{j_L}(\theta))\ket{j_1,\dots j_L}  \label{eq:MPStensor}
\end{align}
where we assume translation invariance for convenience. For fixed $j$ in the on-site basis, $\mathcal{A}_{j}^{\alpha,\beta}$ is a $\chi\times\chi$ matrix on the bond Hilbert space. Ground states of gapped Hamiltonians in 1D can be approximated by MPS \cite{Hastings07}, and analysing symmetry properties of these states can be used to understand the group cohomology classification of 1D SPTs \cite{Turner11,Fidkowski10,Schuch11,Chen11}. One can also consider loops of MPS ground states corresponding to a loop in the space of their corresponding parent Hamiltonians \cite{Shiozaki22,Inamura24}. We study the case where our loops are symmetric, but even without symmetries, the space of MPS states can have non-trivial topological features \cite{Ohyama24,Qi25}.

Note that there is a redundancy in the MPS description $\mathcal{A}_{j}\sim \rme^{\rmi\xi} M\mathcal{A}_{j}M^{-1}$ 
\cite{Cirac21}. This means the tensor will, in general, transform if we apply an on-site symmetry, or if we take a continuous path of MPS returning to the same state \begin{align}\ket{\psi(\theta)}=\ket{\psi(\theta+2\pi)} \Leftrightarrow \mathcal{A}_{j}(\theta) \sim\mathcal{A}_{j}(\theta+2\pi) \ . \end{align}

Given a continuous loop of $\tilde{G}$-symmetric and injective\footnote{A technical condition that we can read as excluding symmetry-breaking. See \cite{Cirac21} for more details.} MPS, this redundancy means that a non-trivial $\tilde{G}$-pump can manifest in different ways. 
In particular, we can always find a gauge where $\mathcal{A}_{j}(\theta) =\mathcal{A}_{j}(\theta+2\pi)$, or a gauge where the fractionalised $\tilde{G}$ symmetry (defined in \cref{eq:Vdef_MPS}) is periodic; however, a non-trivial pump excludes a gauge where both are periodic \cite{Inamura24}. 

If we fix a gauge where the symmetry action is periodic (this will always hold if we generate our family by a symmetric pivot Hamiltonian), then the invariant of the family appears as the following one-dimensional representation of $\tilde{G}$. Since $\mathcal{A}_{j}(2\pi)\sim \mathcal{A}_{j}(0)$, we have 
\begin{align}
    \mathcal{A}_{j}(2\pi)= \rme^{\rmi \xi_W} W \mathcal{A}_{j}(0) W^{-1} \label{eq:Wdef_MPS} \ . 
\end{align}
The on-site symmetry $\tilde{G}$ acts as $U(g) \equiv \prod_j u_j(g)$ and fractionalises periodically as 
\begin{align}
u_{j,j'}^{}(g) \mathcal{A}_{j'}(\theta) = \rme^{\rmi \xi_g }   V(g)^{} \mathcal{A}_{j}(\theta) V(g)^\dagger \ .  \label{eq:Vdef_MPS}
\end{align}
Then, following \cite{Inamura24}, we have
\begin{align}
V(g) W V(g)^\dagger = \rme^{\rmi \chi_g} W    \ , \label{eq:MPS_WV}
\end{align}
where $ \rme^{\rmi \chi_g} $ is a one-dimensional irrep of $\tilde{G}$ that classifies the pump.

Note that we assume injectivity, so  $\ket{\psi(0)}$ is SRE, but it is not necessarily in the trivial phase. In particular, $V(g)$ can form a projective representation of $\tilde{G}$.

\subsubsection{A group cohomology understanding}
A nice interpretation of the above result can be obtained by interpreting the $2\pi$-shift invariance of the MPS tensor as a $\mathbb{Z}$ symmetry for which $W$ is the virtual representation, reminiscent of  Floquet topological phases~\cite{vonKeyserlingkSondhiPhysRevB.93.245145,Else2016PhysRevB.93.201103,Potter17}. Together with the on-site $\tilde{G}$  symmetry, the total symmetry is $G \equiv \tilde{G} \times \mathbb{Z}$.  Then \cref{eq:MPS_WV} tell us that ${G}$ can be realised projectively on the virtual Hilbert space. This is classified by $H^2({G},U(1))$. From the K\"{u}nneth formula \cite{vonKeyserlingkSondhiPhysRevB.93.245145,Else2016PhysRevB.93.201103,Potter17} we have
\begin{equation}
    H^2(\tilde{G} \times \mathbb{Z},U(1)) \cong H^2(\tilde{G} ,U(1)) \times H^1(\tilde{G},U(1))\ . \label{eq:Kunneth}
\end{equation}
The first piece in \cref{eq:Kunneth} classifies standard SPT phases where $V(g)$ are projective representations of $\tilde{G}$. This tells us that non-trivial SPTs can be also considered as `non-trivial' static loops (although these would be contractible, so would be considered trivial in the language of this work). To get something intrinsically loop-like, we consider the second piece that classifies 1d irreps of $\tilde{G}$. This recovers the classification of charge pumps, corresponding to charges $\rme^{\rmi \chi_g}$. By stacking non-trivial SPTs on charge pumps, we get all non-trivial loops in the space of states with symmetry $\tilde{G}$ classified in \cref{eq:Kunneth}.

\section{Strict circular loops in Hamiltonian space}\label{sec:circular}
\begin{figure}
    \centering
    \includegraphics[width=.8\linewidth]{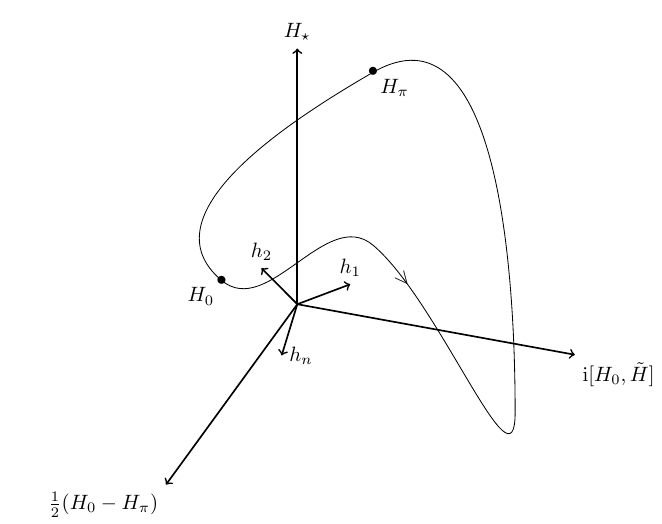}
    \caption{Visualising a generic loop in the space of Hamiltonians (to be contrasted with \Cref{fig:circularloop}). The $h_n$ indicate additional orthogonal directions in the space.}
    \label{fig:genericloop}
\end{figure}

In this section we consider particularly simple paths in parameter space that we call strict circular loops. These are to be contrasted with generic loops that are parameterised by a circle but have an unconstrained image in the space of Hamiltonians---see \Cref{fig:circularloop,fig:genericloop}. Strict circular  loops are paths both parameterised by a circle, and with a circular image in the space of Hamiltonians.
These generalise the pivot loops generated by the Ising ferromagnet, as well as the Onsager ferromagnet \cite{Jones25}, introduced in greater detail below.

\subsection{Strict circular loops are equivalent to the Dolan-Grady relation}\label{sec:circularloop}
To generalise \cref{eq:circleintro}, we consider which pivot Hamiltonians generate analogous strict circular loops.  We show that this is equivalent to the appearance of two Hamiltonians that satisfy the Dolan-Grady relation from integrability \cite{Dolan82}. Note that this is `half an Onsager algebra' \cite{Perk89,Davies91,Perk16}, and is equivalent to the same algebra when there is a duality. While the self-dual Onsager-integrable clock models naturally satisfy this relation (see the next subsection), in \Cref{sec:DG} we will give examples of strict circular loops that do not have an underlying self-dual model.

Consider pivoting $B$ by $A$ and generating such a strict circular loop, then:
\begin{align}
\rme^{-\rmi \theta A}B \rme^{\rmi \theta A}&=\sum_{p=0}^\infty \frac{(\rmi \theta)^p}{p!}\underbrace{\Big[\big[ [B,A],A],\dots A\Big]}_{p-\mathrm{fold}} \nonumber\\&=  \mathcal{X} + \mathcal{Y}\cos(\theta) + \mathcal{Z} \sin(\theta)\ . \label{eq:strictcircular}
\end{align}
The coefficients of $\theta^p$ on each side are immediate, and up to $p=3$ we find
\begin{align*}
\mathcal{X}+\mathcal{Y}&=B \nonumber\\
\mathcal{Z} &= \rmi [B,A] \nonumber\\
\mathcal{Y} &= \big[[B,A],A\big]\nonumber\\
\mathcal{Z} &= \rmi \Big[\big[[B,A],A \big],A\Big].
\end{align*}
Thus, to generate a strict circular loop, $A$ and $B$ must satisfy the Dolan-Grady relation \cite{Dolan82,Perk89,Davies91,Davies91b,Perk16}
\begin{align}
[B,A]=\Big[\big[[B,A],A \big],A\Big] \label{eq:DG}.\end{align}
(Note we use a non-standard normalisation that is convenient for our setting.) This single condition ensures compatibility of all higher terms in $\theta$. If $A$ and $B$ are hermitian, then so are $\mathcal{X},\mathcal{Y}$ and $\mathcal{Z}$.

Using the Dolan-Grady relation and $[\mathcal{X},A]=0$, we have that
\begin{align}
\rme^{-\rmi \theta A}\mathcal{Y} \rme^{\rmi \theta A} =   \mathcal{Y}\cos(\theta) + \mathcal{Z} \sin(\theta)\ ;\label{eq:pivotcircleY}
\end{align}
this can be scaled linearly and we thus have a pivot circle for every radius in the $\mathcal{Y},\mathcal{Z}$ plane, with origin that commutes with $A$. We typically take $\mathcal{X}$ as the origin (since this corresponds to pivoting $B$ by $A$), but we could take, for example, $A$ or $0$ as the origin. We know that $A$ has sectors with integer-spaced eigenvalues \cite{Naudts09,Naudts12}; hence, the origin has a $U(1)$ symmetry (up to scale) if these sectors are commensurate.

As an aside, to justify identifying Eq.~\eqref{eq:strictcircular} as a circle in the space of Hamiltonians, we might want that the different directions are orthogonal and of the same `size'. More precisely, we might want that 
\begin{align}
\langle \mathcal Y, \mathcal Z \rangle_F =0 \quad \textrm{and} \quad \langle \mathcal Y , \mathcal Y \rangle_F = \langle \mathcal Z , \mathcal Z \rangle_F 
\end{align}
for the Frobenius inner product $\langle A,B\rangle_F = \textrm{tr} \left( A^\dagger B \right)$. It is straightforward to see that these relations indeed follow from the above discussion. In particular, $\langle \mathcal Y, \mathcal Y \rangle_F = \langle \mathcal Z , \mathcal Z \rangle_F $ follows directly from Eq.~\eqref{eq:pivotcircleY}. Moreover, the orthogonality follows from Eq.~\eqref{eq:strictcircular} if one uses that the left-hand side implies that $\textrm{tr}(B^\dagger B)$ is independent of $\theta$. Indeed, this implies that 
\begin{align}
\textrm{tr} \left( \left[ \mathcal{X} + \mathcal{Y}\cos(\theta) + \mathcal{Z} \sin(\theta) \right]^2 \right) 
\end{align}
is independent of $\theta$ (in this expression, we use that these operators are hermitian). This also shows that $\mathcal X$ is orthogonal to $\mathcal Y$ and $\mathcal Z$, consistent with \Cref{fig:circularloop}.

\subsection{Onsager-integrable chiral clock family \label{subsec:Onsager}}
As an example of a strict circular loop beyond the Ising pivot, let us consider chains with $N$-state sites, where the $j$th site is acted on by the shift and clock operators $X_j$ and $Z_j$ that satisfy $X_j^N = Z_j^N =1$ and $X_j Z_k = \omega^{\delta_{jk}} Z_k X_j$ for $\omega = \rme^{2\pi\rmi/N}$. In the $Z$-basis $Z_j$ acts on site $j$ as $\sum_{a=0}^{N-1} \omega^{a_j} \ket{a_j}\bra{a_j}$, while $X_j$ acts as $\sum_{a=0}^{N-1}\ket{a_j-1}\bra{a_j}$. We use $L$ to denote the length of the chain. A simple model in the trivial phase is $H_0 = -\sum_j (X_j +X_j^\dagger)$, with ground state $\prod_j \ket{v_{j,0}}$, where $X$ is diagonal in the basis $\ket{v_{j,n}} =N^{-1/2}\sum_{a_j=0}^{N-1} \omega^{-n a_j} \ket{a_j} $.

Recall the definition of the Onsager paramagnet and ferromagnet given in the introduction (now in the standard notation for these operators):
\begin{align}
A_0 &= -\frac{1}{N}\sum_j \sum_{m=1}^{N-1}\alpha_m X_j^m,  \nonumber\\ A_1 &= -\frac{1}{N}\sum_j \sum_{m=1}^{N-1}\alpha_m Z_{j-1}^{-m}Z_j^m, \qquad \alpha_m = \frac{1}{1-\omega^m}\ . 
\end{align}
The Hamiltonians $A_0$ and $A_1$ are Kramers-Wannier dual, and, through commutators, generate an Onsager algebra $\{A_m, G_n\}_{m\in\mathbb{Z},n\in\mathbb{Z}_+}$ \cite{Onsager44}.  This algebra is
\begin{align}
[A_l,A_m]&=G_{l-m}\ ,\nonumber\\
[G_l,A_m]&=\frac{1}{2}(A_{m+l}-A_{m-l})\ ,\nonumber\\ [G_l,G_m]&=0\ ,\label{eq:onsagerrelations}
\end{align}
and is equivalent, given the Kramers-Wannier duality, to the Dolan-Grady relation \cite{Dolan82,Davies91,Davies91b}\begin{align}\Big[\big[[A_1,A_0],A_0\big],A_0\Big]=[A_1,A_0]\ . \end{align} 

For $N=2$, $A_0$ and $A_1$ give the transverse-field Ising model; the other Onsager generators are fermion bilinears. For general $N$, the Hamiltonians $A_0$ and $A_1$ have a $\mathbb{Z}_N$ symmetry generated by $Q=\prod_j X_j$, and an anti-unitary CPT symmetry $CP\mathcal{K}$ where $\mathcal{K}$ is complex conjugation in the $Z$-basis, $P$ is a spatial parity inversion sending site $j$ to site $L+1-j$ and $C=\prod_j C_j$, where $C_j=\sum_{a=0}^{N-1} \ket{a_j}\bra{N-a_j}$ acts as charge conjugation. $A_0$ has a product state ground state and is in the trivial paramagnetic phase, while $A_1$ has a ferromagnetic $\mathbb{Z}_N$ symmetry-breaking ground state.

The second Onsager generator, $A_2$, is of particular interest. It has the following form
\begin{align}
A_2&= -\frac{1}{N} \sum_{j} \sum_{m=1}^{N-1} \alpha_m S_{j-1,j}^{(m)} X_j^mS_{j,j+1}^{(m)} \label{eq:A2def}\ ,\\
 S_{j-1,j}^{(m)} &=  1-\frac{2m}{N} - \frac{2}{N}\sum_{m'=1}^{N-1}\alpha_{m'} (1-\omega^{mm'})Z_{j-1}^{-m'}Z_j^{m'}.   \nonumber \end{align}
 This model is dual to the ferromagnetic $A_{-1}$ \cite{Ahn90,Vernier19}, and has SPT order protected by $D_{2N} = \mathbb{Z}_N^{}\rtimes \cpt$ for even $N$ and $D_{2N}$ RSPT order for odd $N$ \cite{Jones25}. Taking $A_1$ as a pivot Hamiltonian, we have
\begin{align}
 &\rme^{-\rmi \theta A_1}A_0\rme^{\rmi \theta A_1} \nonumber\\
 &= \left(\frac{A_0+A_2}{2} +  \cos(\theta) \frac{A_0-A_2}{2}+ \sin(\theta) \rmi G_{-1}\right)\label{eq:circlepivot}
\ . \end{align}
 This gives us a pivot formula for the Hamiltonian $A_2$ by fixing $\theta=\pi$, and follow from the Dolan-Grady relation and the general results of the previous subsection.

 \subsection{Anomalies and strict circular loops}\label{sec:anomaliesonloops}
 Returning to the general setting of \Cref{sec:circularloop}, let us put $A =\tilde{H}$ and $\mathcal{X}= H_\star = (H_0+H_\pi)/2$. Then, for each fixed $\lambda$, we have the pivot circle $H(\lambda,\theta)$ 
\begin{align}
 \rme^{-\rmi \theta \tilde{H}} \Big(H_\star &+ \lambda \frac{H_0-H_\pi}{2}\Big)  \rme^{\rmi \theta \tilde{H}}\nonumber\\ &= H_\star+ \lambda \cos(\theta) \frac{H_0-H_\pi}{2}+\lambda \sin(\theta)  H'\ .\end{align}
 We will consider
 settings where $\tilde{H}$ is $\tilde{G}$-symmetric, and where $H_0$ and $H_\pi$ may have an enhanced $G$-symmetry, such that $H(\theta)$ is a $G$-equivariant family in the sense of \Cref{sec:equivariant}.

 In the remainder of this section we consider some of the physics of these strict circular loops, discussing some of the connections in \Cref{fig:results}. In particular, we consider criticality along the line $ (1-\lambda)H_0 + \lambda H_\pi$ and expand on \Cref{remark:circ}.

Let us first suppose that we have non-trivial $\tilde{G}$-pump around the unit circle $H(1,\theta)$.
A simple argument that there is a transition inside the circle and along the line  $ (1-\lambda)H_0 + \lambda H_\pi$ is the following. If our strict circular loop is a non-trivial pump, this means it is not contractible and thus contains at least one gapless diabolical point. Since all concentric circles are unitarily equivalent (and thus isospectral), in fact the gapless region (which in general could be quite complicated) is circularly symmetric. This means the simplest cases are a diabolical point at the origin or a single diabolical circle. This circle intersects the horizontal-axis, meaning there is \emph{always} a gapless point on this axis (where this axis often has an enhanced symmetry). This can be a first order transition into a non-trivial phase.

As discussed above, suppose that $H_\star$ is SRE, then $U(\theta)H_\star U(\theta)^\dagger=H_\star$ would be a non-trivial $\tilde{G}$-pump, and hence a non-contractible loop. However, this loop is a point, so this is inconsistent. This means $H_\star$ cannot be SRE. In the $U(1)$ case, since we can rescale our strict circular loop using \cref{eq:pivotcircleY}, we can connect the anomaly of the $\tilde{G}\times U(1)^{\mathrm{pivot}}$-equivariant family to the anomalous symmetry at the origin---see \Cref{sec:equivariant} for details. 

The identical argument works for any operator that commutes with the pivot Hamiltonian, including $\tilde{H}$ itself. Such an operator cannot have an SRE ground state as this would be inconsistent with the non-triviality of the pump. 

Note that for one-dimensional chains with a non-trivial pump, $H_\star$ not being SRE means it is either gapless, or in an SSB phase. In the case where the pivot generates a $U(1)$ symmetry, the ground state must break $\tilde{G}$ \cite{Mermin66,Hohenberg67,coleman73}, and hence the $\tilde{G}$-SSB phase must extend to some finite radius. This is because we cannot have a level-crossing transition between unique symmetric ground states that breaks the symmetry. Making the same argument for the pivot Hamiltonian, this is consistent with the Onsager ferromagnet $A_1$ having a $\mathbb{Z}_N$ symmetry-breaking ground state.

Finally, in some cases $H_\pi$ will be a non-trivial SPT, and one objective of this work is to clarify the relationship between this SPT and any pump around the loop. The following discussion applies to pivot loops and therefore holds for strict circular loops as a special case. First, if $H_\pi$ is a $G$-SPT then $H_\pi$ squares to the trivial phase and the operator $\rme^{-\rmi \pi \tilde{H}}$ is a $\mathbb{Z}_2$ SPT entangler. This SPT entangler necessarily commutes with the group $G$, and so we have a $G\times\mathbb{Z}_2$ symmetry of $H_\star$. Following the argument in Ref.~\onlinecite{Bultinck19}, this SPT entangler cannot act on-site, implying a non-trivial anomaly. See also \Cref{sec:equivariant}, where, as we show in \Cref{proposition}, this implies a non-trivial family around the circle, which  often manifests as a non-trivial pump. Conversely, if we have a pump around the loop, a strict circular loop with an enhanced symmetry on the horizontal axis will often have an additional $\mathbb{Z}_2$ that reflects the circle. This allows us to use \Cref{result:pumpspt,result:pumpantiunitary}, or the  general \Cref{result:Mayer-Vietoris}, to give constraints on possible SPTs at $H_\pi$.

\section{Charge pumps in the Onsager-integrable chiral clock models}\label{sec:onsagerpump}
In this section we apply some of our general results to the Onsager-integrable chiral clock models \cite{Jones25}. Since the corresponding $A_0$ and $A_1$ satisfy the Dolan-Grady relation, pivoting $A_0$ with $A_1$ gives us a strict circular loop. In particular, for $U_1(\theta) = \rme^{-\rmi \theta A_1}$, we have
\begin{align}
&H_{\mathrm{clock}}(\theta,\lambda)= U_1(\theta)\left(\frac{A_0+A_2}{2} + \lambda \frac{A_0-A_2}{2}\right) U_1(\theta)^\dagger \nonumber\\&= \left(\frac{A_0+A_2}{2} + \lambda \cos(\theta) \frac{A_0-A_2}{2}+\lambda \sin(\theta) \rmi G_{-1}\right)\label{eq:circlepivot2}\end{align}
where the Hamiltonian perpendicular to the $A_0$--$A_2$ axis is given by

 \begin{align}
 \rmi G_{-1}&=\frac{\rmi}{N^2}\sum_j \sum_{m,m'}\alpha_m\alpha_{m'}(1-\omega^{m m'}) \mathfrak{h}_{j,j+1} \ , \nonumber
 \\
 \mathfrak{h}_{j,j+1}&= ( X_j^m Z_j^{-m'}Z_{j+1}^{m'}-Z_j^{-m'}Z_{j+1}^{m'} X_{j+1}^m)\ .
\end{align}

\subsection{$A_1$ pumps $\mathbb{Z}_N$ charge}\label{section:A1pump}
\subsubsection{Unitary loop implies a charge pump}
To show that the pivot acts as a charge pump, we can write $U_1(\theta)=\prod_{j} U_{j,j+1}(\theta)$ where the individual gate is given by \begin{align}U_{j,j+1}(\theta)= \exp\left(\frac{\rmi \theta}{N} \sum_m {\alpha_mZ_j^{-m}Z_{j+1}^m}\right).\end{align}
Using the trigonometric identity \begin{align}
\sum_{m=1}^{N-1} \alpha_m \omega^{-mk} &= \frac{(N-1)}{2}-k\qquad &0\leq k \leq N-1 \ ,
\label{eq:trig}\end{align}
we have the action \begin{align}
    &U_{j,j+1}(\theta)\ket{a,b} \nonumber\\
    &=\rme^{\frac{\rmi \theta(N-1)}{2N}}\exp\left(-\frac{\rmi \theta}{N}\Big( a-b \pmod N\Big)\right)\ket{a,b} \ . \label{eq:A1pivot}
\end{align} Hence, $U_{j,j+1}(2\pi)=\omega^{\frac{N-1}{2}} Z_j^{-1} Z_{j+1}^{}$.
From this we conclude that $U_1(2\pi) \propto \mathbb{I}$ where we pivot with the periodic Hamiltonian $A_1$. As discussed in \Cref{sec:pumpMPU}, whether this generates a charge pump in the space of Hamiltonians depends on our initial Hamiltonian. Suppose we have an SRE ground state of a gapped symmetric Hamiltonian, and we apply the $2\pi$-pivot only to a finite region consisting of sites 1 up to $L$. Then, up to a phase, the action is $Z_1^{-1} Z_L^{}$ so we have a localised positive (negative) $\mathbb{Z}_N$ charge at the right (left) edge; alternatively we have pumped a $\mathbb{Z}_N$ charge from left to right.

\subsubsection{Identifying the charge pump from the ground state MPS}
While the above operator-based approach already tells us that this closed loop in Hamiltonian space is a non-trivial pump, we can also see this pump in the bulk wavefunction as follows.
Let us suppose that our starting Hamiltonian is $A_0$, and so we pivot as in \eqref{eq:circlepivot2} for $\lambda =1$. Since we apply a matrix-product operator to a product state, we have an MPS ground state that can be found using the analysis from Ref.~\cite{Jones25}. Let $\ket{\psi(\theta)}$ be the ground state of $U_1(\theta)^{} A_0 U_1(\theta)^\dagger$, then the MPS tensor \eqref{eq:MPStensor} is given by 
\begin{align}
 \mathcal{A}_j^{\alpha,\beta}(\theta) &=  \underbrace{N^{-\frac{1}{2}}~\omega^{j(\beta-\alpha)}  \omega^{\beta/2}}_{\Gamma_j^{\alpha,\beta}} \nonumber\\
&\times\underbrace{N^{-1} ~ { \sin\left(\frac{\theta}{2}\right)}\left({\sin}\left(\frac{\theta+2\pi \beta}{2N}\right)\right)^{-1} }_{\Lambda_\beta} \label{eq:MPS}\ 
\end{align}
for $\theta\neq 2n\pi$. Then, by continuity, $\mathcal{A}_j^{\alpha,\beta}(2n\pi)$ is the limit of this expression, giving the product ground state of $A_0$.
(The $\Gamma$ and $\Lambda$ tensors are given for the usual MPS canonical form; $\Lambda_\beta^2$ gives the entanglement spectrum for a bipartition along the bond \cite{Pollmann12}.)

For all $\theta$, the $\mathbb{Z}_N$ symmetry acts locally on the MPS as
\begin{align}
\sum_{j} X^{}_{j',j}\mathcal{A}_j^{\alpha,\beta}(\theta)   = Z^\dagger \mathcal{A}_j^{\alpha,\beta}(\theta)Z\ .
\end{align}
(Strictly speaking, this action is not uniquely defined at $t=2n\pi$ but we can view this as a `removable singularity'---see below). 
We also have that
\begin{align}
\mathcal{A}_j^{\alpha,\beta}(\theta+2\pi)   = -\omega^{-1/2}X^\dagger \mathcal{A}_j^{\alpha,\beta}(\theta)X\ .
\end{align}
The fractionalised symmetry, $Z^\dagger$, and the pump action, $X^\dagger$, do not commute, meaning we have a non-trivial charge pump. This is necessarily stable within equivalence classes of loops of gapped symmetric Hamiltonians. To compare to the discussion of \Cref{sec:MPS}, let $g$ generate the $\mathbb{Z}_N$. Comparing with \cref{eq:Wdef_MPS,eq:Vdef_MPS}, we see that $W=X^\dagger$, $V(g)=Z^\dagger$ and $\rme^{\rmi \chi_g} = \omega^{-1}$. (This is the charge-one irrep of $\mathbb{Z}_N$, $\chi^{(1)}$, discussed below in \cref{eq:irreps}.) 

Note that strictly speaking, we cannot apply the classification result of Ref.~\onlinecite{Inamura24} since $\mathcal{A}(\theta)$ is continuous on the fixed $N$-dimensional bond space, but fails to be injective for $t=2n\pi$. This means that the symmetry-fractionalisation is not uniquely defined for the state at these $t$ values. However, having chosen a gauge for all other $t$ such that the symmetry-fractionalisation is constant, it is natural to view this as a removable singularity and fix a constant symmetry-fractionalisation. In other words, we can resolve the ambiguity in the symmetry-fractionalisation at an isolated non-injective point along a continuous path of injective MPS by taking it equal to the limit from each side. For a deeper discussion of cases where injective bond-dimension varies with parameters see Ref.~\onlinecite{Qi25}.

\subsection{Physical implications}
Having established that $A_1$ generates a charge pump, we can revisit the physics of these clock models studied in Ref.~\onlinecite{Jones25} and gain new understanding.

\subsubsection{$A_2$ is an SPT for $N$ even and an RSPT for $N$ odd}
Since the pivot loop is of the form 
\begin{align}
& \rme^{-\rmi A_1 \theta}   A_0 \rme^{\rmi A_1 \theta} \nonumber\\&= \begin{cases}
 \rme^{-\rmi A_1 \theta} A_0 \rme^{\rmi A_1 \theta}   & 0\leq \theta\leq \pi\\
CP\mathcal{K} \rme^{-\rmi A_1(2\pi- \theta)}   A_0  \rme^{\rmi A_1 (2\pi- \theta)} CP\mathcal{K}    & \pi\leq \theta\leq 2\pi\\
\end{cases}
\end{align}
and pumps a unit $\mathbb{Z}_N$ charge, we can apply \Cref{result:pumpantiunitary} for $\cpt$ symmetric chains (see also the discussion in \Cref{sec:antiunitary}). For even $N$, a unit charge cannot be written as the sum of two charges, so we conclude that $A_0$ and $A_2$ are in different 
$D_{2N}=\mathbb{Z}_N\rtimes\cpt $ phases. This was explicitly demonstrated using symmetry fractionalisation in Ref.~\onlinecite{Jones25}. For odd values of $N$ the unit charge can be decomposed as  
\begin{align}
    1 = 2 \times \left( \frac{N+1}{2} \pmod {N} \right) \ ,
\end{align}
and so the pump does not exclude a symmetric path. This is consistent, since a singlet in the entanglement spectrum of the $A_2$ ground state implies that this model is not in an SPT phase for any symmetry group. Since the dominant entanglement eigenvalues form a $D_{2N}$ doublet, and this has some stability (although can change without a bulk phase transition), we say $A_2$ is an RSPT for this symmetry~\cite{OBrien2020,Jones25,Verresen25}.
\subsubsection{Bulk criticality inside the circle, and boundary transitions on the circle}
For even $N$, since $A_2$ is an SPT, we know that $A(\lambda) = \sfrac{1}{2}\left((1+\lambda)A_0 + (1-\lambda) A_2\right)$ must go through a bulk phase transition for some $\lambda$. Numerics for $N=3, 4$ indicate that $A(\lambda)$ is gapless for some region $1/2-a < \lambda < 1/2+a$ (corresponding to a gapless disk of radius $a$ in the space $H_{\mathrm{clock}}(\theta,\lambda)$) \cite{Jones25}. This gaplessness is `unnecessary' for $N=3$ since both sides of the transition are in the trivial phase . However, independent of these SPT considerations, since we have a non-trivial pump we can use the argument of \Cref{sec:anomaliesonloops}. The minimal gapless region is either a diabolical point at $\lambda =0$ or a diabolical circle that intersects the $A_0$--$A_2$ line. This means that for all $N$, we have at least a gapless point along $A(\lambda)$ for $0<\lambda<1$, consistent with the numerics. 
We see that using our analysis of pumps in the context of strict circular loops resolves an outstanding puzzle, the presence of this critical point, of the earlier study in Ref.~\cite{Jones25}.

Since $A_2$ is either in the trivial phase, or is an SPT protected by an inversion symmetry, it is not expected to have stable gapless edge modes. However, the non-trivial pump tells us that if we take the family $H(\theta,1)$ and terminate it on the boundary such that the Hamiltonian is $2\pi$ periodic and $\mathbb{Z}_N$ symmetric, then it cannot have a gapped boundary for all $\theta$. If the boundary crossing is unique then it occurs at $\theta=\pi$, i.e., for $A_2$ with some symmetric boundary termination.

If we take $H(\theta,1)$ and remove all terms that have support outside sites 1 to $L$, this expected boundary transition is consistent with the following observation: the Hamiltonian 
\begin{align}
H(\pi,1) =  A_2^{\mathrm{OBC}}&= -\frac{1}{N} \sum_{j=2}^{L-1} \sum_{m=1}^{N-1} \alpha_m S_{j-1,j}^{(m)} X_j^mS_{j,j+1}^{(m)} \label{eq:A2open}
\end{align} is unitarily equivalent to $-\frac{1}{N} \sum_{j=2}^{L-1} \sum_{m=1}^{N-1} \alpha_m  X_j^m$ and thus
has a ground state degeneracy.

\subsubsection{SPT physics of higher Onsager generators}
Higher Onsager generators $A_k$ can be defined from $A_0$ and $A_1$ using \cref{eq:onsagerrelations}, and can be studied as Hamiltonians in their own right. For $N=2$ these correspond to generalised cluster models \cite{Suzuki71}, which are known to have interesting SPT properties \cite{Verresen17}. Following Ref.~\onlinecite{Jones25}, there are pivot formulae for each of the $A_k$:
\begin{align}
A_{2k} &= \rme^{-\rmi \pi A_k}A_0 \rme^{\rmi \pi A_k}   \nonumber\\
A_{2k+1} &= \rme^{-\rmi \pi A_k}A_1 \rme^{\rmi \pi A_k}\ .
\end{align}
Thus for periodic systems, $A_{2k+1}$ has a symmetry-breaking ground state, and $A_{2k}$ has a unique symmetric ground state. A straightforward generalisation is to write $A_{2k} = \rme^{-\rmi \pi A_{k+k'}}A_{2k'} \rme^{\rmi \pi A_{k+k'}}$. 

Let us consider the circular loop generated by $\tilde{H}=A_{2k+1}$ and $H_0 =A_{4k'}$. The pivot unitary satisfies $\rme^{-2\pi \rmi A_{2k+1}} =\rme^{-\pi \rmi A_{k}}\rme^{-2\pi \rmi A_{1}} \rme^{\pi \rmi A_{k}} $. Using the analysis of \Cref{section:A1pump}, for a finite system this pumps a unit-charge local operator $\rme^{-\pi \rmi A_{k}}Z_1^{-1}\rme^{\pi \rmi A_{k}}$ to the left boundary (where we truncate $A_k$ appropriately by removing terms outside the finite system---all remaining terms are symmetric and thus conjugation cannot change the charge). Hence, using \Cref{result:pumpantiunitary}, $A_{4k'}$ and $A_{4(k-k')+2}$ are in distinct $D_{2N}$ SPT phases for $N$ even. Since there are only two such phases, we must have that, for all $k$, $A_{4k}$ is in the trivial phase, while $A_{4k+2}$ is in the non-trivial phase. While we can write down a form of the ground state MPS for each of these Hamiltonians (using $A_0$ and $A_1$ pivots, for example), computing the symmetry fractionalisation requires a further analysis, in comparison to the simplicity of applying \Cref{result:pumpantiunitary}. For $N$ odd all $A_{2k}$ are in the trivial SPT phase.

\subsubsection{Connections to other models}
From a field theory perspective, we note the parallels to the `global inconsistency' or the anomaly in the space of coupling constants for $SU(N)$ gauge theory  \cite{Kikuchi17,Gaiotto17,Cordova20,Cordova20b}. There is an additional time-reversal symmetry when the $\theta$ parameter is 0 or $\pi$. For even $N$ there is a 't Hooft anomaly between the $SU(N)$ and time reversal at $\theta=\pi$, which means that the theory can be thought of as living on the boundary of a higher-dimensional SPT phase, (this is the analogue of the SPT in our case), while for odd $N$ there is no such anomaly. 
This means that for $\theta=0$ or $\theta=\pi$ one may choose counterterms such that the symmetry may be gauged, but the global inconsistency tell us that there is no continuous choice of counterterm that trivialises both $\theta=0$ and $\theta=\pi$. This implies a bulk transition for some value of $\theta$ (corresponding to a boundary transition in our setting \cite{Wen23}). This is analogous to our analysis of the boundary of the RSPT for odd $N$. We note that the anomaly in the space of coupling constants does not rely on the time-reversal symmetry at special points \cite{Cordova20}, just as we still pump a non-trivial $\mathbb{Z}_N$ charge if we break the $\cpt$ symmetry of the clock model.
\subsection{Domain wall pumps}\label{sec:domainwall}
Let us consider pivoting $A_1$ by $A_0$, this is the Kramers-Wannier dual to the picture considered above, and write $U_0(\theta)=\rme^{-\rmi \theta A_0}$. Since $A_1$ is ferromagnetic we are outside the usual domain of applicability of charge pumps, indeed $U_0(\theta)$ is a product of single-site operators so cannot act as a pump. Nevertheless, the loop is non-trivial. As discussed in Refs.~\onlinecite{Hermele_talk,Kuno24}, for symmetry-breaking phases one can find domain-wall pumps. We will show how they arise in this case.

First, in contrast to $A_1$, which has an integer-spaced spectrum on periodic boundaries, $A_0$ has integer-spaced spectrum in each $\prod_jX_j$ symmetry-sector but the sectors are split by steps of $1/N$. This means that $U_0(2\pi)\neq \mathbb{I}$. Indeed $U_0(2\pi)$ is a product of operators that act as
\begin{align}
\rme^{2\rmi\frac{\pi}{N} \sum_{m=1}^{N-1} \alpha_m X^m }\ket{v_{k}} &=\rme^{2\rmi\frac{\pi}{N} \sum_{m=1}^{N-1} \alpha_m \omega^{-mk}} \ket{v_{k}} \nonumber\\
&= \rme^{2\rmi\frac{\pi}{N} ((N-1)/2-k)} \ket{v_{k}} \ ,
\end{align}
where $\ket{v_{k}}=\sum_a \omega^{-a k}\ket{a}$ is the $X$-diagonal basis. Hence $U_0(2\pi)=(-1)^L\omega^{-L/2}\prod_j X_j$. Going round a pivot loop applies this operator that takes us between ferromagnetic (symmetry-broken) ground states of $A_1$. If we apply it to a half-infinite region, in analogy to the usual charge pump, this operator shifts the spins in the region and creates a domain wall at the boundary of the region.

\section{Group cohomology pumps and the Onsager ferromagnet}\label{sec:cohomology}
Work over the past years has shown that various (but not all) symmetry protected topological (SPT) phases can be understood using group cohomology~\cite{Wen_GroupCohomology2013_PhysRevB.87.155114,Kapustin14}.
This formulation can be used to construct topological pumps in any spatial dimension, and we review this in \Cref{app:cohomology}. We apply this to the one-dimensional case here, and derive \Cref{result:cohomology}. In particular, we show that for any abelian group, there exists a choice of basis that reduces the group-cohomology pump construction to a stack of Onsager-integrable chiral clock model pumps, as discussed in the previous section. For non-abelian groups, the pivot Hamiltonian that generates the pump in the trivial phase is related to the Onsager ferromagnet only locally and is unrelated to the Onsager-integrable structure globally. However, we show that the construction reduces to that of a stack of Onsager pumps in a particular spontaneous symmetry  breaking phase.  

\subsection{One-dimensional group cohomology pumps and SPT pivots}
We consider a spin chain where each site has a $\lvert \tilde{G}\rvert$-dimensional Hilbert space with basis labelled by group elements. In this section, we specialise the group cohomology pump \cite{RoyHarperPhysRevB.95.195128} (see \Cref{app:cohomology}) to one dimension. Non-trivial pumps in this case are classified by a choice of one-dimensional irrep of $\tilde{G}$ \cite{Bachmann24}. Let us represent this as 
\begin{align}
    \chi(g) \equiv \exp(\rmi \nu(g)) \ ,
\end{align}
where we fix the branch $0\leq \nu(g)<2\pi $.
Since $\chi(g)$ is a representation, we have 
\begin{align}
    \chi(g)&\chi(h) = \chi(gh) \nonumber \\&\implies \nu(g) + \nu(h)  = \nu(gh) + 2 \pi n(g,h)
\end{align}
where $n(g,h) \in \mathbb{Z}$. For our purpose, we will need an interpolation between the trivial and non-trivial irrep. The following parameterisation is sufficient
\begin{align}
    \mu_\theta(g) = \exp\left(\rmi \frac{\theta}{2\pi}\nu(g) \right). 
\end{align}

To write down the non-trivial family of models, $W_\theta H_0 W_\theta^\dagger$, we consider the on-site Hilbert space to be the regular representation of $\tilde{G}$ with basis states $\ket{g \in \tilde{G}}$ labeled by the elements of $\tilde{G}$. We also choose, without loss of generality, the branching structure (see \Cref{app:cohomologysetting}) on the one-dimensional lattice such that all edges have the same orientation. This gives us the following form for the entangler $W_\theta $:
\begin{align}
    \prod_{j} \sum_{ g_j,g_{j+1}}\exp\left(\rmi \frac{\theta}{ 2\pi}\nu(g_{j}^{-1}g_{j+1})\right) \outerproduct{g_{j},g_{j+1}}{g_{j},g_{j+1}}\ , \label{eq:Entangler_trivial_1d}  \end{align}
which can be written as $ e^{-\rmi\theta \tilde{H} }$ for the pivot Hamiltonian
\begin{align}\tilde{H} &= -\frac{1}{ 2\pi} \sum_j \sum_{ g_j,g_{j+1}}\nu(g_{j}^{-1}g_{j+1}) \outerproduct{g_{j},g_{j+1}}{g_{j},g_{j+1}} \ . \label{eq:Pivot_cohomology}
\end{align}
Note that for $\theta= 2\pi$, each gate in Eq.~\eqref{eq:Entangler_trivial_1d} reduces to $\chi(g_j)^* \chi(g_{j+1})$. Hence, with periodic boundary conditions, we have $W_0 = W_{2\pi} = \mathbb{I}$, whereas with open boundaries, we have $W_0 =  \mathbb{I}$ but $W_{2\pi}$ equals
\begin{align}
\left(\sum_{g_1} \chi(g_1)^* \outerproduct{g_1}{g_1} \right) \otimes \mathbb I_{|\tilde G|}^{\otimes {L-2}} \otimes \left(\sum_{g_L} \chi(g_L) \outerproduct{g_L}{g_L} \right).
\end{align}
This is consistent with the fact that the charge $\chi$ is pumped by the family $W_\theta$.

In group cohomology Hamiltonian constructions \cite{Wen_GroupCohomology2013_PhysRevB.87.155114}, it is usual to start with
\begin{align}
    H_ 0 &= -\sum_{v \in V} \outerproduct{\Omega}{\Omega}_v,\nonumber\\ & \text{ where, }~\ket{\Omega}_v = \frac{1}{\sqrt{|\tilde{G}|}} \sum_{g \in \tilde{G}} \ket{g}_v \label{eq:H_trivial} \ .
\end{align}
This Hamiltonian has a unique ground state that is a product state invariant under the action of $g \in \tilde{G}$ (as $\ket{h}  \mapsto \ket{g^{-1}h}$), given by
\begin{align}
    \ket{\psi_0} = \prod_{v \in V} \ket{\Omega}_v. \label{eq:GS_trivial}
\end{align}
The family produced by pivoting the Hamiltonian \eqref{eq:H_trivial} by \eqref{eq:Pivot_cohomology} we will denote by $H_\theta$ and takes the form
\begin{align}
   - \sum \rme^{ \rmi \frac{\theta}{2\pi} \varphi(h_{j-1},g_j,l_j,h_{j+1})} \outerproduct{h_{j-1},g_j,h_{j+1}}{h_{j-1},l_j,h_{j+1}} \ , \end{align}
   where we sum over sites and group elements $h_{j-1},g_j,l_j, h_{j+1}$.
   The phase 
   $\varphi(h_{j-1},g_j,l_j,h_{j+1})$ 
 is equal to \begin{align}
  \nu(h_{j-1}^{-1} g_j) + \nu(g_j^{-1} h_{j+1}) - \nu(h_{j-1}^{-1} l_j) - \nu(l_j^{-1} h_{j+1}) \ . \end{align}
The wavefunction for the corresponding ground state family is 
\begin{align}
    \psi_\theta(g_1,\ldots,g_L) =\exp\left(\rmi \frac{\theta}{ 2\pi} \sum_j \nu(g_{j}^{-1}g_{j+1})\right)  \ .\label{eq:TrivialgroundstateFamily1d}
\end{align}

Note that we can find several symmetries of the pivot Hamiltonian \eqref{eq:Pivot_cohomology}. Define the unitary charge conjugation $C=\prod_j C_j$, where $C_j = \sum_{g_j} \ket{g_j^{}}\bra{g_j^{-1}}$, $T$ as complex conjugation in the $\ket{g_j}$ basis, and $P$ as a unitary inversion about some fixed site or edge (effectively reversing the orientation of each edge). Then, since for our choice of branch\footnote{Since $\chi(ab)=\chi(a)\chi(b) = \chi(b)\chi(a) = \chi(ba)$ we have $\nu(ab) = \nu(ba) \mod 2\pi$. Since we chose the range $0 \leq \nu(g) < 2\pi$, we have $\nu(ab) = \nu(ba)$.} $\nu(ab)=\nu(ba)$, $\tilde{H}$ has an anti-unitary $\mathbb{Z}_2^T$ time-reversal, a unitary $\mathbb Z_2^\textrm{CP}$ symmetry and the combined 
anti-unitary $\cpt$ symmetry. These different symmetries may be combined with \Cref{result:pumpantiunitary} to identify SPTs at $\pi$, depending on the charge pumped and the symmetries of the pivoted Hamiltonian. Taking $H_0$ above, this has all of the above symmetries, while if we take $A_0$ this is $\cpt$ symmetric only.

Indeed, fixing $H_0$, the Hamiltonian family $H_\theta$ is $\tilde{G}$ symmetric and remains in the trivial $\tilde{G}$-SPT phase throughout. It is, however, not $\cpt$, nor $\mathbb{Z}_2^T$ symmetric, since we have an antiunitary symmetry that reflects $\theta\rightarrow-\theta$. However, the two fixed points $\theta=0,\pi$ do have these symmetries. This means, if $\tilde{G}$ has a (pseudo)-real unitary representation, we can use \Cref{result:pumpantiunitary} to identify $H_\pi$ as a $\tilde{G}\rtimes\cpt$ and as a $\tilde{G}\times \mathbb{Z}_2^T$ SPT in the case where we pump a charge $\chi(g) \neq \tilde\chi(g)^2$ for some irrep $\tilde{\chi}$, i.e., when this charge cannot be decomposed into two identical smaller charges. The entire path is $\mathbb Z_2^\textrm{CP}$ symmetric, so we will not see any distinct SPT phases for this group.

\subsection{$\mathbb{Z}_N$ charge pumps and the Onsager ferromagnet}
Let us consider the particular case of $\tilde{G} \cong \mathbb{Z}_N$, where we label the elements of the group $m = 0,\ldots,N-1$. The $N$ different 1d irreps of $\tilde{G}$ are labelled $k=0,1,\ldots,N-1$ and have the representation
\begin{align}
    \chi^{(k)}(a)& = \exp\left(\frac{2 \pi \rmi }{N} k a\right) \nonumber\\&\implies  \mu^{(k)}_\theta(a) =  \exp\left(\frac{ \rmi\theta }{N} k a\right)\ . \label{eq:irreps}
\end{align}
The entangling operator, found by specialising \eqref{eq:Entangler_trivial_1d}, is
\begin{align}
    W^{(k)}_\theta &= \prod_{j} \sum_{a_j,a_{j+1} = 0}^{N-1}\rme^{-\rmi \theta w_k(a_j,a_{j+1})} \outerproduct{a_{j},a_{j+1}}{a_{j},a_{j+1}}\nonumber\\
    w_k&=\frac{ 1 }{N} k\Big(a_j - a_{j+1} \pmod N\Big)\ 
    . \label{eq:Znentangler}
\end{align}

Comparing to \cref{eq:A1pivot} we see that, up to an unimportant phase factor, $W^{(k)}_\theta = \rme^{-\rmi k \theta A_1}$. Hence, the pivot Hamiltonian \cref{eq:Pivot_cohomology} that generates $W^{(1)}_\theta$ for $\mathbb{Z}_N$, is given by 
\begin{align}
\tilde{H}^{(1)}_{\mathbb{Z}_N} =A_1 + \textrm{const} \ . \label{eq:A1_Zncohomology}
\end{align}

This is a surprising way to arrive at a model which appears as a limiting case of an integrable spin chain. Indeed, $A_1$ was originally described in Ref.~\onlinecite{vonGehlen84} as a generalisation of the Ising model, and most often arises as a Hamiltonian limit of the chiral Potts model in statistical mechanics \cite{Perk89,Davies90}.
A priori there is no connection to pumps constructed via group cohomology. 
One way to understand its appearance is as follows. Up to normalisation, the charge pump in \cref{eq:Znentangler} is generated by a Hamiltonian with integer-spaced eigenvalues \emph{on the bonds}. The appropriate bond-site transformation is Kramers-Wannier duality, or gauging the $\mathbb{Z}_N$ symmetry. The dual Hamiltonian, in this case $A_0$, is on-site and each term is equivalent (again, up to normalisation) to $S_z$ for a spin-$(N-1)/2$ \emph{in the appropriate basis}. $S_z$ is a canonical operator with integer-spaced eigenvalues on a finite Hilbert space, and so the appearance of $A_1$ given this integer spacing is somewhat natural, although the fact that the same basis is picked out by the group cohomology pump is intriguing.

We have established a link between the Onsager-integrable models $A_0$ and $A_1$ with $\mathbb{Z}_N$ pumps. Since all one-dimensional representations of non-abelian groups correspond to representations of an abelian subgroup, and abelian groups decompose as cyclic groups, we now explore to what extent all group-cohomology pumps can be connected to the Onsager-integrable models.

\subsection{Abelian charge pumps as stacks of Onsager ferromagnets}\label{sec:abelian}
\begin{figure}[!ht]
    \centering
\includegraphics[]{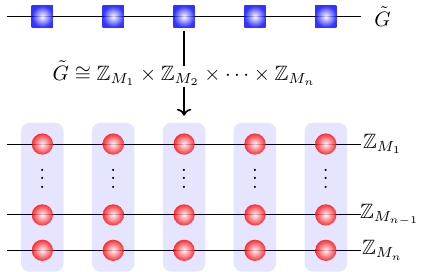}
\caption{Decomposition of the pivot Hamiltonian into disconnected stacks of Onsager ferromagnets when $\tilde{G}$ is an abelian group.}
    \label{fig:abelianchain}
\end{figure}

 To begin with, let us take $\tilde{G}$ to be a finite abelian group. Then $\tilde{G}$ decomposes as a direct product of cyclic groups $\tilde{G} \cong \mathbb{Z}_{M_1} \times  \mathbb{Z}_{M_2}  \times \cdots \times \mathbb{Z}_{M_n} $. Using this, we can write the elements of the group labelling the basis states in \cref{eq:Pivot_cohomology} as well as the 1d irreps $\chi(g)$ as
\begin{align}
    \ket{g } &\equiv \ket{m_1 ,m_2,\cdots, m_n} \text{ for } g \in \tilde{G}, m_k \in \mathbb{Z}_{M_k},\\
    \chi(g) &\equiv \chi_1(m_1) \chi_2(m_2) \cdots\chi_n(m_n)\nonumber\\
    &\implies \nu(g) \equiv \nu_1(m_1) + \nu_2(m_2) + \cdots +\nu_n(m_n) \ .
\end{align}
Here, $\chi_1,\chi_2,\cdots,\chi_n$ are irreps of $\mathbb{Z}_{M_1}, \mathbb{Z}_{M_2}, \cdots, \mathbb{Z}_{M_n}$. We can hence rewrite the pivot Hamiltonian in \cref{eq:Pivot_cohomology} as a sum of disconnected terms acting on disjoint Hilbert spaces,
\begin{align}
    \tilde{H} = \tilde{H}_1 + \tilde{H}_2 + \cdots +\tilde{H}_n \label{eq:Htilde_abelian}
\end{align}
where each term $\tilde{H}_k$ generates a $\mathbb{Z}_{M_k}$ charge pump of the form shown in \cref{eq:A1_Zncohomology}. Hence, $\tilde{H}_k$ can be identified with the $M_k$-state Onsager ferromagnet, that we will denote $A_1[M_k]$ (see \cref{eq:Onsager}). The overall pivot $ \tilde{H} $ corresponds to a stack of these ferromagnets, as schematically shown in \cref{fig:abelianchain}. Note that the initial state $\ket{\psi_0}$ (given in \eqref{eq:GS_trivial}) is the ground state of $\sum_k A_0[M_k]$, the $M_k$-state Onsager paramagnet acting on each chain in the stack. This proves the first part of \Cref{result:cohomology}.

\begin{figure}
    \centering
\includegraphics[]{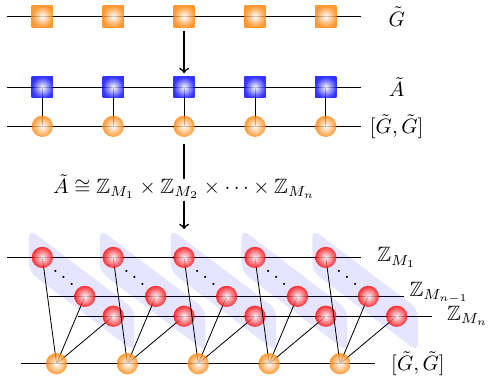}    \caption{Decomposition of the pivot Hamiltonian where $\tilde{G}$ is a non-abelian group. }
    \label{fig:Nonabelianchain}
\end{figure}

\subsection{Charge pumps with non-abelian symmetry}\label{sec:nonabelian}
\subsubsection{Pumps in the trivial phase}
Let us now consider the general case where $\tilde{G}$ is an arbitrary (finite) non-abelian group. Let $[\tilde{G},\tilde{G}]$ denote the commutator subgroup, i.e., the normal subgroup generated by elements in $\tilde{G}$ of the form $ghg^{-1}h^{-1}$. Given any 1d irrep $\chi$ of $\tilde{G}$, we thus have 
\begin{align}
    \chi(\alpha) = 1 ,~ \nu(\alpha) = 0 \qquad\forall \alpha \in [\tilde{G},\tilde{G}]\ . \label{eq:irrep_abeliansubgp}
\end{align}
Then let $\tilde{A}$ be the abelianization of $\tilde{G}$, i.e., the coset
\begin{align}
    \tilde{A} = \tilde{G}/[\tilde{G},\tilde{G}].
\end{align}
Note that this quotient group $\tilde{A}$ is abelian.

Since $\tilde G$ is isomorphic \emph{as a set} to the product of any normal subgroup and the corresponding quotient group, we can label group elements of $\tilde G$ in terms of elements in the commutator subgroup and the abelianization:
\begin{align}
    \ket{g} \equiv \ket{\alpha, a} \text{ where, } g \in \tilde{G},~\alpha \in [\tilde{G},\tilde{G}],~a \in \tilde{A}\label{eq:coset_partition} \ .
\end{align}
Note that although $\tilde A$ is in general not isomorphic to a subgroup of $\tilde G$, any 1d irrep $\chi$ of $\tilde G$ naturally induces a 1d irrep of $\tilde A$, since $\chi(a)$ is independent of the choice of coset representative. It is thus meaningful to write $\nu(a)$. In fact, using the correspondence in \cref{eq:irrep_abeliansubgp,eq:coset_partition}, we see that $\nu(g) = \nu(a)$. We can thus rewrite the Hamiltonian $\tilde{H}$ in \cref{eq:Pivot_cohomology} as follows 
 \begin{align} 
 -  \sum_{j;a \in \tilde{A}} \frac{\nu(a_j^{-1} a_{j+1})}{2\pi} \sum_{\alpha \in [\tilde{G},\tilde{G}]} \outerproduct{\alpha, a_j; \alpha, a_{j+1}}{\alpha, a_j;\alpha ,a_{j+1}}\ . \label{eq:Pivot_Cohomology_nonabelian}
\end{align}
To visualise  \cref{eq:Pivot_Cohomology_nonabelian} , we change basis to decompose the $|\tilde{G}|$ dimensional local Hilbert space into two pieces, one $|\tilde{A}|$ dimensional and the other $|[\tilde{G},\tilde{G}]|$ dimensional
\begin{align}
  \ket{g} =  \ket{\alpha , a} \rightarrow \ket{\alpha} \ket{a}. \label{eq:ladder_basis}
\end{align}
 This views the spin chain as a spin ladder, shown in \cref{fig:Nonabelianchain}. 
 
 The pivot Hamiltonian in this representation can be written as 
 \begin{align} 
 \tilde{H} = -\frac{1}{2\pi}  \sum_{j;a \in \tilde{A}} \nu(a_j^{-1} a_{j+1})  \outerproduct{ a_j,  a_{j+1}}{ a_j,  a_{j+1}}~ C_{j,j+1}. \label{eq:Pivot_Cohomology_nonabelian_ladder}
\end{align}
The local terms are identical to the abelian $\tilde{A}$ pump acting on the $a$ spins, tensored with the term $C_{j, j+1}$. The latter is a two-body operator acting only on the $|[\tilde{G},\tilde{G}]|$-dimensional Hilbert space, and is defined as 
\begin{equation}
    C_{j,j+1} = \sum_{\alpha \in [\tilde{G},\tilde{G}]} \outerproduct{\alpha \alpha}{\alpha \alpha}_{j,j+1}~. \label{eq:Cdef}
\end{equation}
In particular, this term projects onto aligned spins on the $\alpha$ spin chain. If these spins are not aligned, the local pivot term acts trivially on the $a$ spins. At this stage, just as we did for $\tilde{G}$ being an abelian group, we can decompose $\tilde{A}$ into cyclic groups as indicated in \cref{fig:Nonabelianchain}. This reduces the pivot Hamiltonian into stacks of $\mathbb{Z}_m$ Onsager ferromagnets, all of which are coupled to the $|[\tilde{G},\tilde{G}]|$ leg via the $C$ term defined in \cref{eq:Cdef}.

A crucial detail distinguishing abelian from non-abelian pumps obtained from exactly-solvable group cohomology data is seen in the way the pivot Hamiltonian acts on the trivial ground-state for $H_0$, defined in \cref{eq:H_trivial}. 
As discussed above, for the abelian case, the decomposition of the pivot Hamiltonian $\tilde{H}$ into a stack of Onsager ferromagnets $A_1$ is also respected by the ground-state, which is a product of ground-states of Onsager paramagnets $A_0$. Thus, the entire picture is `Onsager reducible'. For non-abelian pumps, we showed that the pivot Hamiltonian can be decomposed to local Onsager ferromagnet pieces coupled to the $[\tilde{G},\tilde{G}]$ chain via the $C_{j,j+1}$ term. However, to be totally reducible in terms of Onsager pivots we would need $C_{j,j+1}$ to act trivially on the ground-state of $H_0$. This ground state includes unaligned spins, so we do not find this reduction. Nevertheless, the only non-trivial action is generated by local terms of the Hamiltonian $A_1$.

\subsubsection{Onsager-reducible pump in a spontaneous symmetry breaking phase}
The decomposition \cref{eq:Pivot_Cohomology_nonabelian_ladder} does suggest an alternative Onsager-reducible pump within a spontaneous symmetry broken phase. Let us consider the the ladder basis defined in \cref{eq:ladder_basis} and choose the cycle-decomposed basis for $\tilde{A}\cong \mathbb{Z}_{M_1} \times \mathbb{Z}_{M_2} \times \ldots \times \mathbb{Z}_{M_n}$. We consider the pivot unitary $W_\theta$ applied to the following Hamiltonian invariant under $\tilde{G}$
\begin{align}
    H_0' = \sum_{k=1}^n A_0[M_k] + A_1[|[\tilde{G},\tilde{G}]|] \ .
    \label{eq:SSB_onsager_reducible}
\end{align}
 The Hamiltonian $H_0'$ represents a phase where $\tilde{G}$ is spontaneously broken to its $\tilde{A}$ subgroup with the following $|[\tilde{G},\tilde{G}]|$ ground states
\begin{align}
    \ket{GS_\alpha} =  \ket{\Omega[\tilde{A}]} \ket{\alpha \alpha \ldots \alpha},~~\alpha \in [\tilde{G},\tilde{G}]\ . 
\end{align}
The state $\ket{\Omega[\tilde{A}]}$ is the symmetry-preserving product-state invariant under $\tilde{A}$, the ground state of $H_0$ for the abelian subgroup. Overall, given  any $\tilde{G}$ charge $\chi$, this set up gives us a $\tilde{G}$-symmetric path through a symmetry-breaking phase, that pumps this charge using Onsager pivots. This completes the proof of \Cref{result:cohomology}.

We expect our analysis will generalise to time-reversal and crystal symmetries but we do not pursue this further here.

\section{Pivot SPT entanglers for $\mathbb{Z}_N\times\mathbb{Z}_N$ models and charge pumps}\label{sec:ZNZN}

In this section we construct pivot Hamiltonians for $\mathbb{Z}_N\times\mathbb{Z}_N$ SPTs. Unlike previous examples, the entangler no longer has a $\mathbb{Z}_2$ action, rather it takes $N$ steps to return to the trivial model. We also have multiple entanglers, related by the second $\mathbb{Z}_N$ symmetry. This means we find a more intricate structure of charge pumps as we combine these entanglers. In fact, unlike the Ising and Onsager examples in the previous sections, these pivots will \emph{not} give rise to a strict circular loop, i.e., the Dolan-Grady relations will not hold. Nevertheless, some of our more general results linking pumps and SPTs will still apply. We first introduce the model, then discuss the charge pumps and relationship to SPT order in these models.

\subsection{$\mathbb{Z}_N \times \mathbb{Z}_N$ models: the cluster model and the dipolar SPT}
There are $N$ SPT phases, labelled by cohomology class $[k]$, for spin chains with $\mathbb{Z}_N \times \mathbb{Z}_N$ symmetry~\cite{Wen_GroupCohomology2013_PhysRevB.87.155114}. For each $N$, the canonical representative for the class $[1]$ is the $\mathbb{Z}_N \times \mathbb{Z}_N$ cluster model \cite{Geraedts14,Santos15} with Hamiltonian $H_C$. This is given by:
\begin{align}
 -\sum_{0\leq j<L/2} \Big( Z^{-1}_{2j-1}X_{2j}^{\vphantom\dagger} Z_{2j+1}^{\vphantom\dagger}+Z^{\vphantom\dagger}_{2j}X^{\vphantom\dagger}_{2j+1} Z_{2j+2}^{-1} + h.c. \Big) \ ,
\end{align}
with symmetry generators $Q_{\mathrm{even}}=\prod_{0\leq j<L/2} X_{2j}$, $Q_{\mathrm{odd}}=\prod_{0\leq j<L/2} X_{2j+1}$. The class $[k]$ is represented by the Hamiltonian $H_C^{(k)}$, given by
\begin{align}
-\sum_{0\leq j<L/2} \Big( Z^{-k}_{2j-1}X_{2j}^{\vphantom\dagger} Z_{2j+1}^{k\vphantom\dagger}+Z^{k\vphantom\dagger}_{2j}X^{\vphantom\dagger}_{2j+1} Z_{2j+2}^{-k} + h.c. \Big)\ .\label{eq:clustermodels}
\end{align}
We discuss pivot entanglers and pumps in the cluster model in \Cref{sec:cluster} below, but we first discuss a related model.
Note that the cluster model is not translation-invariant; in Ref.~\onlinecite{Han23} a translation-invariant cousin of the cluster model was introduced, with a $\mathbb{Z}_N\times\mathbb{Z}_N$ symmetry generated by  $Q=\prod_j X_j$ and $D= \prod_j X_j^j$. The latter is a conventional modulated (in this case dipolar) symmetry for a chain of length $L=0\mod N$ but is a \emph{bundle symmetry} for any length of chain. The SPT Hamiltonian for the class $[k]$ is given by
\begin{align}
H_D^{(k)} = - \sum_j\left( Z_{j-1}^{-k}(Z_j^{k\vphantom\dagger} X_j^{\vphantom\dagger} Z_j^{k\vphantom\dagger}) Z_{j+1}^{-k} +h.c. \right)~. \label{eq:dipolarmodels}
\end{align} 

Next we demonstrate SPT entanglers generated by $\mathbb{Z}_N^{(Q)}$-symmetric pivot Hamiltonians for the $\mathbb{Z}_N\times\mathbb{Z}_N$ SPT Hamiltonians $H_D^{(k)}$ and $H_C^{(k)}$ given in \cref{eq:clustermodels,eq:dipolarmodels}. 
These pivot loops are not strict circular loops, and their structure is more intricate.
From the decorated domain wall picture, such a pivot SPT entangler naturally visits the SPT Hamiltonians at regular angles spaced by $2\pi/N$. We show that for $N$ odd the full loop is not a $\mathbb{Z}_N^{(Q)}$ charge pump, while for even values of $N$ it pumps charge $N/2$. 

\subsection{Dipolar SPT entangler}\label{sec:dipolar}
Our first result gives a family of pivot Hamiltonians for the dipolar SPT (where the protecting symmetry group includes the dipole symmetry $D$).  
\begin{shaded}
\begin{result}[Dipolar SPT entangler]\label{result:dipolar}
Define the $N$ pivot Hamiltonians and corresponding pivot unitaries by
\begin{align}
\tilde{H}_D^{(r)} &=\sum_j \sum_{m=1}^{N-1} \omega^{-mr} \alpha_{ m}^{} \alpha_{- m}^{} Z_j^{-m} Z_{j+1}^{m} \nonumber\\&\equiv\frac{1}{4}\sum_j \sum_{m=1}^{N-1} \frac{\omega^{-mr}}{\sin(m \pi/N)^2}  Z_j^{-m} Z_{j+1}^{m}\nonumber\\& \equiv D^{-r}\tilde{H}_{D}^{(0)}D^{r}
\end{align}
and
\begin{align}
U_D^{(r)}(\theta)&= \rme^{-\rmi \theta \tilde{H}_D^{(r)}}.
\end{align}
Then, for all choices of $0\leq r\leq N-1$, \begin{align}U_D^{(r)}(2\pi k/N)H_D^{(0)} U_D^{(r)}(-2\pi k/N) =H_D^{(k)} \ .\end{align}
\end{result}
\end{shaded}
Note that $\tilde{H}_D^{(r)}$ and $\tilde{H}_D^{(-r)}$ are related by charge conjugation, while each pivot Hamiltonian is time-reversal invariant. Since $\sum_{r=0}^{N-1} \tilde{H}_D^{(r)}=0$, we have $N-1$ independent pivot generators. The proof of \Cref{result:dipolar} is in \Cref{app:pivotentangler}, and relies on the trigonometric identity \eqref{eq:trig} that is key in the analysis of the Onsager-integrable chiral clock model. Moreover, this allows us to show that $\tilde{H}_D^{(r)}$ has integer eigenvalues, and thus generates a $U(1)$, up to rescaling and a constant shift. This follows from noting that for $\varphi_k = -\sum_{m=1}^{N-1} \alpha_m\alpha_{-m} \omega^{-m k}$, we have $\varphi_{k+1}-\varphi_{k} = \sum_{m=1}^{N-1} \alpha_m \omega^{-m k}$, leading to $\varphi_k = k(N-k)/2-(N^2-1)/12$.

\subsection{Dipolar pump}
The pivot unitary $U_D^{(r)}$ can be written as a product of two-site gates \begin{align}
    U^{(r)}_{j,j+1}(\theta) = \exp\left(-\rmi \theta\sum_{m} \omega^{-mr} \alpha_{ m}^{} \alpha_{- m}^{} Z_j^{-m} Z_{j+1}^{m}\right) \ .
\end{align}
For $t=2\pi$ we have, using the formula for $\varphi_k$ above, that 
\begin{align}
U^{(r)}_{j,j+1}(2\pi)=  \begin{cases}
\rme^{2\pi\rmi \varphi_0} \mathbb{I} \qquad & N~\mathrm{odd}\\
\rme^{2\pi\rmi\varphi_0} (-1)^rZ_j^{N/2}Z_{j+1}^{N/2} \qquad & N~\mathrm{even}\ .
\end{cases}\label{eq:dipolarpump}
\end{align}
 Using the discussion in \Cref{sec:pumpMPU}, we then conclude that going around the pivot loop gives a trivial charge pump for $N$ odd, and that we pump charge $N/2$ for $N$ even.

 Each of the pivot Hamiltonians $\tilde{H}_D^{(r)}$ gives a $U(1)$ symmetry of the Hamiltonian $\sum_{k=0}^{N-1} H_D^{(k)}$ (see Appendix \ref{app:u1}). Since we have $N-1$ independent pivots the overall symmetry group includes $\mathbb{Z}_N^Q\times(U(1)^{N-1}\rtimes \mathbb{Z}_N^D)$, as well as $C$, $P$ and $T$ symmetries. 

\subsection{Relation between pumps and SPTs for the dipolar entangler}
\subsubsection{Pump as we go round the full pivot loop gives an SPT for $N=2(2k+1)$}
The above analysis shows that we cannot expect a general relationship between pivot SPT entanglers and charge pumps as we go round a full loop. However, for $N=2(2k+1)$ we can use the non-trivial pump around the $2\pi$ pivot loop to derive SPT order half-way around the loop at $H_D^{(N/2)}$.

Indeed, since the pivot Hamiltonian $\tilde{H}_D^{(0)}$ is real, we can use  \Cref{result:pumpantiunitary} to establish that, for such $N$, $H_D^{(N/2)}$ is an SPT protected by the group $\mathbb{Z}^{}_N\times\mathbb{Z}_2^T$. This is because for $N=2(2k+1)$, $N/2$ cannot be decomposed as two identical charges. The same argument applies for the $\mathbb{Z}^{}_2\times\mathbb{Z}_2^T$ subgroup, since we pump a unit $\mathbb{Z}_2$ charge for these values of $N$. Even without knowledge of the $\mathbb{Z}_N\times\mathbb{Z}_N$ dipolar SPT, it is not surprising to find SPT physics since we can see that
\begin{align}
H_D^{(N/2)}    &= \sum_j Z_{j-1}^{N/2}(X_j +X_j^{-1}) Z_{j+1}^{N/2} \ 
\end{align}
 has the same $\mathbb{Z}_2\times\mathbb{Z}_2^T$ SPT order as the usual spin-1/2 cluster model. Indeed, since $X_j \ket{\psi_0} = \ket{\psi_0}$ in the ground state of 
$H_D^{(0)}$, the ground state of $H_D^{(N/2)}$ has long-range string order $\langle Z_0^{N/2} X_1^{N/2} Z_1^{N/2} \left(\prod_{j=2}^{M-1} X_j^{N/2}\right) Z_M^{N/2}X_M^{N/2} Z_{M+1}^{N/2}\rangle $. This has Hermitian end-point operator $\rmi Z_0^{N/2} X_1^{N/2} Z_1^{N/2}$, which is odd under $\mathbb{Z}_2^T$ \cite{Smacchia11,Pollmann12}.

Note that if $N$ is divisible by four we cannot conclude using \Cref{result:pumpantiunitary} that $H_D^{(0)}$ and $H_D^{(N/2)}$ are in distinct SPT phases (importantly this does not mean they are in the same SPT phase, just that being in the same phase would not lead to an inconsistency). Indeed, the ground state of $H_D^{(N/2)}$ has long-range order in 
 $\langle Z_0^{N/2} X_1^{} Z_1^{N/2} \left(\prod_{j=2}^{M-1} X_j^{}\right) Z_M^{N/2}X_M^{} Z_{M+1}^{N/2}\rangle $, which has $\mathbb{Z}_2^T$ charged end-points, and thus we have $\mathbb{Z}_N^{}\times\mathbb{Z}_2^T$ SPT order. However, we no longer have a $\mathbb{Z}^{}_2\times\mathbb{Z}_2^T$ SPT for $N=4k$, since we also have long-range order in $\langle  \prod_{j=1}^{M} X_j^{N/2} \rangle $. 
\subsubsection{Pump as we go round a loop that visits a single SPT}
From the decorated domain wall picture \cite{Bultinck19}, we cannot have a pivot Hamiltonian that generates a $\mathbb{Z}_2$ entangler unless the corresponding SPT is such that a double stack of the SPT is trivial. Equivalently, in our setting, if $U(\theta)$ entangles the SPT $H_D^{(1)}$ then $U(2\theta)$ necessarily entangles $H_D^{(2)}$. Hence, if we want a loop that takes us from $H_D^{(0)}$ to $H_D^{(1)}$ and back, it cannot be generated by a pivot Hamiltonian.

We can, however, easily construct such loops from pairs of pivot Hamiltonians defined above. In particular consider the loop 
\begin{align}
&U^{(r,s)}(\theta) \nonumber\\
&= \begin{cases}
   U_D^{(r)}(2\theta/N) \qquad &0\leq \theta \leq \pi\\
     U_D^{(s)}(2(\pi-\theta)/N)   U_D^{(r)}(2\pi/N)   \qquad &\pi \leq \theta \leq 2\pi
\end{cases} \ . \label{eq:smallloop}
\end{align}
For $r=s$ this is a trivial contractible loop. For $r\neq s$ we can consider the gate $U^{(r,s)}_{j,j+1}(2\pi)$, which is given by
  \begin{align}
&\rme^{-\frac{2\pi\rmi}{N}\sum_{m}  \alpha_{ m}^{} \alpha_{- m}^{}\left(\omega^{-mr} Z_j^{-m} Z_{j+1}^{m}- \omega^{-ms} Z_j^{-m} Z_{j+1}^{m}\right)} \nonumber\\
&=\rme^{\rmi \theta} Z_j^{-(r-s)}Z_{j+1}^{r-s} \end{align}
where\begin{align}\rme^{\rmi \theta} = (-1)^{r-s}\omega^{\frac{s^2-r^2}{2}} \  .
\end{align} 
Thus each path  $U^{(s+1,s)}(2\pi)$ fixes $H_D^{(k)}$ and pumps a unit charge.

Following Ref.~\onlinecite{Bultinck19}, we note that the operator $ U_D^{(0)}(2\pi/N)\mathcal{K}$ is a $\mathbb{Z}_2^T$ SPT entangler (as is $ U_D^{(r)}(2\pi/N)\mathcal{K}$ for each $r$), exchanging $H_D^{(0)}$ and $H_D^{(1)}$.
For even $N$, since $H_D^{(1)}$ is not the square of another SPT, there is a mixed anomaly between $\mathbb{Z}_N\times\mathbb{Z}_N$ and this entangler. This gives rise to an anomalous symmetry at the point $H_D^{(r)}+H_D^{(r+1)}$, which therefore cannot have a unique gapped ground state. For odd $N$, there is no anomaly since $H_D^{(1)}$ is in the same phase as  two copies of $H_D^{(\frac{N+1}{2})}$---moreover, $U_D^{(0)}(2\pi/N)\mathcal{K}$ is a symmetry of $H_D^{(\frac{N+1}{2})}$ which has a unique gapped ground state. We know on general SPT grounds that $(1-\lambda)H_D^{(r)}+\lambda H_D^{(r+1)}$ has a phase transition for some value of $\lambda$ (and if there is a unique transition it occurs at the self-dual $\lambda=1/2$). In any case, and for all $N$, there is some obstruction to gapped paths along this line. 

Since $U^{(0,1)}(\theta)$ is a non-trivial pump, this obstruction must continue into the space of Hamiltonians with explicitly broken $\mathbb{Z}_N^D$ symmetry. In fact, it must split symmetrically into $N$ equivalent obstructions, related by the $\mathbb{Z}_N^D$ symmetry. Indeed $U^{(s,s+1)}(\theta)=D^{-s}U^{(0,1)}(\theta)D^{s}$, and so enclosing any single one of these obstructions corresponds to pumping a unit charge  (this is essentially the argument we use to prove \Cref{result:pumpspt}). The loop $U^{(r,s)}(\theta)$ then encloses $s-r$ of these obstructions (the sign indicates orientation of the loop) and therefore pumps charge $s-r$, just as we found by direct calculation.

We can also use these pump results to argue that $H_D^{(k)}$ are in $N$ distinct SPT phases. Indeed, the loop $H(\theta)=V(\theta)H_D^{(r)}V(\theta)^\dagger $, for $V(\theta)$ given by
\begin{align}
\begin{cases}
   U_D^{(0)}(2s \theta/N) \quad &0\leq \theta \leq \pi\\
     U_D^{(1)}(2s (\pi-\theta)/N)   U_D^{(0)}(2s \pi/N)   \quad &\pi \leq \theta \leq 2\pi
\end{cases} \ ,
\end{align}
is a non-trivial pump of the form in \Cref{result:pumpspt} ,where $H_0=H_D^{(r)}$ and $H_\pi =H_D^{(r+s)}$. Hence, no pair of $H_D^{(k)}$ can be connected by a $D$-symmetric path.
\subsubsection{Connecting these two pumps}
The individual obstructions found above can also show that the $2\pi$ pivot $U_D^{(s)}(2\pi)$ pumps charge $0$ ($N/2$) for $N$ odd (even). The key is that since $\sum_r \tilde{H}_D^{(r)}=0$ and all $\tilde{H}_D^{(r)}$ commute, we have 
\begin{align}
\prod_{r=0}^{N-2}U_D^{(N-1)}\left(-\frac{2\pi }{N}\right)&= U_D^{(N-1)}\left(-\frac{2\pi (N-1)}{N}\right)\nonumber\\
&= \prod_{r=0}^{N-2}U_D^{(r)}\left(\frac{2\pi (N-1)}{N}\right) \end{align}
which can be rewritten
\begin{align} 
\prod_{r=0}^{N-2} \underbrace{\left(U_D^{(r)}\left(\frac{2\pi }{N}\right) U_D^{(N-1)}\left(-\frac{2\pi }{N}\right)\right)}_{\mathrm{each~pumps~charge}~(N-1-r)} = \underbrace{\prod_{r=0}^{N-2}U_D^{(r)}\left(2\pi \right)}_{N-1~2\pi~ \mathrm{pivots}} .\end{align}
Since (using the $D$ symmetry) each of the $2\pi$ pivots pumps the same charge, $q$, we find that $N(N-1)/2 = (N-1)q\mod N$, and so $q=0$ for $N$ odd, and $q=N/2$ for $N$ even. 

We can partially invert this logic. Suppose that the small loop $U^{0,1}(2\pi)$ pumps charge $k$, where $k$ is not known. For $N$ even, the fact that $U_D^{(r)}\left(2\pi \right) = N/2$ tells us that $k$ is an odd integer, and, in particular, that it must be non-trivial. We cannot make this conclusion for $N$ odd.

\subsection{Cluster SPT entanglers and pumps}\label{sec:cluster}
Our results for the cluster models are entirely analogous to the dipolar SPT, and identical discussion applies. The results are collected in this subsection.
\begin{shaded}
\begin{result}[Cluster entangler]
Define the $N$ pivot Hamiltonians and corresponding pivot unitaries by
\begin{align}
\tilde{H}_C^{(r)} &=\sum_j \sum_{m=1}^{N-1} \omega^{(-1)^jmr} \alpha_{ m}^{} \alpha_{- m}^{} (-1)^{j+1} Z_j^{-m} Z_{j+1}^{m} \nonumber\\&\equiv Q_{\mathrm{even}}^{-r}\tilde{H}_{C}^{(0)}Q_{\mathrm{even}}^{r}\end{align}
and\begin{align}
U_C^{(r)}(\theta)&= \rme^{-\rmi \theta \tilde{H}_C^{(r)}}.
\end{align}
Then, for all choices of $r$, \begin{align}U_C^{(r)}(2\pi k/N)H_C^{(0)} U_C^{(r)}(-2\pi k/N) =H_C^{(k)} \ .\end{align}
\end{result}
\end{shaded}
The proof uses the analysis of the dipolar SPT and is given in Appendix \ref{app:pivotentangler}. Then, fixing $r=0$ for simplicity and using \cref{eq:dipolarpump}, we can write 
\begin{align}
U_C^{(0)}(2\pi)=\prod_{j} \underbrace{U^{(0)}_{2j-1,2j}(2\pi)U^{(0)}_{2j,2j+1}(-2\pi)}_{V_{2j-1,2j,2j+1}}\\
V_{2j-1,2j,2j+1}=  \begin{cases}
 \mathbb{I} \qquad & N~\mathrm{odd}\\
Z_{2j-1}^{N/2}Z_{2j+1}^{N/2} \qquad & N~\mathrm{even}\ ,
\end{cases}
\end{align}
that is, we have the analogous charge pump behaviour around a full pivot loop.

For the small loops, the analogues of \cref{eq:smallloop}, we have that
$U^{(r,s)}_{j,j+1}(2\pi) = \rme^{(-1)^{j+1}\rmi \theta}Z_j^{-(r-s)}Z_{j+1}^{r-s} $. The discussion is then entirely analogous, and we can conclude that $H_C^{(r)}$ are distinct SPTs using \Cref{result:pumpspt}. On the line $(1-\lambda)H_C^{(r)}+\lambda H_C^{(r+1)}$ we have a phase transition, and note that Ref.~\onlinecite{Tsui17} show that the point $\lambda=1/2$ is a continuous transition for $N\leq 4$, while there is an extended gapless interval for $N>4$. 
\section{Dolan-Grady and the Onsager algebra}\label{sec:DG}
As stated above, the Dolan-Grady relation \eqref{eq:DG} is `half-way' to an Onsager algebra \cite{Perk89,Davies91,Perk16}. To generate a full Onsager algebra, we would also need $[A,B]=\Big[\big[[A,B],B \big],B\Big]$. Having both is automatic in the case where $A$ and $B$ are self-dual (in particular, if there is a linear operation that exchanges $A$ and $B$). Here, we will give an example of a strict circular pivot loop where the Dolan-Grady relation holds, but the initial and pivot Hamiltonian do not generate an Onsager algebra.

\subsection{Dolan-Grady without the Onsager algebra}
In the Onsager-integrable chiral clock models, where we take $A=A_1$, $B=A_0$, we do have an underlying Onsager algebra and recover the usual strict circular loop. A non self-dual spin chain giving rise to an Onsager algebra is found in Ref.~\onlinecite{Davies91b}, but we are not aware of any other examples. Finding non self-dual $A$ and $B$ that satisfy one Dolan-Grady relation is more straightforward. Indeed, several examples are given in \cite{Naudts09,Naudts12}, but we will derive a new solution from the chiral clock family below.

Consider complex conjugation in the $Z$-basis, denoted $\mathcal{K}$. Then $\mathcal{K}A_1\mathcal{K}=A_1$, while $\overline{A}_0=\mathcal{K}A_0\mathcal{K}    \neq A_0$ for $N>2$. Complex conjugation preserves the Dolan-Grady relations satisfied by $A_0$ and $A_1$, and so $\overline{A}_0$ and $A_1$ themselves generate an Onsager algebra. By linearity it follows that $B=\alpha A_0+ \beta \overline{A}_0$ and $A=A_1$ will satisfy the Dolan-Grady relation \eqref{eq:DG} for all $\alpha,\beta$. The dual relation is non-linear in $B$ and so we do not expect it to satisfy the dual Dolan-Grady relation. Fixing $B=A_0+\overline{A}_0$, and using that $[A_0,\overline{A}_0]=0$, we find that the dual Dolan-Grady relation can be written
\begin{align}
\Big[\big[[A_1, &A_0+  \overline{A}_0], A_0 +  \overline{A}_0\big], A_0+  \overline{A}_0\Big]\nonumber\\
&=[A_1,A_0+  \overline{A}_0]
+3\Big[\big[[A_1,A_0],{A}_0\big],\overline{A}_0\Big]
\nonumber\\&\qquad+3\Big[\big[[A_1,A_0],\overline{A}_0\big],\overline{A}_0\Big]\nonumber\\
&=_? \gamma^2 [A_1,A_0+  \overline{A}_0]\ , \label{eq:dualDG}
\end{align}
for some normalisation $\gamma$.
For $N=2$, since $A_0=\overline{A}_0$, the right-hand-side is $8[A_1,A_0]$, and this gives the usual dual-relation.
For $N>2$, $A_0\neq \overline{A}_0$ and we do not expect this to hold for any $\gamma$. We prove this in Appendix \ref{app:DG}.

\subsection{A strict circular loop without the Onsager algebra---pivoting the Potts model with the Onsager ferromagnet}\label{sec:potts}

Having established that this example satisfies the Dolan-Grady relation and not the full Onsager algebra, we now study in more detail the resulting pivot loop (which is naturally a strictly circular loop per Sec.~\ref{sec:circularloop}).
The Hamiltonian $A_0+\overline{A}_0$ is equal to the Potts Hamiltonian ${H}_{\mathrm{Potts}}= -\frac{1}{N} \sum_{j} \sum_{m=1}^{N-1} X_j^m$. Diagonalisation is straightforward and the paramagnetic ground state coincides with the ground state of $A_0$. Pivoting with $A_1$ we have a strict circular loop passing $H_2 = A_2 +\overline{A}_2$ at the half-way point. This Hamiltonian has the form
\begin{align}
H_2&= -\frac{1}{N} \sum_{j} \sum_{m=1}^{N-1} S_{j-1,j}^{(m)} X_j^mS_{j,j+1}^{(m)}\ ,   \nonumber \end{align}
with $S$ defined in \Cref{eq:A2def}.
Both Hamiltonians have a $\mathbb{Z}_N\times \mathbb{Z}_2^T$ symmetry, while the circular loop is $\mathbb{Z}_N$ symmetric.
\subsubsection{$H_2$ is an SPT for even $N$ using the pump invariant}
In fact, the circular loop is of the form:
\begin{align}
 \rme^{-\rmi A_1 \theta}   {H}_{\mathrm{Potts}} \rme^{\rmi A_1 \theta}\hspace{5cm}\nonumber\\ = \begin{cases}
 \rme^{-\rmi A_1 \theta}   {H}_{\mathrm{Potts}} \rme^{\rmi A_1 \theta}  \qquad & 0\leq \theta\leq \pi\\
 \mathcal{K} \rme^{-\rmi A_1(2\pi- \theta)}   {H}_{\mathrm{Potts}} \rme^{\rmi A_1 (2\pi- \theta)} \mathcal{K}   \qquad & \pi\leq \theta\leq 2\pi\\
\end{cases}
\end{align}
where we use that $\rme^{-\rmi 2\pi A_1} \propto \mathbb{I}$. Since the $\mathbb{Z}_N$ symmetry is real, this fits the statement of \Cref{result:pumpantiunitary}. 

From \Cref{section:A1pump}, we know that this loop pumps a single $\mathbb{Z}_N$ charge. For $N$ even, this cannot be written as the sum of two identical charges. Hence, by \Cref{result:pumpantiunitary} and that ${H}_{\mathrm{Potts}}$ is trivial, $H_2$ is a non-trivial $\mathbb{Z}_N^{}\times\mathbb{Z}_2^T$ SPT. For $N$ odd, $1=2 \frac{(N+1)}{2} \pmod{N}$, and so we cannot conclude $H_2$ is a non-trivial SPT. However, there will be a boundary transition around the loop, and, if it is unique, then $H_2$ will have a gapless boundary.

\subsubsection{$H_2$ is an SPT for even $N$ using symmetry fractionalisation}
Since the ground state of ${H}_{\mathrm{Potts}}$ coincides with that of $A_0$, the ground state of $H_2$ coincides with the ground state of $A_2$.  

To study the symmetry fractionalisation of $\mathbb{Z}_N^{}\times\mathbb{Z}_2^T$, recall that $
\ket{\psi_2}$ is an MPS with tensor $\mathcal{A}_j^{\alpha,\beta}$ given by \begin{align}\underbrace{N^{-\frac{1}{2}}~\omega^{j(\beta-\alpha)}  \omega^{\beta/2}}_{\Gamma_j^{\alpha,\beta}}  ~ \underbrace{N^{-1} ~\left(\sin\left(\frac{\pi(2\beta+1)}{2N}\right)\right)^{-1}}_{\Lambda_{\beta}}\label{eq:MPSpsi2}.
\end{align}
We know \cite{Jones25} that the $\mathbb{Z}_N$ symmetry fractionalises as
\begin{align}
\sum_{b=0}^{N-1} X_{a,b} \, \Gamma_{b} 
=Z^\dagger \Gamma_a Z \ .
\end{align} For the time-reversal symmetry we have:
\begin{align}
\overline\Gamma^{\alpha,\beta}_j = \rme^{\rmi \varphi} V ~\Gamma^{\alpha,\beta}_j~ V^\dagger
\end{align}
for some $\varphi$ and $V$ that commutes with $\Lambda$. Since $\overline\Gamma^{\alpha,\beta}_j = N^{-\frac{1}{2}}~\omega^{j(\alpha-\beta)}  \omega^{-\beta/2}$, this holds for $V = V^\dagger = \sum_{j=0}^{N-1} \ket{j}\bra{N-1-j}$ and  $\rme^{\rmi \varphi}=\omega^{-\frac{N-1}{2}}$.

Analogous to the $\cpt$ case, for $N$ even we have a $\mathbb{Z}_2^{}\times \mathbb{Z}_2^T$ subgroup that is realised projectively on the bond space. Note that for this group there is more than one non-trivial SPT phase, due to the time reversal symmetry. In addition to the sign found above, we have that $V^2 = \pm 1$. For all $N$ this is equal to $+1$ in this model.

\section{SPT phases implied by pumps}\label{sec:pumpSPT}
In the previous sections, we have encountered a myriad of examples of pivots, where the presence of a pump often (but not always) went hand-in-hand with a non-trivial SPT at high-symmetry points. Here, we return to the general question of when we can conclude that we have a non-trivial SPT due to a non-trivial pump, deriving \Cref{result:pumpspt,result:pumpantiunitary}. We note that our analysis can also be applied in cases beyond those summarised in those results.

\subsection{Symmetric paths between Hamiltonians forbid certain pump invariants}
Consider $H_{\tilde{G}}(\theta)$, a gapped $\tilde{G}$-symmetric path between Hamiltonians $H_0$ and $H_\pi$ (for $\theta=0$ and $\theta=\pi$ respectively). We suppose that these end-point Hamiltonians have a strictly larger symmetry group described by $G$, of which $\tilde{G}$ is a normal subgroup. Since $\tilde{G}\subsetneq G$ then $H_0$ and $H_\pi$ can be in distinct $G$-SPT phases. (The strict subgroup condition implies the path $H_{\tilde{G}}(\theta)$ is not accidentally $G$-symmetric, meaning distinct $G$-SPT phases are possible.) Our aim is to establish a link to pump invariants, enabling us in some cases to prove that two Hamiltonians are in distinct SPT phases by analysing only pumps.
\subsubsection{Unitary symmetries}
Let us first assume that all group elements are represented as on-site unitaries acting on our system, and that our $\tilde{G}$-symmetric Hamiltonians are $d$-dimensional. 
To make a link to pump invariants, we need to construct a closed loop from our path. Let us fix some non-trivial element $g_0\in G/\tilde{G}$, then we have a new gapped $\tilde{G}$-symmetric path $g_0^{}H_{\tilde{G}}(\theta)g_0^\dagger$ from $H_0$ to $H_\pi$.
Then we can define a loop
\begin{align}
H_A(\theta) = \begin{cases}
    H_{\tilde{G}}(\theta) \qquad & 0\leq \theta \leq \pi  \\
    g_0^{}H_{\tilde{G}}(2\pi-\theta)g_0^\dagger & \pi \leq \theta \leq 2\pi
\end{cases}  \label{eq:Aloop}
\end{align}
where this loop may pump a $(d-1)$-dimensional $\tilde{G}$-SPT corresponding to a cocycle $\rme^{\rmi\nu_A(\tilde{g}_1,\dots,\tilde{g}_d)}$. 

Now suppose that $H_0$ and $H_{\pi}$ are in the same $G$-SPT phase. This means there is a gapped $G$-symmetric path, $H_{{G}}(\theta)$, connecting them---illustrated in Fig.~\ref{fig:paths}--- and we can consider two further loops
\begin{align}
H_B(\theta) &= \begin{cases}
    H_{\tilde{G}}(\theta) \qquad & 0\leq \theta \leq \pi  \\
    H_G(2\pi -\theta) & \pi  \leq \theta \leq 2\pi 
\end{cases}  \nonumber\\ H_C(\theta) &= \begin{cases}
    g_0^{} H_{\tilde{G}}(\theta)g_0 ^\dagger \qquad & 0\leq \theta \leq \pi   \\
    H_G(2\pi -\theta) & \pi  \leq \theta \leq 2\pi 
\end{cases}  \ .
\end{align}
These loops pump $\rme^{\rmi\nu_B(\tilde{g}_1,\dots,\tilde{g}_d)}$ and $\rme^{\rmi\nu_C(\tilde{g}_1,\dots,\tilde{g}_d)}$, and, since we have a normal subgroup, they satisfy 
$\rme^{\rmi\nu_C(\tilde{g}_1,\dots,\tilde{g}_d)} \equiv \rme^{\rmi\nu_B(g_0^{-1}\tilde{g}_1^{}g_0^{},\dots,g_0^{-1}\tilde{g}_d^{}g_0^{})}$.
Moreover, noting that the original loop $H_A$ corresponds to 
\begin{align}
H_A(\theta) = \begin{cases}
   H_B(\theta) \qquad & 0\leq \theta \leq \pi  \\
  H_C(2\pi -\theta) & \pi  \leq \theta \leq 2\pi  \ ,
\end{cases}  
\end{align}
which is homotopic to following $H_B$ (over its full range $0 \leq \theta \leq 2\pi$) and then the reversed $H_C$ (the $G$-symmetric part cancels),
we have that \begin{align}\rme^{\rmi\nu_A(\tilde{g}_1,\dots,\tilde{g}_d)} &=\rme^{\rmi(\nu_B(\tilde{g}_1,\dots,\tilde{g}_d)-\nu_C(\tilde{g}_1,\dots,\tilde{g}_d)}\nonumber\\& =\rme^{\rmi(\nu_B(\tilde{g}_1,\dots,\tilde{g}_d)-\nu_B(g_0^{-1}\tilde{g}_1^{}g_0^{},\dots,g_0^{-1}\tilde{g}_d^{}g_0^{})} \ .\end{align}

This is a restriction on the allowed pumps $\nu_A$ given that $H_0$ and $H_\pi$ are the same SPT phase. 
A key case of interest is when $g_0^{}\tilde{g}g_0^{-1} =\tilde{g}$ (i.e., $g_0$ is in the centraliser $C_G(\tilde{G})$). In this case, we find $\nu_A=0$, i.e., $H_A(\theta)$ must be a trivial pump. This gives \Cref{result:pumpspt}; i.e., the contrapositive says that if the pump is non-trivial (in this setting), then $H_0$ and $H_\pi$ must be in distinct SPT phases. The case where $g_0$ is a $\mathbb{Z}_2$ element is proved in greater generality in the next section, leading to \Cref{result:Mayer-Vietoris}.

Another interesting case is charge conjugation in $\tilde{G}$, where $g_0^{}\tilde{g}g_0^{-1} =\tilde{g}^{-1}$.
If we have a one-dimensional pump, then $\nu_B(\tilde{g}^{-1})=-\nu_B(\tilde{g}^{})$. Hence we find   $\nu_A(\tilde{g})=2 \nu_B(\tilde{g})$, so we must have that the $0$-dimensional $\tilde{G}$-charge pumped around $H_A(\theta)$ can be decomposed into a stack of two identical smaller such charges---otherwise $H_0$ and $H_\pi$ must be in distinct SPT phases. 

\subsubsection{Anti-unitary symmetries}\label{sec:antiunitary}
Let $\mathcal{K}$ be complex conjugation in a convenient basis. Just as in the classification of SPTs \cite{Cirac21}, if this is a symmetry, then this operation adds an additional twist relative to the on-site unitary case. Let us consider the case where $G=\tilde{G}\times \mathbb{Z}_2^T$ and where $\mathcal{K}$ acts in a basis such that $\tilde{g}$ is real (this is possible if we have a real or pseudo-real representation of $\tilde{G}$). Taking $g_0=\mathcal{K}$ in \cref{eq:Aloop}, we have that $\nu_C=-\nu_B$; and thus $\nu_A = 2\nu_B$. If $H_0$ and $H_{\pi}$ are in the same SPT phase, then the $(d-1)$-dimensional $\tilde{G}$-SPT pumped around $H_A(\theta)$ can be decomposed into a stack of two smaller such SPTs.

Restricting now to one-dimensional chains, for the Onsager-integrable clock models, we have a $\cpt$ action that decomposes as a product of unitary charge conjugation, inversion and time reversal (see Sec.~\ref{sec:circular}). Time reversal and inversion each conjugate $\nu_B(\tilde{g})$, while charge conjugation takes $\tilde{g}\rightarrow \tilde{g}^{-1}$. Overall we again find $\nu_C(\tilde{g}) = -\nu_B(\tilde{g})$ and so $\nu_A(\tilde{g}) = 2\nu_B(\tilde{g})$. Thus, we conclude in both cases that if $\nu_A(\tilde{g})$ cannot be decomposed into two smaller identical charges, then $H_0$ and $H_\pi$ are distinct SPTs. 

\subsubsection{Example: $\mathbb{Z}_2^3$ SPT in 2D}\label{sec:2dexample}
Above, we have a number of applications of these results to 1D bulk systems. Here we apply this to a two-dimensional bulk and consider the example of constructing the $\mathbb{Z}_2^3$ SPT using a pivot Hamiltonian, as described in Refs.~\onlinecite{Tantivasadakarn22,Tantivasadakarn23}. First, take qubits on the vertices of the three-colourable Union-Jack lattice, where each nearest-neighbour triangle, $\triangle_{abc}$, contains all three colours $A,B,C$. We then define 
\begin{align}
H_0 &= -\sum_j X_j \qquad\qquad 
\tilde{H}=-\frac{1}{8}\sum_{\triangle_{abc}}Z_aZ_bZ_c
\end{align}
such that the path $H(\theta) = \rme^{-\rmi \tilde{H}\theta}H_0\rme^{\rmi \tilde{H}\theta}$ is $\tilde{G}=\mathbb{Z}_2\times\mathbb{Z}_2$ symmetric with generators $P_{AB}=\prod_{j\in A \cup B}X_j$ and $P_{BC}=\prod_{j\in B \cup C}X_j$.

In Ref.~\onlinecite{Tantivasadakarn22} it is shown that $\rme^{-2\pi\rmi \tilde{H}}$ acts as identity in the bulk of the system, and as the 1+1D cluster model entangler on the boundary spins (if present). Hence, using the reasoning in \Cref{sec:pumpMPU}, we conclude that the path $H(\theta)$ is a non-trivial $\tilde{G}$-pump.
Let us now consider $P_{A}$ where $P_S=\prod_{j\in S}X_j$; this anticommutes with $\tilde{H}$ and so we can write
\begin{align}
H(\theta) = \begin{cases}
   \rme^{-\rmi \tilde{H}\theta}H_0\rme^{\rmi \tilde{H}\theta} \qquad & 0\leq \theta \leq \pi  \\
    P_A \rme^{-\rmi \tilde{H}(2\pi-\theta)}H_0\rme^{\rmi \tilde{H}(2\pi-\theta)}P_A & \pi  \leq \theta \leq 2\pi  \ .
\end{cases}  
\end{align}
Clearly $H_\pi$ is $P_A$ symmetric, and $H_0$ and $H_\pi$ share a $\mathbb{Z}_2^3$ symmetry generated by $\{P_A,P_{AB},P_{BC}\}$ or, equivalently, $\{P_A,P_B,P_C\}$. It then follows from \Cref{result:pumpspt} that $H_\pi$ is a non-trivial 2D SPT protected by $\mathbb{Z}_2^3$.

Interestingly, it is known that this SPT is protected by $\tilde{G}\times\mathbb{Z}_2^T$, and by the $\mathbb{Z}_2$ generated by $\prod_{j}X_j$ (the Levin-Gu phase) \cite{Levin12,Tantivasadakarn22}. The non-triviality of $H_\pi$ for $\tilde{G}\times\mathbb{Z}_2^T$ can be argued analogously using \Cref{result:pumpantiunitary}. However, proving that we have a non-trivial $\mathbb{Z}_2$-SPT using only our pump results seems very challenging. This is because there is only one symmetry generator, and if this were used to relate two paths between $H_0$ and $H_\pi$, then $\tilde{G}$ would be trivial and there could not be a pump. See also related comments in Ref.~\onlinecite{Chen14}, and in \Cref{sec:equivariantfamily}.

\subsubsection{Decorated domain walls}\label{sec:decorateddomainwall}
Focusing on the $G$-symmetric Hamiltonians $H_0$ and $H_\pi$, our analysis naturally picks out a subgroup of $G$ formed by $\tilde{G}$ and $g_0$. In the simplest case this gives rise to $\tilde{G}\times \mathbb{Z}_n$ where $n$ is the order of $g_0$.

For such product groups, there is a natural connection to the decorated domain wall construction of SPTs \cite{Chen14}. Decomposing the product group using the K\"{u}nneth formula, some $\tilde{G}\times\mathbb{Z}_n$ SPTs in $d$-dimensions come from $\tilde{G}$ SPTs in $(d-1)$-dimensions that are attached to $\mathbb{Z}_n$ domain walls in a consistent way. Then we can write $H_{\mathrm{SPT}}=U H_0 U^\dagger$ where $U$ is an SPT entangler that decorates the $\mathbb{Z}_n$ domain walls in the ground state of $H_0$. One way of writing this is 
\begin{align}
U \left(\prod_{j\in A} g_{j}\right) U^\dagger = \left(\prod_{j\in A} g_{j}\right) ~ \tilde{U}_{\partial{A}}\ , \label{eq:domainwall}
\end{align}
for some subregion $A$. Here the global symmetry $g_0 = \prod_j g_j$, and $\tilde{U}_{\partial{A}}$ entangles the $(d-1)$-dimensional SPT on the boundary of $A$.

Our pump approach starts with a gapped path of Hamiltonians $H(\theta)$ between $H_0$ and $H_\pi$. Let us consider the related path in the space of ground states, then there is a local unitary evolution between them \cite{Zeng19}. If we denote this by $U_P$, then \Cref{result:pumpspt} tells us that, on a finite system $M$, $U_P(g_0 U_P g_0)^\dagger =\tilde{U}_{\partial M} $. This is a simple rearrangement of \cref{eq:domainwall}, so we conclude that \Cref{result:pumpspt} presents a complementary perspective
on the decorated domain wall construction.

Note that the decorated domain wall construction allows us to identify which SPT class $H_\pi$ is in relative to $H_0$ (from the corresponding term in the K\"{u}nneth formula). Our approach shows only whether we have a trivial or non-trivial difference in the SPT phase of $H_0$ and $H_\pi$, more work is required to identify the actual SPT phases in the non-trivial case. However, see \Cref{sec:equivariant} where we more quantitatively match the invariants of pumps with the invariants of (candidate) SPTs in the same dimension.

\subsection{Boundary transitions, SPTs and pumps}
The pump invariant manifests itself in different ways when we write our system with open boundary conditions. If we consider a set-up with a $\tilde{G}$-symmetric and $2\pi$ periodic Hamiltonian, then the ground state cannot be $2\pi$ periodic---there must be (at least) a gapless boundary diabolical point along our path \cite{Hsin20,Shiozaki22}. In fact, the boundary phase diagram is such that boundary transition lines terminate at the bulk diabolical point that is responsible for the pump \cite{Hsin20,Prakash24}.

The location of this boundary transition is a priori unconstrained. If we deduce that $H_0$ and $H_\pi$ are in different $G$-SPT phases, then, from general SPT considerations (assuming internal symmetries), there will be a gapless boundary for at least one of these two Hamiltonians.
If we take a path $H_{\tilde{G}}(\theta)$ with a symmetric boundary, and construct the loop \eqref{eq:Aloop}, then, by construction, the spectrum is symmetric about $t=\pi$. If we pump a  $(d-1)$-dimensional $\tilde{G}$-SPT, and we assume the minimal case of a single boundary diabolical point, then it is located at either $H_0$ or $H_\pi$. This completes the proof of \Cref{result:pumpantiunitary}.
The same reasoning applies in the cases involving a charge conjugation.

\section{On the classification of equivariant families}
\label{sec:equivariant}
In this final section we place the ideas discussed so far in a general setting, justifying the expectation that pumps can be related to SPTs with minimal assumptions. We prove an analogue of \Cref{result:pumpspt} for an additional $\mathbb{Z}_2$ reflection symmetry in the setting of equivariant families, and discuss anomalies of the family as we go around a pivot loop.
\subsection{Generalities}

\begin{definition}\label{def:equivariant}
An anomaly-free $G$-equivariant family over a parameter space $M$ is specified by
\begin{enumerate}
    \item a family of $d$-spatial-dimensional Hamiltonians, $H(\theta)$, that depend on a parameter $\theta \in M$ such that each $H(\theta)$ has gap $\Delta(\theta) \ge \Delta_0 > 0$ for some uniform lower bound $\Delta_0$ (the family is uniformly gapped),
    \item a group $G$ with a map $\sigma:G \to \mathbb{Z}_2$ identifying the time-reversing elements,
    \item a collection of tensor product unitary operators $U_g$ such that $U_g K^{\sigma(g)}$, where $K$ is complex conjugation, is a representation of $G$,
    \item an action of $G$ on $M$, such that
    \begin{align} U_g K^{\sigma(g)} H(\theta) K^{\sigma(g)} U_g^{-1} = H(g \cdot \theta)\ .\end{align}
    (Points $\theta$ such that $g \cdot \theta = \theta$ have $g$ as a symmetry, otherwise $g$ is duality-like.)
\end{enumerate}
\end{definition}
Deformations from one such family, $H_0(\theta)$, to another, $H_1(\theta)$, are given by $H(\theta,s)$ for $s \in [0,1]$. For each fixed $s$, $H(\theta,s)$ is a $G$-equivariant family, and we have that $H(\theta,0) = H_0(\theta)$, $H(\theta,1) = H_1(\theta)$, and that the gap $\Delta(\theta,s)$ has a uniform nonzero lower bound.

Deformation classes of anomaly-free $G$-equivariant families are thought to be classified by a certain equivariant cobordism group of $M$ (see Ref.~\onlinecite{debray2024longexactsequencesymmetry} and references therein). The type depends on whether we study Hilbert spaces of spins or bosons (oriented cobordism) or Hilbert spaces of fermions (spin cobordism). In the former case, which is our case of interest, we study closed $(d+1)$-dimensional manifolds $X^{d+1}$ equipped with
\begin{enumerate}
    \item a principal $G$-bundle, which we can think of as a homotopy class of maps $A:X^{d+1} \to BG$,
    \item together with $\sigma$, $A$ defines a real line bundle $A^*\sigma$, and our manifolds are also equipped with an orientation on $TX^{d+1} \oplus A^*\sigma$.
    \item with the action of $G$ on $M$ it also defines an associated $M$-bundle $E$, and our manifolds are equipped with a section, $\phi$, of this bundle, which can be thought of as a background field of varying parameters.
\end{enumerate}
Deformation classes of anomaly-free $G$-equivariant families of bosons are supposed to be classified by (Anderson-dual) cobordism invariants of these manifolds, which form an abelian group that we denote
\begin{align}
\Omega^{d+1}_{SO,G,\sigma}(M)\ .\end{align}

We can think of these cobordism invariants as the partition functions of the infinite IR limit of our family, coupled to background gauge field $A$ and spatially and temporally varying parameters $\phi$, which form a section of a bundle rather than an ordinary function, because $G$ acts on the parameter space. The cobordism invariants above turn out to have a very explicit form, namely
\begin{align}(X,A,\phi) \mapsto \rme^{\rmi \int_X \omega(X,A,\phi)} \ , \end{align}
where $\omega(X,A,\phi)$ is a combination of characteristic classes of the tangent bundle of $X$, such as Stiefel-Whitney classes and Pontryagin classes/gravitational Chern-Simons terms, as well as classes in the \emph{twisted equivariant cohomology}
\begin{align} H^{d+1}_G(M,U(1)^\sigma)\ .\end{align}
The simplest cobordism invariants come from elements of this group, which are $U(1)$-valued $(d+1)$-cocycles $\omega(A,\phi)$ depending only on $A$ and $\phi$, with no explicit dependence on the topology of $X$. These are analogous to the group cohomology SPTs.

On a general $M$, we can define pump invariants around any cycle $i:S^1 \hookrightarrow M$ by restriction, which corresponds to the pullback
\begin{align} i^*:\Omega^{d+1}_{SO,G,\sigma}(M) \to \Omega^{d+1}_{SO,G,\sigma}(S^1)\ .\end{align}
If $i(S^1)$ is contractible in $M$, this pullback must be trivial. Thus, a non-trivial pump is an extension of an $S^1$ family to a contractible one. In a phase diagram containing such a pump circle, there must always be a diabolical locus where the gap closes (in the sense of violating condition one of \Cref{def:equivariant}) somewhere inside the circle, preventing it from contracting.

\subsubsection{The Thouless pump}
    As an example, we have that the bosonic 1+1D Thouless pump with $M = S^1$ corresponds to a generator of
\begin{align}
\Omega^2_{SO,U(1),\text{triv}}(S^1) &= H^2_{U(1)}(S^1,U(1))\nonumber\\& = H^2(BU(1) \times S^1,U(1)) \nonumber\\& = H^1(BU(1),U(1)) = \mathbb{Z}     \ .
\end{align}
In this case $U(1)$ acts trivially on $S^1$, so each $H(\theta)$ has $U(1)$ symmetry, and this is a \emph{symmetric} family. We can express the background field $\phi$ as just a $2\pi$-periodic scalar, and the cobordism invariant partition function on a surface $X$ equipped with such a $\phi$ and a $U(1)$ gauge field $A$ is
\begin{align}Z(X,A,\phi) = \rme^{\rmi \int_X A \wedge \frac{\rmd\phi}{2\pi}}\ .\end{align}
More generally, if we have a $G$-symmetric family, we can use
\begin{align}H^{d+1}_G&(S^1,U(1)^\sigma) \nonumber\\&= H^{d+1}(BG,U(1)^\sigma) \oplus H^d(BG,U(1)^\sigma)\ ,\end{align}
which corresponds to the partition function
\begin{align} Z(X,A,\phi) = \rme^{\rmi\int_X \omega_{d+1}(A) + \omega_d(A) \frac{\rmd\phi}{2\pi}}\ .\end{align}
The class $\omega_{d+1}$ is an SPT class which characterizes the $G$-symmetric phase of each $H(\theta)$. The class $\omega_{d}$ is an SPT class which characterizes a $d$-dimensional SPT which gets pumped to the boundary when we complete a loop around the circle, which can be derived from the above by considering varying $\phi$ on a space $X$ with boundary.

\subsection{Equivariant circular families with reflection}

Suppose now that we have an equivariant circular family (i.e., $X=S^1$), with some elements of $G$ acting by a fixed reflection, $\theta \mapsto  2\pi-\theta$, described by a map $\rho:G \to \mathbb{Z}_2$ (we fix the branch so $\theta\in[0,2\pi)$). This action has two fixed points at $\theta = 0$, $\pi$ which together describe a zero-sphere $i:S^0 \hookrightarrow S^1$. We can consider the restriction
\begin{align} i^*:\Omega^{d+1}_{SO,G,\sigma}(S^1) \to \Omega^{d+1}_{SO,G,\sigma}(S^0)\ .\end{align}
Since over $S^0$ we have a $G$-symmetric family, we can compute the latter group separately over each point
\begin{align} \Omega^{d+1}_{SO,G,\sigma}(S^0) = \Omega^{d+1}_{SO,G,\sigma}(\{0\}) \oplus \Omega^{d+1}_{SO,G,\sigma}(\{\pi\})\ .\end{align}
The two terms on the right-hand side correspond to the $G$-SPT class at the two fixed points $0$ and $\pi$, respectively.

Let $\tilde{G}$ be the kernel of $\rho$, so that $\tilde{G}$ is the subgroup of $G$ which acts as a symmetry for all $\theta$. Thus, the two $G$-SPTs we obtained above must be the same as $\tilde{G}$-SPTs.

On the other hand, suppose both these two $G$-SPTs are trivial. Let us choose a $G$-symmetric trivialization of each one. Then, we may consider the upper path from $0$ to $\pi$ as defining a \emph{loop} of $\tilde{G}$-symmetric states, and associate to it a $(d-1)$-dimensional $\tilde{G}$-SPT which gets pumped to the boundary. The pump along the lower path must be related to this one by the $\rho$ action so we get one $\tilde{G}$-SPT invariant out of this. However, this invariant is ambiguous, because we could have chosen different trivializations of the $G$-SPTs at $0$ and $\pi$. These again amount to $(d-1)$-dimensional $G$-SPTs which we can choose at either point. The $\tilde{G}$-SPT which gets pumped is ambiguous by the difference between these two $G$-SPTs when considered as $\tilde{G}$-SPTs.

To summarise, the data of the family is equivalent to the data of the two $G$-SPTs at $0$ and $\pi$, subject to the condition that they agree as $\tilde{G}$-SPTs, plus the lower-dimensional $\tilde{G}$-SPT `half-pump', subject to the ambiguity above. 

This argument is formalised and proved by the Mayer-Vietoris sequence\footnote{This in fact holds for generalized cohomology theories, so these arguments apply to fermions/spin cobordism, $K$ theory, etc.).} given by

\begin{align}
    &\cdots \to \Omega^{d}_{SO,G,\sigma}(\{0\}) \oplus \Omega^{d}_{SO,G,\sigma}(\{\pi\}) \to \Omega^d_{SO,\tilde{G},\sigma}  \nonumber\\ &\to \Omega^{d+1}_{SO,G,\sigma}(S^1) \xrightarrow{i^*} \Omega^{d+1}_{SO,G,\sigma}(\{0\}) \oplus \Omega^{d+1}_{SO,G,\sigma}(\{\pi\}) \nonumber\\ &\to \Omega^{d+1}_{SO,\tilde{G},\sigma} \to \cdots \ .
\end{align}
This sequence comes from considering the circle as the union of two $G$-symmetric intervals, one around 0 and one around $\pi$ (each interval equivariantly contracts onto its associated fixed point), with an intersection consisting of two disjoint intervals on which $G/\tilde{G} = \mathbb{Z}_2$ acts freely.

Given a $G$-equivariant family as above, we can consider the restriction to the $\tilde{G}$-symmetric circle family
\begin{align}j^*:\Omega^{d+1}_{SO,G,\sigma}(S^1) \to \Omega^{d+1}_{SO,\tilde{G},\sigma}&(S^1) \nonumber\\&= \Omega^{d+1}_{SO,\tilde{G},\sigma} \oplus \Omega^d_{SO,\tilde{G},\sigma}\ ,\end{align}
which is characterized by a $d$-dimensional $\tilde{G}$-SPT (first summand) and a $(d-1)$-dimensional $\tilde{G}$-SPT which gets pumped (second summand). We can relate this pump invariant to the characterisation above:
\begin{shaded}
\begin{result}\label{result:Mayer-Vietoris}
     If  both $G$-SPT invariants are trivial, i.e. the class of the family satisfies $i^*\alpha = 0$, then the restricted family $j^*\beta$ is a $\tilde{G}$-SPT pump labelled by an element in $\Omega^d_{SO,\tilde{G},d}$ of the form
     \[\omega - \omega^{\rho},\]
     where $\omega \in \Omega^d_{SO,\tilde{G},d}$ and $\omega^{\rho}$ is the action on $\omega$ of the $\mathbb{Z}_2$ outer automorphism coming from $\tilde{G}$ being a normal subgroup of index 2 in $G$. Therefore, if the pump is not of this form, then we must have $i^*\alpha \neq 0$. In fact, in that case the two $G$-SPT invariants must be \underline{distinct}, because tensoring with a $G$-SPT does not change this pump.
\end{result}\end{shaded}
    \begin{center}
    \begin{figure*}[t!]
        \begin{tikzcd}
\Omega^{d}_{SO,\tilde{G},\sigma} \arrow[r,"\delta"] \arrow[d,"f",] 
\arrow[dr, phantom , very near start, color=black]
& \Omega^{d+1}_{SO,G,\sigma}(S^1) \arrow[r,"i^*"] \arrow[d,"j^*"] & \Omega^{d+1}_{SO,G,\sigma}(\{0\})\oplus \Omega^{d+1}_{SO,G,\sigma}(\{\pi\})  \arrow[d] \arrow[r] & \Omega^{d+1}_{SO,\tilde{G},\sigma} \arrow[d,"f"] \\
\Omega^{d}_{SO,\tilde{G},\sigma} \oplus \Omega^{d}_{SO,\tilde{G},\sigma}  \arrow[r,"g"] & \Omega^{d+1}_{SO,\tilde{G},\sigma} \oplus \Omega^d_{SO,\tilde{G},\sigma} \arrow[r,"h"] & \Omega^{d+1}_{SO,\tilde{G},\sigma}(\{0\})\oplus \Omega^{d+1}_{SO,\tilde{G},\sigma}(\{\pi\}) \arrow[r,"k"] & \Omega^{d+1}_{SO,\tilde{G},\sigma} \oplus \Omega^{d+1}_{SO,\tilde{G},\sigma}\\
\end{tikzcd}    
\caption{Commutative diagram used in the proof of \Cref{result:Mayer-Vietoris}.}
\label{fig:commutative}
\end{figure*}\end{center}
\begin{proof}
    Since the Mayer-Vietoris sequence is functorial, we have the commutative diagram given in \Cref{fig:commutative}, where the vertical maps come from restriction along the inclusion $\tilde{G} \hookrightarrow G$. The outer-most maps are
\[f(\omega) = (\omega,\omega^\rho)\ ,\]
where $\omega^\rho$ is obtained from $\omega$ by applying the automorphism on $\tilde{G}$ associated with it being a normal subgroup of index 2. We also have
\begin{align}
  g(\omega,\omega') &= (0,\omega-\omega') \nonumber\\
  h(\alpha,\omega) &= (\alpha,\alpha) \nonumber\\
  k(\alpha,\beta) &= (\alpha-\beta,\alpha-\beta)\ .
\end{align}

Now suppose $i^*\alpha = 0$. Then $\alpha = \delta \omega$ for some $\omega$ by exactness. We have by commutativity and the formulas above
\begin{align}j^* \delta \omega = g f \omega = (0,\omega - \omega^\rho)\ ,\end{align}
which is what we wanted to prove.

\end{proof}

\subsubsection{SPTs, pumps and equivariant contractible families} \label{sec:equivariantfamily}   We cannot prove a converse to \Cref{result:Mayer-Vietoris}, as there may not, in general, be a suitable subgroup $\tilde{G}$ which fixes the family and can measure a pump. Indeed, suppose that $G = \mathbb{Z}_2$ and $\tilde{G}$ is trivial. Then the Mayer-Vietoris sequence reads
    \begin{align}
    \Omega^d_{SO} \to& \Omega^{d+1}_{SO,\mathbb{Z}_2,\sigma}(S^1) \to \nonumber\\&\Omega^{d+1}_{SO,\mathbb{Z}_2,\sigma}(\{0\}) \oplus \Omega^{d+1}_{SO,\mathbb{Z}_2,\sigma}(\{\pi\}) \to \Omega^{d+1}_{SO} \ ,
    \end{align}
    where the terms at the ends are invertible states which require no symmetry protection. Let us take $d = 2$, and $\sigma$ to be trivial. We obtain
    \begin{align}0 \to &\Omega^{3}_{SO,\mathbb{Z}_2}(S^1) \nonumber\\&\to \Omega^{3}_{SO,\mathbb{Z}_2}(\{0\}) \oplus \Omega^{3}_{SO,\mathbb{Z}_2}(\{\pi\}) = \mathbb{Z}_2 \oplus \mathbb{Z}_2 \to \mathbb{Z}\ .\end{align}
    Since all maps $\mathbb{Z}_2 \oplus \mathbb{Z}_2 \to \mathbb{Z}$ are zero, we find
    \begin{align} \Omega^{3}_{SO,\mathbb{Z}_2}(S^1) = \mathbb{Z}_2 \oplus \mathbb{Z}_2\ .\end{align}
    The diagonal class in this group can be thought of as the constant family coming from the 2+1d $\mathbb{Z}_2$ SPT (Levin-Gu phase). However, the other class is a non-trivial family phase, which is not obvious how to describe via pumps. It is a $\mathbb{Z}_2$ equivariant family on $S^1$, where $\mathbb{Z}_2$ acts as reflection fixing $0$ and $\pi$. If we restrict to these points, we find they differ by the Levin-Gu phase.
    
    One way to construct this family is to choose a (not-$\mathbb{Z}_2$-symmetric) uniformly gapped path $H(s)$ such that $H(0)$ is the trivial phase and $H(\pi)$ is the Levin-Gu phase. Then we consider, for $\pi < s < 2\pi$,
    \begin{align}H(s) = gH(2\pi-s)g^{-1}\end{align}
    where $g$ is the $\mathbb{Z}_2$ symmetry. Because $H(\pi)$ and $H(0)$ are symmetric, this is a $2\pi$-periodic family. Furthermore, it is $\mathbb{Z}_2$-equivariant, with the action $g \cdot s = 2\pi - s$, which corresponds to reflection across the axis connecting $0$ and $\pi$. By construction, $H(0)$ and $H(\pi)$ differ by the Levin-Gu SPT.
    
    Suppose this family was equivariantly contractible, so that there was a uniformly-gapped 2-parameter family $H(s,r)$ such that $H(s,0) = H_0$ is a constant family, $H(s,1) = H(s)$, and $g H(s,r) g^{-1} = H(g \cdot s,r)$. Then $H(0,r)$ for $0 \le r < 1$ and $H(\pi,r)$ for $0 \le r < 1$ together form a gapped symmetric path from the trivial to the Levin-Gu SPT, which is impossible. Therefore, we have a non-trivial family. 
    
    This does not contradict the fact that there is no pump, and that the loop can be contracted if we do not demand equivariance under the $\mathbb{Z}_2$ symmetry.

\subsection{Anomalous families and pivots}
\begin{center}
\begin{figure*}[t!]
        \begin{tikzcd}
 \Omega^{d+1}_{SO,G,\sigma}(S^n) \arrow[r,"{\rm Ind}_\rho"] 
& \Omega^{d+1-n}_{SO,G,\sigma} \arrow[r,"{\rm Def}_\rho"] & \Omega^{d+2}_{SO,G,\sigma}   \arrow[r,"{\rm Res}_\rho"] & \Omega^{d+2}_{SO,G,\sigma}(S^n)
\end{tikzcd}    
\caption{A piece of the symmetry breaking long exact sequence, used in the description of anomalous families.}
\label{fig:commutative2}
\end{figure*}\end{center}

It is possible for an anomalous symmetry to give rise to a uniformly gapped equivariant family, relaxing the condition that the operators $U_g$ are tensor product operators (i.e., condition three of \Cref{def:equivariant}). We will call this more general family simply a $G$-equivariant family. Such families are common when we consider pivot symmetries. For example, the `axial symmetry' in the Thouless pump acts by rotation on the $S^1$ parameter, and has a mixed anomaly with the $U(1)$ charge symmetry.

In general, a $d$-dimensional such family has an anomaly class $\omega \in \Omega^{d+2}_{SO,G,\sigma}$ as well as a parameter space $M$ with a $G$ action, such that for each parameter value $\theta \in M$, the stabilizer group
\begin{align} H_\theta = \{g \in G\ |\ g \cdot \theta = \theta\}\ \ ,\end{align}
which acts as symmetries of $H(\theta)$, is an anomaly-free subgroup of $G$.

The full obstruction for an anomaly to have a uniformly gapped family over $M$ is the `residual family anomaly' \cite{debray2024longexactsequencesymmetry},
\begin{align}{\rm Res}_M(\omega) \in \Omega^{d+2}_{SO,G,\sigma}(M)\ .\end{align}
This class must vanish for the anomaly to be compatible with being uniformly gapped over $M$. When $G$ does not act transitively on $M$, this obstruction may be more general than the vanishing of the anomalies for each $H_\theta$, and may even be non-trivial when all $H_\theta = 1$ \cite{debray2024longexactsequencesymmetry}. When this obstruction vanishes, families with a given anomaly form a torsor over the anomaly-free families $\Omega^{d+1}_{SO,G,\sigma}(M)$.

When $M = S^n$ is a sphere and $G$ acts via an orthogonal representation $\rho:G \to O(n+1)$, we can take further advantage of the symmetry breaking long exact sequence (SBLES, a version of the Gysin sequence), see Fig. \ref{fig:commutative2}. This says that $\omega$ is in the image of the `defect map'
\begin{align}{\rm Def}_\rho:\Omega^{d+1-n}_{SO,G,\sigma \oplus \rho} \to \Omega^{d+2}_{SO,G,\sigma}\ .\end{align}
This means there is a class $\alpha \in \Omega^{d+1-n}_{SO,G,\sigma \oplus \rho}$ with
\begin{align}\label{eqndefanommatching}{\rm Def}_\rho(\alpha) = \omega\ .\end{align}
In fact, $\alpha$ is an invariant of the anomalous family. It is a type of pump invariant, describing the anomaly of a codimension-$n$ defect around which the parameter winds symmetrically over $S^n$. This defect may be given a twisted $G$ symmetry, restoring broken symmetry generated by combining them with rotations and CPT transformations, which changes their type (from $\sigma$ to $\sigma \oplus \rho$ in our notation) \cite{Hason_2020,debray2024longexactsequencesymmetry}. The invariant $\alpha$ gives the anomaly of this twisted symmetry\footnote{It was shown in Ref.~\onlinecite{debray2024longexactsequencesymmetry} that the ambiguity in $\alpha$, given $\omega$, is given by tensoring with an anomaly-free $G$-equivariant family. Our equivalence relation on families does not allow this, so $\alpha$ is an invariant.}. In group cohomology, the defect map is easily computed, given by taking the cup product of $\alpha$ with the (twisted) Euler class of $\rho$ \cite{debray2024longexactsequencesymmetry}.

For $n = 0$, $M = S^0$ is two points, and $\Omega^{d+1}_{SO,G,\sigma \oplus \rho}$ represents the $G$ anomaly of the domain wall between the theories at the two points. Here the twisting is shifted by $\rho$, indicating that elements of $G$ which exchange the two points must be combined with CRT symmetry to obtain a symmetry on the wall.

In the case that the anomaly is trivial, we can classify our family by some $\zeta \in \Omega^{d+1}_{SO,G,\sigma}(S^n)$. In this case we can also consider the codimension-$n$ defect above, whose anomaly is given in terms of $\zeta$ by the `index map' \cite{debray2024longexactsequencesymmetry}, and goes
\begin{align}{\rm Ind}_\rho: \Omega^{d+1}_{G,SO,\sigma}(S^n) \to \Omega^{d+1-n}_{G,SO,\sigma\oplus\rho}\ .\end{align}
We have in this case
\begin{align}\alpha = {\rm Ind}_\rho(\zeta)\ ,\end{align}
and generally
\begin{align}{\rm Def}_\rho \circ {\rm Ind}_\rho = 0\ .\end{align}
For group cohomology classes, the index map is given by integration (slant product) over the sphere $S^n$, mapping to parameter space by the identity $S^n \to S^n$.

In the case $n = 0$, $M = S^0$, if $G$ is an anomaly-free symmetry,
\[\zeta = (\zeta_1,\zeta_2) \in \Omega^{d+1}_{SO,G,\sigma}(S^0) = \Omega^{d+1}_{SO,G,\sigma} \oplus \Omega^{d+1}_{SO,G,\sigma}\]
is classified by the SPT class of each point. The class
\[{\rm Ind}_\rho(\zeta) = \zeta_1 - \zeta_2\]
gives the relative SPT class, which is the anomaly on the domain wall between the two points.

More generally, the anomaly of the codimension-$n$ defect must be consistent when we consider an anomaly-free subgroup $H \le G$, which yields the following:
\begin{shaded}

\begin{result}\label{proposition}
If $H$ is any anomaly-free subgroup, we can regard our theory as an anomaly-free $H$-equivariant family, classified by some $\zeta_H \in \Omega^{d+1}_{SO,H,\sigma}(S^n)$. There is a pump invariant associated to this family, defined by the index map \cite{debray2024longexactsequencesymmetry}
\begin{align}{\rm Ind}_\rho: \Omega^{d+1}_{H,SO,\sigma}(S^n) \to \Omega^{d+1-n}_{H,SO,\sigma\oplus\rho} \ .\end{align}
This must match the above according to
\begin{align}{\rm Ind}_\rho(\zeta_H) = i_H^* \alpha \ , \end{align}
where $i^*:\Omega^{d+1-n}_{G,SO,\sigma\oplus\rho} \to \Omega^{d+1-n}_{H,SO,\sigma\oplus\rho}$ is restriction from $G$ to $H$. In particular, if the right-hand-side is non-zero, we find we must have a non-trivial $H$-equivariant family.

\end{result}
\end{shaded}
Note that in particular, if $H = \tilde G$, the anomaly-free subgroup of $G$ acting as a symmetry at each point of $M$ (i.e., $\tilde G = \bigcap_{\theta \in M}H_\theta$), $\zeta$ is the class of this $\tilde G$-symmetric family, and $i: \tilde G\hookrightarrow G$, then
\[{\rm Ind}_\rho(\zeta) = i^*\alpha \in \Omega^{d+1-n}_{\tilde G,SO,\sigma}\]
is the $\tilde G$-SPT pumped over $S^n$, and must be non-trivial if $i^*\alpha \neq 0$. Applying this requires calculation, we give examples below.

\subsubsection{Revisiting the Ising pivot ($n=0$)}
As an illustration, let us reconsider the Ising pivot of the introduction. 

First of all, let us fix $n=0$ and consider an equivariant $S^0$ family. In particular, let $G = \mathbb{Z}_2^3$ be the symmetries of a spin-$1/2$ chain generated by $U_1 = \prod_j X_{2j+1}$, $U_2 = \prod_j X_{2j}$, $E = \prod_j \rme^{\frac{\rmi\pi}{4} Z_j Z_{j+1}}$. We consider the two Hamiltonians\footnote{Since $H$ denotes the anomaly-free subgroup, we denote Hamiltonians by $\hat{H}$ in this section.}
\begin{align}\hat{H}_0 &= - \sum_j X_j \nonumber\\
\hat{H}_\pi &= \sum_j Z_j X_{j+1} Z_{j+2}\ .
\end{align}
These each enjoy the anomaly-free $H = \mathbb{Z}_2^{U_1} \times \mathbb{Z}_2^{U_2}$ symmetry and are gapped. The symmetry $E$ acts as an entangler, mapping $\hat{H}_0$ to $\hat{H}_\pi$ and vice versa. We regard them together as a $G$-equivariant $S^0$ family. 

There is a mutual anomaly
\begin{align}\omega = \pi A_E \cup A_1 \cup A_2\end{align}
which is understood as $E$ changing the SPT from trivial ($\hat{H}_0$) to cluster $\pi A_1 \cup A_2$ ($\hat{H}_\pi$). In the above, the associated class of the domain wall is
\begin{align}\alpha = \pi A_1 \cup A_2 \in \Omega^2_{G,SO,\sigma \oplus \rho}\ .\end{align}
The defect map in these group cohomology classes is given by cup product with the Euler class of $\rho$, which in this case is $A_E$. Hence, we find that \cref{eqndefanommatching} holds. Crucially, the above is a non-trivial SPT when restricted to the anomaly-free subgroup $H$. Thus, we find the anomaly of this form must act as an SPT entangler, such that any gapped $H$-symmetric Hamiltonians related by $E$ have relative SPT class $\pi A_1 \cup A_2$.

\subsubsection{Revisiting the Ising pivot ($n=1$)}

For $n = 1$, $M = S^1$, the anomaly class is in the image of some $\alpha \in \Omega^{d}_{G,SO,\sigma \oplus \rho}$ under the defect map. For an anomaly-free subgroup $H$ acting as a symmetry, this restricts to the $H$-charge pump around the family.

    We extend the previous example to an $S^1$ family by considering the pivot Hamiltonian $\tilde{H} = - \frac{1}{4} \sum_j Z_j Z_{j+1}$:
    \begin{align}\hat{H}(\theta) = \rme^{-\rmi \theta \tilde{H}} \hat{H}_0 \rme^{\rmi \theta \tilde{H}}\ .\end{align}
    This family is clearly uniformly gapped, and is $G$-equivariant, with $U_1$ and $U_2$ both acting as reflections $\theta \mapsto -\theta$. The Euler class is
    \begin{align} e(\rho) = (A_1 + A_2 + A_E) \cup A_E \in H^2(B\mathbb{Z}_2^3,\mathbb{Z}^{A_1,A_2})\ .\end{align}
    The associated class of the domain wall must, by \cref{eqndefanommatching}, therefore be
    \begin{align}\alpha = \pi A_1 \in H^1(B\mathbb{Z}_2^3,U(1)^{A_1,A_2})\ ,\end{align}
    with both $U_1$ and $U_2$ acting now as anti-unitary symmetries (note $A_1 = A_2$ in this twisted cohomology, so this class is symmetric), since they reflect the $S^1$ ($E$ is still unitary in this case, being a rotation). Furthermore, if we consider the anomaly-free symmetry group $H = \mathbb{Z}_2^{U_1 U_2}$ given by the combined symmetry $U_1 U_2$, $\alpha$ restricts to the nontrivial charge class $\pi A_{12} \in H^1(B\mathbb{Z}_2^{U_1 U_2},U(1))$. This must come from the class of the $\mathbb{Z}_2$ charge pump
    \begin{align}\zeta_H = \pi \frac{\rmd\theta}{2\pi} \cup A_{12} \in H^1(S^1 \times B\mathbb{Z}_2^{U_1 U_2},U(1))\ ,\end{align}
    which has
    \begin{align}{\rm Ind}_\rho(\zeta_H) = \int_{S^1}  \pi \frac{\rmd\theta}{2\pi} \cup A_{12}  = \pi A_{12}\ .\end{align}

\section{Outlook}
In this work we have studied the connections between pivot loops in exactly solvable models, charge pumps, anomalies and SPT phases. We showed that the Dolan-Grady relation is necessary and sufficient to have a strict circular loop in the space of Hamiltonians, and that these loops have particularly nice properties. A key example of such a loop was in the Onsager-integrable chiral clock family, where we understood some unusual features of the phase diagram from Ref.~\onlinecite{Jones25} as being direct consequences of a non-trivial charge pump, particularly in the RSPT case. We also developed a range of pivot examples beyond strict circular loops. Motivated by these examples, we examined the relationship between pumps and SPTs, showing that a non-trivial pump can put constraints on the SPT phase diagram, and exclude the possibility of symmetric gapped paths between Hamiltonians.

A natural question is to what extent one can recover classification results for $d$-dimensional SPT phases using arguments based on pumps, such as \Cref{result:pumpspt}, and whether this is useful perspective for classification purposes in some cases. For example, in one-dimension, arguments based on the area law and MPS are typical \cite{Hastings07,Chen11,Schuch11}. There has been recent work, see Ref.~\onlinecite{Kapustin21} and references therein, where the `split property' is used \cite{Ogata21} rather than MPS. This property can be derived from the area law \cite{Matsui13}. If we were to instead begin with a classification of 1D charge pumps, as in Ref.~\onlinecite{Bachmann24}, it is not clear how the area law enters the classification. It would be interesting to understand how an argument built from classifying non-trivial pumps, that in turn imply no symmetric path between Hamiltonians, may implicitly use the area law. It would also be interesting to see if we can recover the exact SPT class from this kind of argument, rather than simply arguing that two Hamiltonians are in different phases.
In any case, as discussed above, it is difficult to see how one could find the non-trivial Levin-Gu $\mathbb{Z}_2$-SPT using only a pump argument, and thus we expect this perspective to be useful only in a sub-class of models.

In comparison to SPTs, the physics of RSPTs is relatively unexplored. An interesting direction to pursue is the connection between pumps and RSPTs in a general setting, and whether we can sharpen the constraints on the boundary phase diagram. 

For $N=2$, the Onsager-integrable clock models reduce to Ising-cluster models, Jordan-Wigner dual to free-fermion models. While the many-body analysis applies, in this case the charge pump relates to a non-contractible loop in a particular class of symmetric Fredholm operators.  We can probe more deeply by making connections to the Wiener-Hopf decomposition of the matrix symbol corresponding to the Hamiltonian, as well as related techniques that illuminate the boundary phase diagram \cite{Widom74,Basor18,Jones23,Alase23}. We will analyse this perspective in a forthcoming work.

In \Cref{sec:equivariant} we focused on equivariant families over a circle (or sphere). The $\mathbb{Z}_N\times\mathbb{Z}_N$ examples we considered in \Cref{sec:ZNZN} have a more interesting equivariant structure consisting of many loops, with several non-trivial pumps (related by symmetry). It would be interesting to understand this case in greater detail, along with an understanding of anomalies of the large symmetry group of the Hamiltonian $H_S$ at the centre of this structure.  

It would be very interesting to explore how our interpretation of the Dolan-Grady relation as the generator of a strict circular loop could lead to new solutions of the Onsager algebra. For models that satisfy Dolan-Grady, a natural candidate for further exploration is to look at whether any interesting (R)SPT physics arises when pivoting in the Hubbard model \cite{Naudts09,Naudts12}.

\section*{Acknowledgements}
\noindent We are grateful to Dan Borgnia, Paul Fendley, Po-Shen Hsin, Kansei Inamura, Nathanan Tantivasadakarn, Marvin Qi and Mallika Roy for illuminating discussions.

\appendix
\section{The $U(1)$ radius and anomalies}
\label{app:periodicity}
In this appendix we clarify that a $U(1)$ pivot symmetry must be normalised to be $2\pi$-periodic in order to use a non-trivial pump to conclude that we have an anomalous symmetry at $H_\star$. 

To see this, consider the following pivot Hamiltonian:
\begin{align}
\tilde{H} =  -  \frac{1}{4} \sum_j Z_{2j-1} X_{2j} Z_{2j+1}  
\end{align}
and the following family of states:
\begin{align}
\ket{\psi(\theta)} = \rme^{-\rmi \theta \tilde{H}} \ket{+}^{\otimes N} \ .
\end{align}
Alternatively, we can consider the loop of Hamiltonians generated by pivoting $H_0 = -\sum_j X_j$ with $\tilde{H}$.

The family is $2\pi$-periodic, i.e., $\ket{\psi(\theta+2 \pi)} = \ket{\psi(\theta)}$. Indeed, we can see that
\begin{align}
\ket{\psi(\theta)} = \rme^{ \frac{\rmi \theta}{4} \sum_j Z_{2j-1} Z_{2j+1}} \ket{+}^{\otimes N} 
\end{align}
which is the Ising pivot from \Cref{sec:firstresult} applied to only the odd sites. We therefore pump a $\mathbb Z_2$ charge as $\theta$ goes from 0 to $2\pi$.

However, as an operator $U(\theta) = \rme^{-\rmi \theta \tilde{H}}$, we have
\begin{align}
U(2\pi) = \prod_{j} X_{2j}\ .
\end{align}
This operator has period $4\pi$, and so it is in fact $2H_p$ that generates a $U(1)$ symmetry with period $2\pi$. This generator does \emph{not} have a mutual anomaly with the $\mathbb Z_2$ symmetry $\prod_n X_n$. 

Indeed, to see it is non-anomalous, note that applied to a finite region we have
\begin{align}
U(2\pi) = Z_{2m-1} X_{2m} X_{2m+2} \cdots X_{2n-2} X_{2n} Z_{2n+1} .
\end{align}
This means that the truncated operator also has period $4\pi$---the same as the bulk untruncated operator. There is thus no anomaly. Relatedly, pivoting through the whole period, $U(4\pi)$ pumps two $\mathbb Z_2$ charges. 
\section{Group cohomology pumps}\label{app:cohomology}
\begin{figure}[!h]
    \centering
    \includegraphics[width=.8\linewidth]{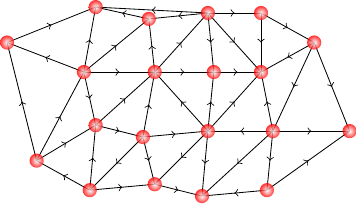}
    \caption{Triangulation of a two-dimensional lattice with branching structure.}
    \label{fig:simplex}
\end{figure}
 In this section, we review how the group-cohomology formulation of SPTs can be used to construct topological pumps in any spatial dimension. The corresponding expressions have appeared in previous studies, originally in the context of classifying Floquet unitaries~\cite{RoyHarperPhysRevB.95.195128}, and also in the context of topological pumps~\cite{Shiozaki22,OhyamaShiozakiSato_GeneralizedThouless}. We apply this to the one-dimensional case in Section \ref{sec:cohomology}.

\subsection{Review of exactly solvable SPT Hamiltonians}\label{app:cohomologysetting}
A straight-forward way to understand group-cohomology SPT phases is through exactly solvable models. We review this here to fix notation. Recall that for bosonic systems with on-site symmetry $\tilde{G}$, we can produce exactly solvable models of non-trivial SPT phases in $d$  spatial dimensions classified by the group cohomology group $H^{d+1}(\tilde{G},U(1))$ as shown in Refs.~\onlinecite{Wen_GroupCohomology2013_PhysRevB.87.155114,BUERSCHAPER_SptTensorNetwork2014}. The data required for this are as follows\footnote{This can be further generalised, e.g., to include anti-unitary and spatial symmetries as well. We will not consider these for simplicity.}:
\begin{enumerate}
    \item A set of representative cocycles, $\omega(g_1,\ldots,g_{d+1}) \in Z^{d+1}(\tilde{G},U(1))$  satisfying $\delta\omega(g_1,\ldots,g_{d+2}) = 1$ where 
    \begin{align}
\delta\omega(g_1,&\ldots,g_{d+2})=\frac{\omega(g_2,g_3,\ldots,g_{d+2})}{\omega(g_1,g_2,g_3,\ldots,g_{d+1})^{(-1)^{d+1}}}  \nonumber \\& 
        \times\prod_{j=1}^{d+1}\omega(g_1,\ldots,g_{j-1},g_{j}g_{j+1},\ldots,g_{d+2})^{(-1)^j}
        \label{eq:delta_coderivative_definition} \ .
    \end{align}
    Subject to the equivalence relation $\omega(g_1,\ldots,g_{d+1}) \sim \omega(g_1,\ldots,g_{d+1}) \delta \mu(g_1,\ldots,g_{d+1})$, where $\mu(g_1,\ldots,g_{d})$ is a $d$-cochain and $\delta \mu$ is a coboundary, we get the class $[\omega]$ which takes values in the abelian group  $[\omega] \in H^{d+1}(\tilde{G},U(1))$ and classifies the SPT phase.
    \item An orientable $d$-dimensional spatial manifold $\mathcal{M}$ with triangulation and a branching structure. See \cref{fig:simplex} for an example for $d=2$.
    \item A $|\tilde{G}|$-dimensional Hilbert space attached to the vertex of the triangulation transforming as the regular representation of $\tilde{G}$ whose bases can be chosen by the elements of the group $\ket{g \in \tilde{G}}$ on which symmetry acts by left action as 
    \begin{align}
        U(g) \ket{h} = \ket{g^{-1}h}. \label{eq:Symmetry action}
    \end{align}
\end{enumerate}
Using the above data, we can furnish an exactly solvable SPT Hamiltonian $H[\{\omega\}]$ classified by $[\omega] \in H^{d+1}(\tilde{G},U(1))$ whose ground state  can be written as~\cite{Wen_GroupCohomology2013_PhysRevB.87.155114,BUERSCHAPER_SptTensorNetwork2014} 
\begin{align}
    \ket{\psi_\omega} = \sum_{g_1,\ldots,g_V}    \psi_\omega(g_1,\ldots,g_V) \ket{g_1,\ldots,g_V}\label{eq:GroupCohomologygroundstate}\end{align}where \begin{align}
    \psi_\omega(g_1,\ldots,g_V) = \prod_{\triangle_d} \omega(g_{v_1}^{},g_{v_1}^{-1}g_{v_2}^{},\ldots,g_{v_d}^{-1}g_{v_{d+1}}^{})^{\sigma(\triangle_d)} \ .
\end{align}
Here, $V$ is the total number of vertices, $\triangle_d$ represents the set of $d$ simplices, $\{v_1,\ldots,v_{d+1}\} \in \triangle_d$ lists the $d+1$ vertices in this simplex ordered using the branching structure and $\sigma$ represents the orientation inherited from the branching structure and the underlying orientable structure of the manifold. 

The Hamiltonian whose unique ground state is \cref{eq:GroupCohomologygroundstate} is constructed as follows. Recall that we begin with a Hamiltonian belonging to the trivial phase \cref{eq:H_trivial}, with ground state \cref{eq:GS_trivial}.
To get the non-trivial SPT Hamiltonian, we use the following entangler,
\begin{align}
    W_\omega = \prod_{\triangle_d} \sum_{g_{v_1},\ldots,g_{v_{d+1}}} \omega(g_{v_1}^{},g_{v_1}^{-1}g_{v_2}^{},\ldots,g_{v_d}^{-1}g_{v_{d+1}}^{})^{\sigma(\triangle_d)}\nonumber\\ \outerproduct{g_{v_1},g_{v_2},\ldots,g_{v_d},g_{v_{d+1}}}{g_{v_1},g_{v_2},\ldots,g_{v_d},g_{v_{d+1}}}  \label{eq:SPT_Entangler_ddim}
\end{align}
which is a finite-depth circuit that can be expressed as a product of commuting operators with support on the $d$-simplices. It is easy to check that $H_{\omega} =  W_\omega^{} H_0 W_\omega^{\dagger}$ is equal to
\begin{align}
     - \sum_{v \in V} &\prod_{\triangle_d \in v} \sum_{\substack{h_{v_1},\ldots,\\g_v,l_v,\ldots ,h_{v_{d+1}}}} \left(\mathfrak{h}_{h_{v_1},\ldots,g_v,l_v,\ldots ,h_{v_{d+1}}}\right)^{\sigma(\triangle_d)}\nonumber\\& \outerproduct{h_{v_1},\ldots,g_v,\ldots h_{v_{d+1}}}{h_{v_1},\ldots,l_v,\ldots h_{v_{d+1}}} \label{eq:SPT_Hamiltonian}\,\end{align}
     where
     \begin{align}
    \mathfrak{h}= \frac{\omega(h_{v_1},h_{v_1}^{-1}h_{v_2},\ldots,h_{v-1}^{-1}g_v,g_v^{-1}h_{v+1},\ldots,h_{v_{d}}^{-1} h_{v_{d+1}})}{\omega(h_{v_1},h_{v_1}^{-1}h_{v_2},\ldots,h_{v-1}^{-1}l_v,l_v^{-1}h_{v+1},\ldots,  h^{-1}_{v_{d}} h_{v_{d+1}})}\  .
\end{align}
This has the ground state 
\begin{align}
    \ket{\psi_\omega} = W_\omega \ket{\psi_0} 
\end{align}
defined in \cref{eq:GroupCohomologygroundstate}. $\triangle_d \in v$ refers to all simplices containing the vertex $v$, while $\{v_1,\ldots,v,\ldots,v_{d+1}\}$ refer to the vertices of the simplex $\triangle_d$, which necessarily includes $v$.

\subsection{Pumps within group cohomology SPTs}
 We now give a recipe to produce a non-trivial loop within the fixed class. Recall that given a set of cocycles $\omega(g_1,\ldots,g_{d+1})$, the class $[\omega]$ is preserved if we deform the former by a coboundary
\begin{align}
    \omega(g_1,\ldots,g_{d+1}) \sim \omega(g_1,\ldots,g_{d+1}) ~\delta \mu(g_1,\ldots,g_{d+1})\ .
\end{align}
When $\mu(g_1,\ldots,g_d)$ is itself a non-trivial $d$ cocycle, we have $\delta \mu = 1$ and we get the same $d+1$ cocycle representative $\omega$. Thus, we can generate a one-parameter loop of cocycle representatives $\omega_\theta$ within the same cohomology class using a reference $d=1$ cocycle $\omega$ as follows
\begin{align}
    \omega_\theta(g_1,\ldots,g_{d+1}) = \omega(g_1,\ldots,g_{d+1})~\delta \mu_\theta(g_1,\ldots,g_{d+1}) \label{eq:GroupCohomology_general_loop} \ ,
\end{align}
where $\mu_\theta$ is an interpolation between $\mu_0$ and $\mu_{2\pi}$ such that
\begin{align}
    \mu_0(g_1,\ldots,g_d) = 1 \ ,\end{align}  and  $\mu_{2\pi}$  is a solution to  \begin{align}\delta\mu_{2\pi}(g_1,\ldots,g_{d+1}) = 1\ . 
\end{align}

Using this we can produce a loop of Hamiltonians and ground states as in \cref{eq:GroupCohomologygroundstate}. The loop is classified by the $d$-cocycle $\mu_{2\pi}(g_1,\ldots,g_d)$ and thus by $[\mu_{2\pi}] \in H^d(\tilde{G},U(1))$. This agrees with the expectation that $d$-dimensional invertible loops are classified by $(d-1)$-dimensional invertible phases \cite{vonKeyserlingkSondhiPhysRevB.93.245145,Else2016PhysRevB.93.201103,Potter17,RoyHarperPhysRevB.95.195128} . 

\subsection{Specialisation to the trivial phase}
\label{sec:Trivial phase pump general}
For the analysis in \Cref{sec:cohomology}, we focus on pumps within the trivial phase. In $d$-dimensions this looks as follows. In the trivial phase, we have
\begin{align}
    \omega_\theta(g_1,\ldots,g_{d+1}) = \delta \mu_\theta(g_1,\ldots,g_{d+1})\ .
\end{align}

We can produce exactly-solvable families of models using the above recipe for this restricted case. In the absence of open boundaries, the entangler in \cref{eq:SPT_Entangler_ddim} takes on a simpler form 
\begin{align}
      W_\theta &= \prod_{\triangle_d} \sum_{g_{v_1},\ldots,g_{v_{d+1}}}\mu_\theta(g_{v_1}^{-1}g_{v_2},\ldots,g_{v_d}^{-1}g_{v_{d+1}})^{\sigma(\triangle_d)}\nonumber\\ &\outerproduct{g_{v_1},g_{v_2},\ldots,g_{v_d},g_{v_{d+1}}}{g_{v_1},g_{v_2},\ldots,g_{v_d},g_{v_{d+1}}}\ . \label{eq:Entangler_trivial_ddim}
\end{align}
For this, the ground state amplitudes correspond to 
\begin{align}
    \psi_\theta(g_1,\ldots,g_V) = \prod_{\triangle_d} \mu_\theta(g_{v_1}^{-1}g_{v_2},\ldots,g_{v_d}^{-1}g_{v_{d+1}})^{\sigma(\triangle_d)}\ . \label{eq:TrivialgroundstateFamily}
\end{align}
Unlike \cref{eq:GroupCohomologygroundstate}, phase factors assigned to the simplices of \cref{eq:TrivialgroundstateFamily} transform trivially under symmetry action of \cref{eq:Symmetry action} for both periodic and open boundary conditions consistent with the fact that we are in the trivial phase. 
\section{Calculations with $\mathbb{Z}_N\times\mathbb{Z}_N$ pivots}
\label{app:pivotentangler}
\subsection{Dipolar SPT}
\subsubsection{Pivot Hamiltonians entangle the SPT}
For the dipolar model, we analyse the pivot by writing \begin{align}
U_D^{(r)}(\theta)= \rme^{-\rmi \theta \tilde{H}_D^{(r)}} =\prod_{j} U^{(r)}_{j,j+1}(\theta).
\end{align}
$U_D^{(r)}(\theta)$ commutes with $Z_j$ for all $\theta$, and so the action of $U_D$ on the Hamiltonian can be deduced from the action on the $X_j$. We have 
\begin{align}
&U^{(r)}_{j,j+1}(-2\pi/N )X^{}_j U^{(r)}_{j,j+1}(2\pi/N)
\nonumber\\
&= \sum_{a_j,a_{j+1}}  \omega^{a_{j+1}-a_j -r-\frac{N-1}{2} }\ket{a_j-1,a_{j+1}}\bra{a_j,a_{j+1}}\nonumber\\
&= \omega^{-\frac{N-1}{2} } \omega^{-r-1}Z^{-1}_{j} Z_{j+1}^{} X_{j}\ .
\end{align}
To derive this, we use the identity
\begin{align}
&\sum_{m} \omega^{mr} \alpha_m^{} \alpha_{-m} \Big(\omega^{-m(a_{j+1} - (a_j-1))} - \omega^{-m(a_{j+1} - a_j)} \Big) \nonumber\\&=  \sum_{m} \alpha_m^{} \Big( \omega^{-m(a_{j+1} - a_j-r)} \Big)
\end{align}
and then simplify using \cref{eq:trig}. Then, since inversion of the chain exchanges $\tilde{H}_D^{(r)}$ and $\tilde{H}_D^{(-r)}$ we deduce \begin{align}
U^{(r)}_{j,j+1}&(-2\pi/N )X_{j+1} U^{(r)}_{j,j+1}(2\pi/N)\nonumber\\
&= \omega^{-\frac{N-1}{2} } \omega^{r-1}Z^{}_{j} Z^{{-1}}_{j+1} X_{j+1}.
\end{align}
Hence, \begin{align}U^{(r)}(-2\pi/N)X_jU^{(r)}(2\pi/N) = Z_{j-1}^{}Z_j^{-1} X_j Z_j^{-1} Z^{}_{j+1}\end{align} while \begin{align}U^{(r)}(2\pi/N)X_jU^{(r)}(-2\pi/N) = Z_{j-1}^{{-1}}Z_j^{} X_j Z_j^{} Z^{{-1}}_{j+1}\ .\end{align} This gives the result.
\subsubsection{$U(1)$ symmetries}\label{app:u1}
First, let us write $h_{2j-1} = X_j$ and  $h_{2j} = Z_j^{-1}Z_{j+1}^{}$, such that $\omega h_k h_{k+1}=h_{k+1}h_k$.
Consider the commutator 
\begin{align}
  & [\tilde{H}_D^{(0)},h_{2j-2}^k h_{2j-1}^{} h_{2j}^{-k}] \nonumber \\
  &= h_{2j-2}^k\Big(\sum_m\alpha_m\alpha_{-m}(h_{2j}^mh_{2j-1}^{}-h_{2j-1}^{}h_{2j}^m \nonumber\\&\qquad+h_{2j-2}^mh_{2j-1}^{}-h_{2j-1}^{}h_{2j-2}^m)\Big)h_{2j}^{-k} \nonumber\\
&=  \sum_{m} \alpha_m \Big(Z_{j-1}^{-k}Z_j^{k-m} X_j^{}  Z_j^kZ_{j+1}^{-k+m} \nonumber\\& \qquad\qquad \qquad\qquad+Z_{j-1}^{-k+m}Z_j^{k-m} X_j Z_j^kZ_{j+1}^{-k}\Big)\ .
\end{align}
We can then sum over $k$ and change variables in the second sum to find $[\tilde{H}_D^{(0)},\sum_{k=0}^{N-1}Z_{j-1}^{-k}Z_j^k X_j^{} Z_j^kZ_{j+1}^{-k}] $ is equal to
 \begin{align}
 \sum_{k=0}^{M-1}\sum_{m=1}^{N-1} \alpha_m\omega^k \Big(Z_{j-1}^{-k}&Z_j^{2k-m} X_j^{}  Z_{j+1}^{-k+m} \nonumber\\&+ Z_{j-1}^{-k+m}Z_j^{2k-m} X_j Z_{j+1}^{-k}\Big)\ .\end{align}
 This in turn equals
 \begin{align}
\sum_{k=0}^{M-1}\sum_{m=1}^{N-1} \omega^k(\alpha_m+\alpha_{-m}\omega^{-m}) &Z_{j-1}^{-k}Z_j^{2k-m} X_j^{}  Z_{j+1}^{-k+m}\ ,
\end{align}
which vanishes 
since $\alpha_m+\alpha_{-m}\omega^{-m}=0$. Hence, $\tilde{H}_D^{(0)}$ commutes with the Hamiltonian $H_S=\sum_{k=0}^{N-1} H_D^{(k)}$. Moreover, since this Hamiltonian has $\mathbb{Z}_N\times\mathbb{Z}_N$ symmetry generated by $D$ and $Q$, and $\tilde{H}_D^{(r)}=D^{-r}\tilde{H}_D^{(0)}D^r$, each of the $\tilde{H}_D^{(r)}$ are also symmetries of $H_S$. Since $\sum_r \tilde{H}_D^{(r)}=0$, we have that the group $\mathbb{Z}_N^Q\times(U(1)^{N-1}\rtimes \mathbb{Z}_N^D)$ commutes with $H_S$. This Hamiltonian also has discrete $C$, $T$ and $P$ symmetries, where these act non-trivially on other symmetry generators.
\subsection{Cluster SPT}
To understand the cluster entangler, we study \begin{align}
U_C^{(0)}(\theta)= \rme^{-\rmi \theta \tilde{H}_C^{(0)}} =\prod_{j} V_{j,j+1}(\theta)\ .
\end{align}
Consider first even sites, $j$, then we have
\begin{align}
V_{j,j+1}(2\pi/N) X_j& V_{j,j+1}(-2\pi/N)\nonumber\\&=U^{(0)}_{j,j+1}(-2\pi/N )X^{}_j U^{(0)}_{j,j+1}(2\pi/N)\nonumber\\
&=\omega^{-\frac{N-1}{2} } \omega^{-1}Z^{-1}_{j} Z_{j+1}^{} X_{j}\\
V_{j-1,j}(2\pi/N) X_j& V_{j-1,j}(-2\pi/N)\nonumber\\&=U^{(0)}_{j-1,j}(2\pi/N )X^{}_j U^{(0)}_{j-1,j}(-2\pi/N)\nonumber\\
&=\omega^{\frac{N-1}{2} } \omega^{}Z^{-1}_{j-1} Z_{j}^{} X_{j}\ ,
\end{align}
so that \begin{align}U_C^{(0)}(2\pi/N)X_{2j} U_C^{(0)}(-2\pi/N) = Z_{2j-1}^{-1}X_{2j} Z_{2j+1} \ . \end{align} An analogous calculation gives \begin{align}X_{2j-1}\rightarrow Z_{2j-2}^{}X_{2j-1}^{} Z_{2j}^{-1} \ ,\end{align} and so $U_C^{(0)}$ is an entangler for the cluster model. 

Since each of the Hamiltonians $H_C^{(k)}$ is invariant under $Q_\mathrm{even}$, we have that 
\begin{align}
H_C^{(k)} &= Q_\mathrm{even}^{-r} H_C^{(k)} Q_\mathrm{even}^r \nonumber\\&= Q_\mathrm{even}^{-r} U_C^{(0)}(2\pi k/N)H_C^{(0)}U_C^{(0)}(-2\pi k/N)  Q_\mathrm{even}^r  \nonumber\\
&= U_C^{(r)}(2\pi k/N)H_C^{(0)}U_C^{(r)}(-2\pi k/N) \ ,
\end{align}
and hence $\tilde{H}_C^{(r)}$ are pivot Hamiltonians for the cluster SPT.
\section{No dual Dolan-Grady for the Potts model}\label{app:DG}
In this appendix we show that the dual Dolan-Grady relation, \Cref{eq:dualDG}, does not hold for the Potts model $A_0+\overline{A}_0$ and the Onsager ferromagnet $A_1$. Hence, there is no normalisation of $A_0+\overline{A}_0$ that, together with $A_1$, generates an Onsager algebra.

As well as $h_j$ defined above, we also write $\alpha_{a,\ah}=\alpha_a(1-\omega^{a\ah})$ as in Ref.~\onlinecite{Jones25}, and use a number of results from Appendix A of that paper.
In particular $\big[A_1,\,A_0 \big]$ is equal to
\begin{align}
-N^{-2} \sum_{j=1}^L \sum_{a,\hat{a}=1}^{N-1} \alpha_{a,\ah}\alpha_\ah\,\big({h}_{2j-1}^a {h}_{2j}^\ah - {h}_{2j}^\ah {h}_{2j+1}^a\big) \end{align}and $\Big[\big[A_1,\,A_0 \big],A_0\Big]$ is given by
\begin{align}
 -N^{-3}  &\sum_{j=1}^L\Bigg( -2\sum_{a,\hat{a},b=1}^{N-1} \alpha_{a,\ah}\alpha_\ah\alpha_{b,\ah}\,{h}_{2j-1}^a {h}_{2j}^\ah  {h}_{2j+1}^b\nonumber\\
&+
\sum_{a,\hat{a}=1}^{N-1}\alpha_{a,\ah}\alpha_\ah\,(N-2\ah) \big({h}_{2j-1}^a {h}_{2j}^\ah + {h}_{2j}^\ah {h}_{2j+1}^a\big)\nonumber\\
&+2\sum_{\ah=1}^{N-1} \ah(N-\ah)\alpha_\ah {h}_{2j}^\ah\Bigg)\ . \label{eq:triplecommutator}
\end{align}
It is important to note that in $[A_1,A_0]=\overline{[A_1,\overline{A_0}]}$, each term contains products of two $h_k$ operators. If \cref{eq:dualDG} holds for some $\gamma$, then all products of three $h_k$ operators in 
\begin{align}
  & \Big[\big[[A_1,A_0],{A}_0\big],\overline{A}_0\Big]
+\Big[\big[[A_1,A_0],\overline{A}_0\big] ,\overline{A}_0\Big] \nonumber\\&= \Big[\big[[A_1,A_0],{A}_0\big],\overline{A}_0\Big]+ \overline{\Big[\big[[A_1,A_0],{A}_0\big],\overline{A}_0\Big]} \label{eq:quadruple}
\end{align}
must cancel.

We will find the coefficient of the term ${h}_{2j-1}^{} {h}_{2j}^{} h_{2j+1}^{N-1}$ in  \cref{eq:quadruple}. To do this, we can ignore the last term in \cref{eq:triplecommutator}, which will generate products of two $h_k$. Then
\begin{widetext}
\begin{align}
\Bigg[\Big[\big[A_1,\,A_0 \big],A_0\Big] ,\overline{A}_0\Bigg]&=N^{-4}  \sum_{j=1}^L\Bigg( 2\sum_{a,\hat{a},b,d=1}^{N-1} \alpha_{-d}(1-\omega^{\ah d})\alpha_{a,\ah}\alpha_\ah\alpha_{b,\ah}\,({h}_{2j-1}^{a+d} {h}_{2j}^\ah  {h}_{2j+1}^b-{h}_{2j-1}^{a} {h}_{2j}^\ah  {h}_{2j+1}^{b+d})\cr
&-
\sum_{a,\hat{a},d=1}^{N-1} \alpha_{-d}(1-\omega^{\ah d})\alpha_{a,\ah}\alpha_\ah\,(N-2\ah) \big(h_{2j-1}^d  {h}_{2j}^\ah {h}_{2j+1}^a-{h}_{2j-1}^a {h}_{2j}^\ah h_{2j+1}^d\big)\Bigg)+ \dots \ .\label{eq:quadruple2}
\end{align}
The coefficient, $c_1$, of ${h}_{2j-1}^{} {h}_{2j}^{} h_{2j+1}^{N-1}$ in \cref{eq:quadruple2} is given by
\begin{align}
   N^{4}c_1 &= 2 \left( \sum_{a=2}^{N-1} \alpha_{-(N+1-a)} (1-\omega^{(1-a)})\alpha_{a,1}\alpha_1 \alpha_{N-1,1} -  \sum_{b=1}^{N-2} \alpha_{-(N-1-b)} (1-\omega^{-(1+b)})\alpha_{1,1}\alpha_1 \alpha_{b,1}\right)\nonumber\\
&\qquad - (N-2)\left(\alpha_{-1}(1-\omega)\alpha_{N-1,1}\alpha_1 - \alpha_{-(N-1)}(1-\omega^{-1})\alpha_{1,1}\alpha_1 \right) = -N(1+\omega^{-1});\label{eq:c1}
\end{align}\end{widetext}
the coefficient $c_{N-1}$ of ${h}_{2j-1}^{} {h}_{2j}^{N-1} h_{2j+1}^{N-1}$ in \cref{eq:quadruple2} is computed similarly and we find $c_{N-1} = \overline{c}_1$. From this we deduce that the coefficient of ${h}_{2j-1}^{} {h}_{2j}^{} h_{2j+1}^{N-1}$ in \cref{eq:quadruple} is given by $c_1 + \overline{c}_{N-1}= -2N^{-3}\left(1+\omega^{-1}\right)$. \cref{eq:c1} holds for $N>2$ and is non-vanishing. Hence, \cref{eq:dualDG} cannot be satisfied for $N>2$.

\end{document}